\documentclass[reprint,showpacs,twocolumn,superscriptaddress,aps,nofootinbib]{revtex4-2}

\usepackage{amsmath,amsthm,amssymb}
\usepackage{graphicx}  
\usepackage{xcolor}
\usepackage{mathtools}
\usepackage{xfrac}
\usepackage{bbm}
\usepackage{subfigure}
\usepackage{dcolumn}
\usepackage{braket}
\usepackage{lipsum}
\usepackage{bm}
\usepackage{comment}
\usepackage{wasysym}
\usepackage{pifont}
\usepackage{soul}
\usepackage{dsfont}
\usepackage{physics}
\usepackage[normalem]{ulem}

\usepackage{tikz}
\usepackage{multirow}
\usepackage{makecell}

\newcommand{\avg}[1]{\left\langle #1 \right\rangle}
\newcommand{\R}[1]{{\color{red} #1}}

\newcommand{\B}[1]{{ #1}}

\definecolor{newblue}{RGB}{0,145,232}

\definecolor{dodgerblue}{HTML}{1E90FF}
\usepackage[colorlinks=true,citecolor=dodgerblue,linkcolor=dodgerblue,urlcolor=dodgerblue,pdftitle={loops}]{hyperref}

\begin{document}

\title{Stability and Loop Models from Decohering Non-Abelian Topological Order}

\author{Pablo Sala}%
\email{psala@caltech.edu}
\affiliation{Department of Physics and Institute for Quantum Information and Matter, California Institute of Technology, Pasadena, CA 91125, USA}
\affiliation{Walter Burke Institute for Theoretical Physics, California Institute of Technology, Pasadena, CA 91125, USA}
\author{Ruben Verresen}
\email{verresen@uchicago.edu}
\affiliation{Pritzker School of Molecular Engineering, University of Chicago, Chicago, IL 60637, USA}
\affiliation{Department of Physics, Harvard University, Cambridge, MA 02138, USA}
\affiliation{Department of Physics, Massachusetts Institute of Technology, Cambridge, MA 02139, USA}

\date{\today}

\begin{abstract}
Decohering topological order (TO) is central to the many-body physics of open quantum matter and decoding transitions.
We identify statistical mechanical models for decohering non-Abelian TO, which have been crucial for understanding the error threshold of Abelian stabilizer codes.
The decohered density matrix can be described by loop models, whose topological loop weight $N$ is the quantum dimension of the decohering anyon---reducing to the Ising model if $N=1$.
In particular, the R\'enyi-$n$ moments correspond to $n$ coupled O$(N)$ loop models. Moreover, by diagonalizing the density matrix at maximal error rate, we connect the fidelity between two logically distinct ground states to random O$(N)$ loop and spin models.
We find a remarkable stability to quantum channels which proliferate non-Abelian anyons with large quantum dimension, with the possibility of critical phases for smaller dimensions. 
Intuitively, this stability is due to non-Abelian anyons not admitting finite-depth string operators.
We confirm our framework with exact results for Kitaev quantum double models, and with numerical simulations for the non-Abelian phase of the Kitaev honeycomb model.
Our work opens up the possibility of non-Abelian TO being robust against maximally proliferating certain anyons, which can inform error-correction studies of these topological memories.
\end{abstract} 

\maketitle

Topological order (TO) \cite{Wen_book,Sachdev_2023} is of significant interest due to its `anyonic' quasiparticles with generalized exchange statistics \cite{Leinaas_77,Goldin_81,Wilczek_82}. This leads to a topological ground state degeneracy \cite{Einarsson90} within which one can store quantum information, and anyon braiding enacts logical gates \cite{Kitaev_2003,Freedman_2000gwh, Freedman_2006,Nayak_08,Terhal_15}. For \emph{Abelian} TO (where anyons have no internal structure), the stability of the quantum memory below a decoherence threshold is well-understood in terms of an effective statistical mechanical (`stat-mech') model \cite{Dennis_2002}.
This error-correction transition was recently revisited from the perspective of various order parameters and higher moments of the density matrix
\cite{fan2023diagnostics,bao2023mixedstate,LeeYouXu2022,chen2023separability, Renorm_QECC_23,wang2023intrinsic,Mong_24,ellison2024classificationmixedstatetopologicalorders,sohal_24,Mong_24, tapestry_24,chen2024unconventional,Hauser_24,lyons24,2024_sala_SSSB,Markov_length_24,TshungCheng_24,lee2024exactcalculationscoherentinformation}.

However, there is a richer landscape of \emph{non-Abelian} TO \cite{Goldin_85,Wen_91_FQH,MOORE1991362, Moore1989}, where (i) anyons host a topological degeneracy (the quantum dimension $d$ of the anyon) within which one can store information, and (ii) braiding can even implement a universal set of gates in certain TOs \cite{Kitaev_2003,Mochon03,Mochon04,Freedman_2000gwh,Freedman_2006,Nayak_08}. Error-correction thresholds have been explored for certain non-Abelian states~\cite{Wootton_14,Brell_14,Wootton_16,Burton_2017,Dauphinais_2017,Schotte_22a,Schotte_22b}, but stat-mech models were left as an open question. More recently, studies of mixed-state non-Abelian TO consider either \emph{Abelian} anyon channels \cite{sohal_24,Mong_24,ellison2024classificationmixedstatetopologicalorders} or new \emph{classical} phases which can emerge \cite{Mong_24,ellison2024classificationmixedstatetopologicalorders}. Fundamental questions thus remain, which are timely since non-Abelian TOs are being explored in quantum platforms, such as a 27-qubit $D_4$ TO \cite{iqbal2023creation} and a 9-qubit Fibonacci TO \cite{minev24}.

The first key result of our work is that stat-mech `loop models' \cite{Nienhuis_81,Nienhuis_82,peled2019lectures} provide a natural description for the trajectories of decohering non-Abelions in two dimensions. The decoherence rate corresponds to the string tension of the loops, and the TO phase is destroyed when loops proliferate---akin to anyon condensation \cite{Bais09,Burnell_2018}. In particular, if we incoherently poison a TO with anyons $a$, loop models arise if the fusion $a \times \bar a$ does not contain $a$ itself; in the more general case we can expect `net' models consistent with fusion rules (see our companion work for a case study \cite{long_paper}).
The second key result is that by utilizing the literature on loop models, we find that non-Abelian TOs are remarkably robust against decoherence. In particular, if the quantum dimension of the decohering anyon is sufficiently large, we find that the TO can \emph{persist even to the maximal-decoherence limit}.

\begin{figure}
    \centering
    \begin{tikzpicture}
    \node at (0,0){
    \includegraphics[scale=0.36]{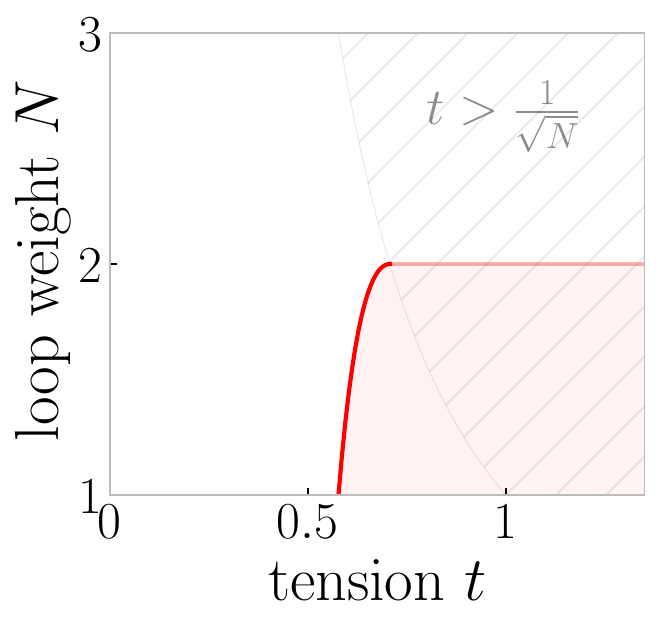}
    };
    \node[opacity=1] at (-0.5,0.4){
    \includegraphics[scale=0.18]{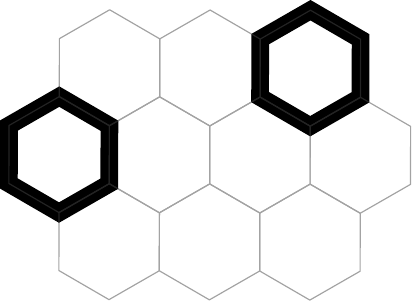}
    };
    \node[opacity=1] at (1.1,-0.4){
    \includegraphics[scale=0.18]{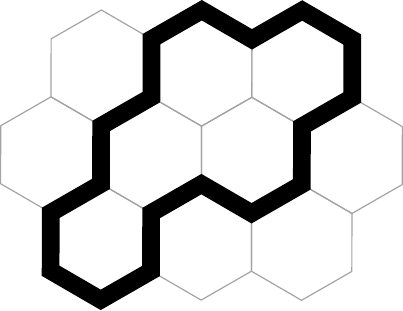}
    };
    \node at (4.1,0){
    \includegraphics[scale=0.36]{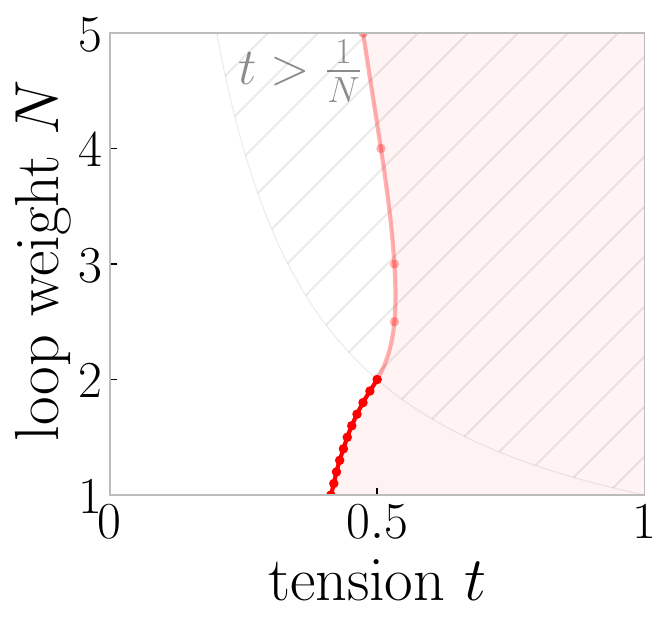}
    };
    \node[opacity=1] at (3.55,0.45){
    \includegraphics[scale=0.15]{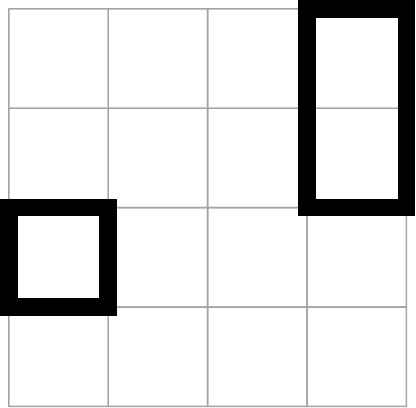}
    };
    \node[opacity=1] at (5.25,0.45){
    \includegraphics[scale=0.15]{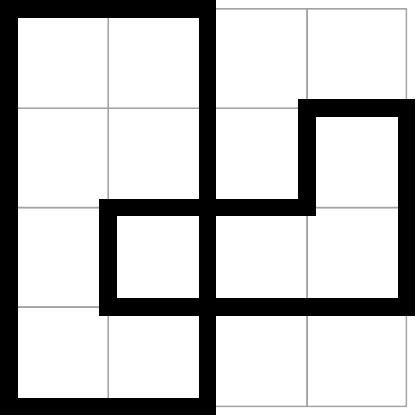}
    };
    \node at (-1,1.4) {(a)};
    \node at (3,1.4) {(b)};
    \node at (0,-3) {\includegraphics[scale=0.55]{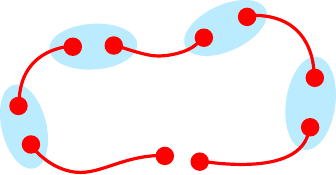}};
    \node at (-1.6,-2.2) {(c)};
    \node[red] at (0.2,-2.9) {$a \times a = 1$};
    \node[red] at (1.8,-2.85) {$a$};
    \node[newblue] at (1.8,-3.15) {$\times$};
    \node[red] at (1.8,-3.45) {$a$};
    \node[newblue,right] at (2,-3.15) {$=1 \; + \; \cdots$};
    \node[newblue] at (4.5,-3.5) {\footnotesize weight $\sfrac{1}{d_a}$};
    \draw[->,newblue] (3.8,-3.15) -- (5.2,-3.15) node[right] {1};
    \node at (4,-2.5) {$\bra{\psi} \prod_{n \in L} U_n \ket{\psi} = \frac{d_a}{d_a^{|L|}}$};
    \end{tikzpicture}
    \caption{\textbf{Loops and anyons.} (a) Phase diagram of the O$(N)$ loop model on the honeycomb lattice (Eq.~\eqref{eq:loopmodel}) \cite{Nienhuis_82}: loops proliferate in the red shaded region; a local spin model representation is known outside the hashed region for integer $N$. (b) Similarly for the square lattice, where loops can have even-degree intersections \cite{Deng_2007}; $C(L)$ in Eq.~\eqref{eq:loopmodel} is the number of faces enclosed by the graph $L$, also known as the cyclomatic number \cite{Essam70}. (We do not show additional transitions within the proliferated phase \cite{Chayes_2000,Deng_2007}.) (c) Loop weights naturally arise for wavefunction overlaps \eqref{eq:f_nonAbelian} of non-Abelian anyon pairs along a loop $L$, leading to a loop weight given by the quantum dimension.}
    \label{fig:loops}
\end{figure}
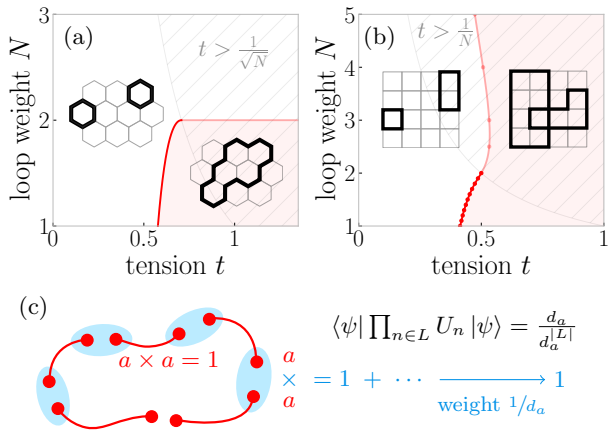

\textbf{Loop models.} Let us briefly review the salient features of O$(N)$ loop models. These are stat-mech models of closed loops:
\begin{equation} \label{eq:loopmodel}
       \mathcal Z(t;N) = \sum_{\textrm{loops } L} t^{|L|} N^{C(L)}
\end{equation}
where $t \in \mathbb R^+$ is the string tension for a loop of length $|L|$, with an additional topological loop weight $N \in \mathbb R^+$ for the number of loops $C(L)$. On a trivalent lattice, the latter is simply the number of components, since intersections do not arise. While $N=1$ is equivalent to the Ising model~\cite{SM}, this model has been studied for continuous $N$. For $0 \leq N \leq 2$ it has a transition for a string tension $t_c^{-1} = \sqrt{2+\sqrt{2-N}}$ on the honeycomb lattice \cite{Nienhuis_82}, beyond which it enters the large-loop phase (with algebraic string correlations for $1<N \leq 2$ \cite{Nienhuis_84}), whereas for $N>2$ loops never proliferate. We show the phase diagram in Fig.~\ref{fig:loops}, including a square lattice variant, where `loops' can now intersect and can be thought of as Eulerian graphs (i.e., all vertices have even degree) \cite{BLOTE19841,Chayes_2000,Guo06,Deng_2007}. Proliferation on the square lattice is possible since for lattices where loops can cross, having many components can coexist with having many occupied bonds.
As we will discuss, decohering non-Abelions with quantum dimension $d$ naturally leads to effective loop models with weight $N=d$. We note that proliferated phases of loop and net models have been used to build deconfined quantum phases \cite{Kitaev_2003,Freedman2003,Freedman04,Freedman05a,Freedman05b,Fidkowski06,Fendley_2008,Troyer08,Fendley13}.
Here, instead, loops (or nets) will describe the \emph{errors} of the TO, i.e., the worldlines of the decohering anyons.

\textbf{Set-up.} In this work we consider an initial state $\ket{\psi}$ which
randomly experiences a unitary error $\ket{\psi} \to U_n \ket{\psi}$ for any site $n$ (for simplicity we consider a site-independent unitary). We follow Ref.~\onlinecite{Dennis_2002} and call the subset of affected sites the `error chain' $E$, with the corrupted state $\ket{E} = \prod_{n \in E} U_n \ket{\psi}$. If the error chain $E$ occurs with probability $p_E$, the state devolves into a mixture:
    \begin{equation}
     \rho = \mathcal E(\ket{\psi} \bra{\psi} ) = \sum_{E} p_E \ket{E} \bra{E}. \label{eq:rho}
    \end{equation}
It would be easy to ascertain which error occurred if the states $\{ \ket{E} \}$ were orthogonal, which is however not the case.
We will find that the overlap $\langle E|E'\rangle$ between error-corrupted states carries great physical significance. We will take\footnote{Physically, it will create anyons which are their own antiparticle, i.e., $a \times a = 1 + \cdots$.} $U_n^2=1$, i.e., $U_n^\dagger = U_n$, such that the overlap depends only on the symmetric difference\footnote{To wit, the symmetric difference of two sets contains all elements which are only in \emph{one} of the two sets.} of $E$ and $E'$:
\begin{equation} \label{eq:fL}
f(E \oplus E') \equiv \langle E | E' \rangle = \bra{\psi} \prod_{n \in E \oplus E'} U_n \ket{\psi},
\end{equation}
where $\oplus$ is the symmetric difference.

\textbf{Abelian TO.} As a warm-up, we illustrate the importance of the overlap function \eqref{eq:fL} for Abelian topological order. A broad class is that of CSS codes, which have Pauli stabilizers of the form $Z^{\otimes n}$ or $X^{\otimes n}$ \cite{CS_96,S_96,nielsen2010quantum}, such as the toric code \cite{Kitaev_2006}. For CSS states, a Pauli matrix $U_n \in \{ Z_n, X_n\}$ creates a pair of Abelian anyons, and more generally $\ket{E}$ supports anyons at the endpoints $\partial E$ \cite{Dennis_2002}.
Naturally, $\ket{E}$ and $\ket{E'}$ are indistinguishable if and only if the error chains have the same boundary $\partial E = \partial E'$, where we work on a planar geometry\footnote{This way we do not have to distinguish topological sectors, but the discussion extends straightforwardly.}. The overlap \eqref{eq:fL} is thus an indicator function for closed loops $L$; more precisely, denoting $L \equiv E \oplus E'$,
    \begin{equation} \label{eq:f_Abelian}
    f_{\rm CSS}(L) = \bra{\psi} \prod_{n \in L} U_n \ket{\psi} = \left\{ \begin{array}{cl}
    1 & \textrm{if }\partial L = 0, \\
    0 & \textrm{if }\partial L \neq 0.
    \end{array} \right.
    \end{equation}
Hence, Eq.~\eqref{eq:rho} is diagonalized \cite{LeeYouXu2022} by simply grouping together all states with the same endpoints: 
    \begin{equation} \label{eq:Z_a}
        \rho = \sum_{\textrm{anyons } \bm a} Z_{\bm a} \ket{\bm a} \bra{\bm a}, \quad Z_{\bm a} = \sum_{E: \partial E = {\bm a}} p_E.
    \end{equation} 
The probability $Z_{\bm a}$ of obtaining the anyon configuration ${\bm a}$ is a stat-mech model of strings ending on $\bm a$. As we increase the probability $p_E$, there is a proliferation transition of these strings. This can be detected by, e.g., a singularity in the von Neumann entropy, which equals the average free energy of this ensemble $\{ Z_{\bm a} \}$ of partition functions:
    \begin{equation}
    S = - \sum_{\bm a} Z_{\bm a} \ln Z_{\bm a} = [ F_{\bm a}], \quad F_{\bm a} = -\ln Z_{\bm a}.
    \end{equation}
For uncorrelated probabilities $p_E = p^{|E|} (1-p)^{N-|E|}$, the strings experience a tension $t= \frac{p}{1-p}$ (which is dual to a random-bond Ising model \cite{Dennis_2002} as reviewed in App.~\ref{app:from_ON_to_Ising}), and for the square lattice toric code these proliferate at $p_c \approx 0.109$ \cite{Dennis_2002}.

\textbf{Non-Abelian TO.} What if the initial state $\ket{\psi}$ is instead a non-Abelian TO, and $U_n$ creates two non-Abelian $a$ anyons with quantum dimension $d_a$?
Since anyons are orthogonal to the ground state, $\bra \psi U_n \ket \psi = 0$, and likewise for any error chain with unpaired non-Abelions.
Hence, for the overlap function \eqref{eq:fL} to be nonzero, $E \oplus E'$ must still form a closed loop $L$, so that the $a$'s pairwise annihilate (see Fig.~\ref{fig:loops}c).
However, the difference with the Abelian case is that the nonzero values are no longer just unity: the fusion $a \times a = 1 + b + \cdots$ only leads to a nonzero overlap with the ground state if we fuse into `1', which for two anyons that are independently created occurs with probability $\frac{1}{d_a^2}$ \cite{Shi_2020,Preskill_LN}, namely one over the number of possible outcomes of the fusion.
Hence, the wavefunction overlap \eqref{eq:fL} will incur a suppression factor $\frac{1}{d_a}$ for each pair of $a$ anyons we fuse, \emph{except} the last pair, since these are no longer causally disconnected.
In conclusion, the wavefunction overlap \eqref{eq:fL} is nonzero only for loop configurations $L$, and the result factorizes for each loop component $\ell$:
\begin{equation} \label{eq:f_nonAbelian}
    f(L) = \bra{\psi} \prod_{n \in L} U_n \ket{\psi} = \prod_{\ell \in L} \frac{1}{d_a^{|\ell|-1}} = \frac{d_a^{C(L)}}{d_a^{|L|}}
\end{equation}
where $C(L)$ is the number of loops\footnote{For trivalent lattices this is the number of components; more generally it is the number of bonds one has to cut to remove all loops, called the `cyclomatic number' \cite{Essam70}.}. Since the assumption of causal disconnectedness might only hold for a coarse-grained picture, it is more accurate to expect $f(L) = t_a^{|L|} d_a^{C(L)}$ for a non-universal tension $t_a$. The topological part ($d_a^{C(L)}$) is robust for large loops, which we will see is responsible for much of the resulting physics.

\textbf{Purity.} For the Abelian case, the overlap function \eqref{eq:f_Abelian} was simple enough to allow for a direct diagonalization of $\rho$. We do not expect this to be the case for general non-Abelian TO. However, the overlap \eqref{eq:f_nonAbelian} provides significant insight into $\rho$. Let us first consider the purity $\textrm{tr}(\rho^2)$. From Eq.~\eqref{eq:rho} and the above overlap, we directly find
\begin{equation} \label{eq:purity}
\textrm{tr}(\rho^2) = \sum_{E,E'} p_{E} p_{E'} f(E \oplus E')^2 = \sum_L \tilde p_L t_a^{2|L|} d_a^{2C(L)}
\end{equation}
where $\tilde p_L = \sum_E p_E p_{E \oplus L}$ contributes to the tension of the loop $L$. E.g., for uncorrelated probabilities, $\tilde p_L \propto \big( \frac{p-p^2}{p^2 - p + \sfrac{1}{2}}\big)^{|L|} $. More compactly, $\textrm{tr}(\rho^2) = \sum_L t_{\rm eff}^{|L|} d_a^{2C(L)}$, which takes the form of an O$(N)$ loop model \eqref{eq:loopmodel} with loop weight $N= d_a^2$. This means that for anyons whose worldlines live on the honeycomb lattice, the purity cannot undergo a phase transition if $d_a > \sqrt{2}$ (see Fig.~\ref{fig:loops}a), and can enter a critical phase for $1<d_a\leq \sqrt{2}$. In fact, even on the square lattice we would not expect a transition for large enough $d_a$, since $t_a$ is expected to rapidly decrease with $d_a$ (as in Eq.~\eqref{eq:f_nonAbelian}), not reaching the red region in Fig.~\ref{fig:loops}b. We will explicitly confirm this in a family of quantum double models where $t_a = \frac{1}{d_a}$ for integer $d_a$. This is in striking contrast to the Abelian case; e.g., for the square lattice toric code the purity undergoes a transition into a short-range phase at $p_c^{(2)} \approx 0.178$ \cite{fan2023diagnostics,LeeYouXu2022}.

\begin{figure}
    \centering
    \begin{tikzpicture}
    \node at (0,0) {\includegraphics[scale=0.33]{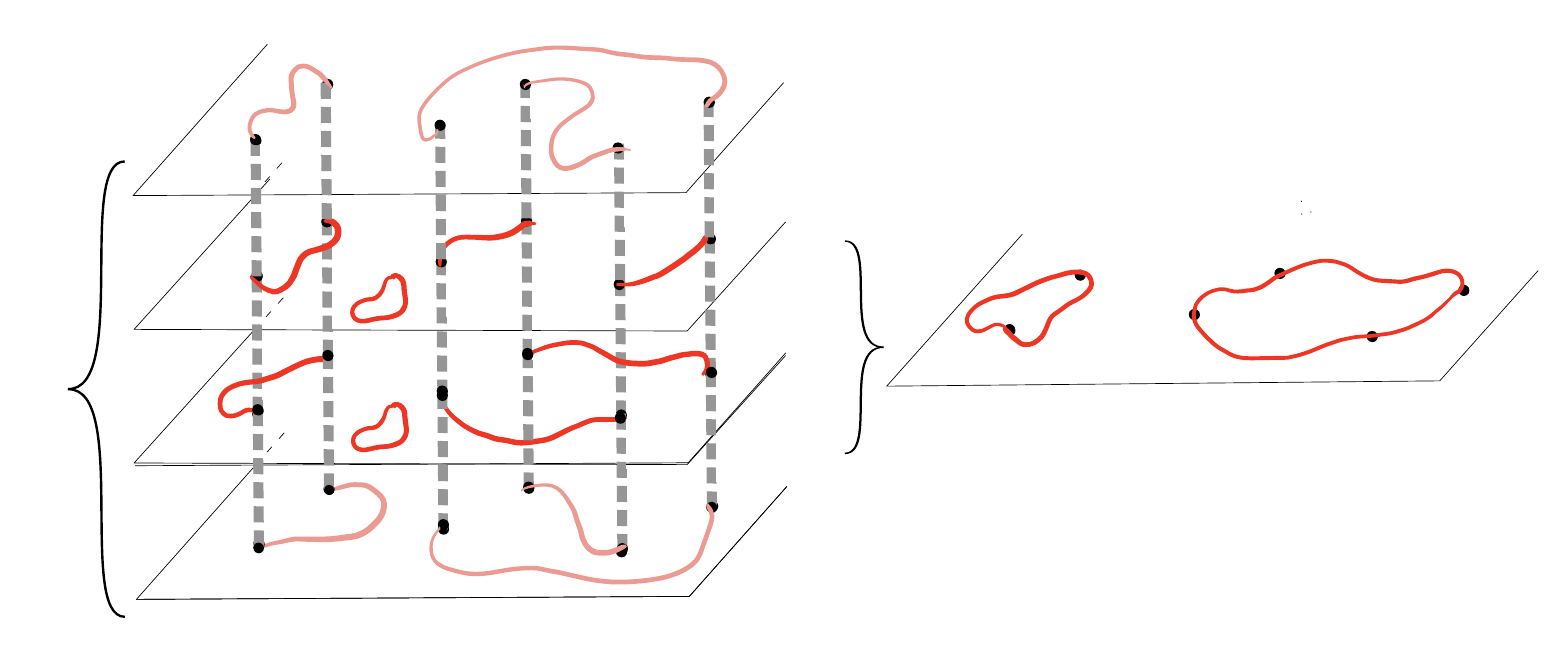}};
    \node at (-4.15,-0.38) {$n$};
    \node at (0,0.95) {$E^{(4)}$};
    \node at (0,0.23) {$E^{(3)}$};
    \node at (0,-0.55) {$E^{(2)}$};
    \node at (0,-1.3) {$E^{(1)}$};
    \node at (2.6,1.6) {$\vdots$};
    \node at (2.8,0.8) {$L^{(2)} = E^{(2)} \oplus E^{(3)}$};
    \node at (2.5,-0.7) {$\propto t_a^{|L^{(2)}|} \; d_a^{C(L^{(2)})}$};
    \node at (2.6,-1.35) {$\vdots$};
    \node at (-3.8,1.5) {(a)};
    \node at (1,1.5) {(b)};
    \end{tikzpicture}
    \caption{
    \textbf{Loop models from $\textrm{tr}(\rho^n)$.} (a) Higher moments of the density matrix are stat-mech models of strings \eqref{eq:tr_rhon}. Error chains $E^{(s)}$ are constrained such that $E^{(s)}\oplus E^{(s+1)}$ is a closed loop configuration $L^{(s)}$. Hence, their boundaries $\partial E^{(s)}$ need to match. (b) This results in $n$ coupled $O(d_a)$ loop models \eqref{eq:coupledloop} if we use the transitional variables $L^{(s)}$.
    }
    \label{fig:tr_rhon}
\end{figure}
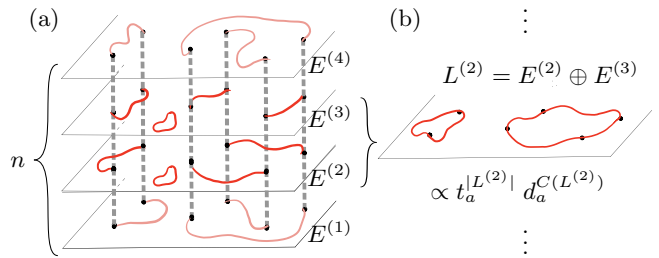

\textbf{Higher moments.} More generally, the moments $\textrm{tr}(\rho^n)$ are of interest \cite{fan2023diagnostics,bao2023mixedstate,LeeYouXu2022}. In fact, the full spectrum of $\rho$ can be reconstructed from these moments (for integer $n$) by Specht's theorem. Similarly to the purity, we can compute higher moments from Eq.~\eqref{eq:rho}:
\begin{equation} \label{eq:tr_rhon}
\textrm{tr}(\rho^n) \propto \sum_{ \{E^{(s)}\} } \prod_{s=1}^n p_{E^{(s)}} \prod_{s=1}^n f(E^{(s)} \oplus E^{(s+1)})
\end{equation}
where $E^{(n+1)} \equiv E^{(1)}$. The only nonzero contributions are where $L^{(s)} \equiv E^{(s)} \oplus E^{(s+1)}$ form closed loops (see Fig.~\ref{fig:tr_rhon}), leading to $n$ coupled $O(d_a)$ loop models:
\begin{equation} \label{eq:coupledloop}
\textrm{tr}(\rho^n) \propto \sum_{\{L^{(s)}\}} \tilde p_{\{L^{(s)}\}} \prod_{s=1}^n t_a^{|L^{(s)}|} d_a^{C({L^{(s)}})}.
\end{equation}
This is simplest to analyze in the maximal-decoherence limit, where $p_E$ is a constant, such that the only coupling is the global constraint $L^{(1)}\oplus L^{(2)} \oplus \cdots \oplus L^{(n)} = \emptyset$. While this is a strong coupling for small $n$, its effect diminishes as $n$ increases, suggesting that $n \to \infty$ is described by an O$(d_a)$ loop model, which we will momentarily prove via an explicit diagonalization. Hence, $\textrm{tr}(\rho^n)$ for large $n$ also fails to proliferate the decohering anyons if the quantum dimension $d_a$ is sufficiently large (e.g., $d_a>2$ for the honeycomb lattice; see Fig.~\ref{fig:loops}).

\textbf{Maximal decoherence.} Surprisingly, the above analysis suggests that non-Abelian TO can potentially persist even if we maximally decohere an anyon $a$ with quantum dimension $d_a$ of sufficient size. We can put this on firmer ground by diagonalizing $\rho$. If $p_E = \frac{1}{2^{\mathcal N}}$, Eq.~\eqref{eq:rho} is a mixture of random projected states:
\begin{equation} \label{eq:rho_eta}
\rho = \sum_{\bm \eta} \ket{\bm \eta} \bra{\bm \eta} \; \quad \textrm{with }\ket{\bm \eta} = \prod_{n} \frac{1 + \eta_n U_n}{2} \ket{\psi},
\end{equation}
where $\bm \eta = \{ \eta_n = \pm 1 \}$ is defined on all decohered sites. The projected states $\ket{\bm \eta}$ are orthogonal, such that the eigenvalues of $\rho$ are:
\begin{equation} \label{eq:Zeta}
Z_{\bm \eta} = \langle \bm \eta | \bm \eta \rangle = \frac{1}{2^{\mathcal N}} \sum_L \Big( \prod_{n \in L} \eta_n t_a \Big) \; d_a^{C(L)}.
\end{equation}
These are randomly signed O$(d_a)$ loop models, with the largest eigenvalue (describing $\textrm{tr}(\rho^n)$ for large $n$) indeed being the clean loop model. If $d_a$ is large enough to prevent anyons from proliferating in the largest eigenvalue, Eq.~\eqref{eq:Zeta} suggests \emph{all} eigenvalues likely remain in the short-loop phase since the destructive interference will not encourage proliferation. Alternatively, we note Eq.~\eqref{eq:Zeta} can be thought of as the high-temperature expansion of a disordered O$(n)$ spin model, which on the honeycomb lattice is exact for the interaction $H = \sum_{\langle i,j \rangle} \ln \left( 1 + \eta_{ij} d_a t_a \bm S_i \cdot \bm S_j \right) $ for a $d_a$-dimensional unit vector $\bm S_i$ \cite{Nienhuis_81,Nienhuis_82}. If the clean ferromagnetic (FM) model $\eta_{ij} = 1$ fails to order, it is suggestive that the random FM/AFM model will also remain in the disordered (i.e., short-loop) phase.

\textbf{Anyons, fidelity, and memory.} If all eigenvalues of the maximally-decohered state \eqref{eq:rho_eta} remain in the short-loop phase, can we conclude $\rho$ remains in the same TO phase as $\ket{\psi}$? We show this by using the eigenstates $\ket{ \bm \eta}$. Firstly, TO means anyonic excitations are orthogonal to the vacuum: if $\ket{\psi_{x,x'}}$ has $a$ anyons at locations $x$ and $x'$, then $|\langle \psi|\psi_{x,x'}\rangle| \sim e^{-|x-x'|/\xi}$. If $\rho = \mathcal E(\ket{\psi}\bra{\psi})$ and $\rho_{x,x'} = \mathcal E( \ket{\psi_{x,x'}} \bra{\psi_{x,x'}})$, one can show (see End Matter) that the quantum fidelity $| F(\rho, \rho_{x,x'})| \leq \big( \sum_{\bm \eta} |\langle \bm \eta| {\bm \eta}_{x,x'} \rangle| \big)^2$, where the left-hand-side holds for any $p_E$, and $\ket{ \bm{\eta}_{x,x'}}$ is defined as in Eq.~\eqref{eq:rho_eta}. Similar to the derivation of Eq.~\eqref{eq:f_nonAbelian}, we see that $\frac{|\langle \bm \eta| {\bm \eta}_{x,x'} \rangle|}{|\langle \bm \eta| {\bm \eta} \rangle|}$ is the expectation value of a string between $x$ and $x'$ in the (random) O$(d_a)$ loop model \eqref{eq:Zeta} (in the spin model this maps to a two-spin correlation function~\cite{Nienhuis_84}). In the short-loop phase, this decays exponentially with distance, such that $\rho$ and $\rho_{x,x'}$ remain orthogonal---the anyon fails to proliferate.

We can intuitively relate this stability to the fact that non-Abelian anyons do not admit tensor product string
operators \cite{Shi_2019}: even though at maximal decoherence one can freely insert (onsite) $U_n$, this cannot be used to generate the non-Abelian anyon string operators, unlike the more familiar Abelian case.
The same approach works for characterizing the non-Abelian quantum memory: if $\ket{\psi}$ and $\ket{\varphi}$ are two logical states on the torus, related by an $a$ anyon loop, the quantum fidelity of the decohered states must vanish if the expectation value of a large non-contractible loop vanishes. In summary, the non-Abelian quantum memory remains robust if the random O$(d_a)$ loop models \eqref{eq:Zeta} are in the short-loop phase.

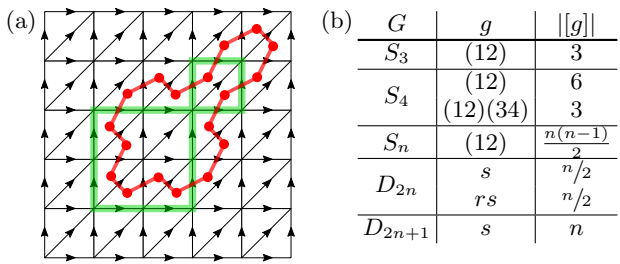
\begin{figure}
    \centering
\begin{tikzpicture}
\node at (0.25,-0.1){
\includegraphics[scale=0.33]{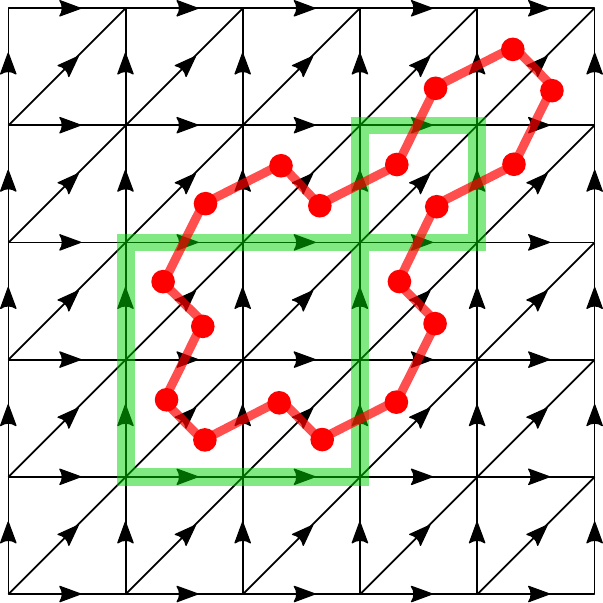}
};
\node at (4.5,0){ 
\begin{tabular}{ c|c|c } 
$G$ & $g$ & $|[g]|$ \\
\Xhline{2\arrayrulewidth}
\multirow{1}{*}{$S_3$} & $(12)$ & 3 \\ 
\hline
\multirow{2}{*}{$S_4$} & $(12)$ & 6 \\
 & $(12)(34)$ & 3 \\
\hline
\multirow{1}{*}{$S_n$} & $(12)$ & $\frac{n(n-1)}{2}$ \\ 
\hline
\multirow{2}{*}{$D_{2n}$} & $s$ &  $\sfrac{n}{2}$ \\ 
 & $rs$ &  $\sfrac{n}{2}$ \\ 
 \hline
\multirow{1}{*}{$D_{2n+1}$} & $s$ &  $n$ \\
\end{tabular}
};
\node at (-1.7,1.4) {(a)};
\node at (2.5,1.4) {(b)};
\end{tikzpicture}
\caption{\textbf{Decohering quantum doubles.}  (a) Example of a closed loop of flux anyons (red) on the triangular lattice, with its `shadow' (green) on the square lattice. (b) Decohering $g$-fluxes in a quantum double for group $G$ gives rise to loop models with loop weight given by the size of the conjugacy class $|[g]|$. We tabulate order-two elements: $(12)$ corresponds to a transposition; $r,s$ correspond to a rotation and mirror, respectively.}
\label{fig:quantumdoubles}
\end{figure}
\textbf{Quantum doubles.} As a first example of the above framework, we consider Kitaev's quantum double models \cite{Kitaev_2003}. These are spin models with an emergent gauge symmetry $G$, whose charges and fluxes have anyonic statistics due to the Aharonov-Bohm effect \cite{Aharonov59,Preskill_LN}.
The Hilbert space consists of qudits labeled by group elements $\{ \ket{g} | g \in G \}$ on the bonds of a lattice.
We consider the ground state $\ket \psi$ of the equal-weight superposition of all flux-free configurations, i.e., $B_p \equiv \prod_{n \in p} g_n^{s_n}=e$ for any plaquette $p$, where $s_n=\pm 1$ refers to bond orientations as in Fig.~\ref{fig:quantumdoubles}a. We choose\footnote{Such an element exists if and only if $|G|$ is even.} $g\in G$ with $g=g^{-1}$, and with probability $p$ on every site, we randomly apply $U = X_g$ defined as $X_g \ket{h} = \ket{gh}$. Note that a single $X_g$ creates two fluxes $B_p \neq e$ in the conjugacy class $[g]$, where $|[g]|$ is the corresponding quantum dimension \cite{Kitaev_2003}. To understand the resulting mixed state $\rho$, we first consider the wavefunction overlap \eqref{eq:fL}. The flux hops between plaquettes, and we find $f(L)$ is nonzero only for closed loops $L$ (red loop in Fig.~\ref{fig:quantumdoubles}a). The nonzero value coincides with Eq.~\eqref{eq:f_nonAbelian}, i.e., $f(L) = \frac{d^{C(\mathsf L)}}{d^{|\mathsf L|}}$ where $d = |[g]|$ and $\mathsf L$ is a `shadow' loop on the square lattice (green loop in Fig.~\ref{fig:quantumdoubles}a), \B{which is in 1-to-1 correspondence with a red loop}; see the End Matter for a derivation.

We can now apply our general results. The purity \eqref{eq:purity} is $\textrm{tr}(\rho^2) = \sum_{\mathsf{L}} t^{|\mathsf L|} N^{C(\mathsf L)}$ with\footnote{More precisely, this is for maximal decoherence $p=\sfrac{1}{2}$ on the diagonal bonds, otherwise $|L| \neq |\mathsf L|$.} $t=\frac{p-p^2}{(p^2 - p + \sfrac{1}{2})|[g]|^2}$ and $N = |[g]|^2$. This is the O$(N)$ loop model on the square lattice, so from Fig.~\ref{fig:loops}b we see the purity does not have a transition for any non-Abelian flux (i.e., $|[g]| \geq 2$). Moreover, for maximal decoherence, the largest eigenvalue \eqref{eq:Zeta} of $\rho$ is the $O(N)$ loop model with $N = |[g]|$ at tension $t=\frac{1}{|[g]|}$. While this is critical for $|[g]|=2$, it is in the short-loop phase for larger $|[g]|$ (Fig.~\ref{fig:loops}b). As discussed, all eigenvalues are then likely in the short-loop phase, indicating that if $|[g]| \geq 3$, the non-Abelian topological memory is robust to maximal $X_g$ noise! This applies, e.g., to $G=S_3$ (Fig.~\ref{fig:quantumdoubles}b). This can be numerically tested in future work by checking whether the random-bond $|[g]|$-state face-cubic spin model $\mathcal Z = \prod_{\langle i,j\rangle} \big( 1 + \eta_{ij} \bm S_i \cdot \bm S_j \big)$ is disordered since its exact high-temperature expansion coincides with Eq.~\eqref{eq:Zeta} \cite{BLOTE19841,Chayes_2000,Guo06,Deng_2007} (see in particular~\cite{SM}).

\begin{figure}
    \centering
    \begin{tikzpicture}
    \node at (0,0) {\includegraphics[scale=0.28]{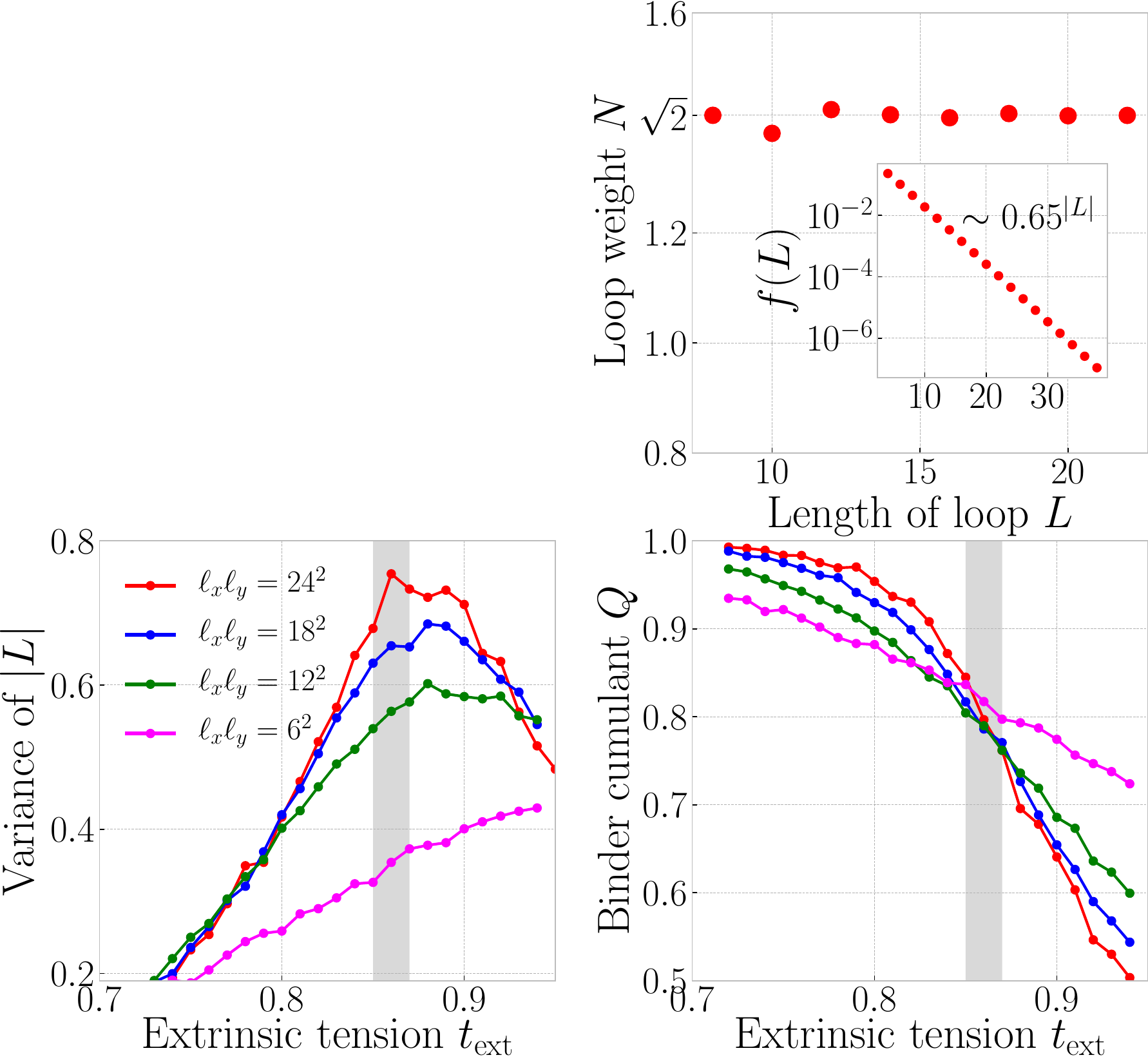}};
    \node at (-2.1,2.1) {\includegraphics[scale=0.33]{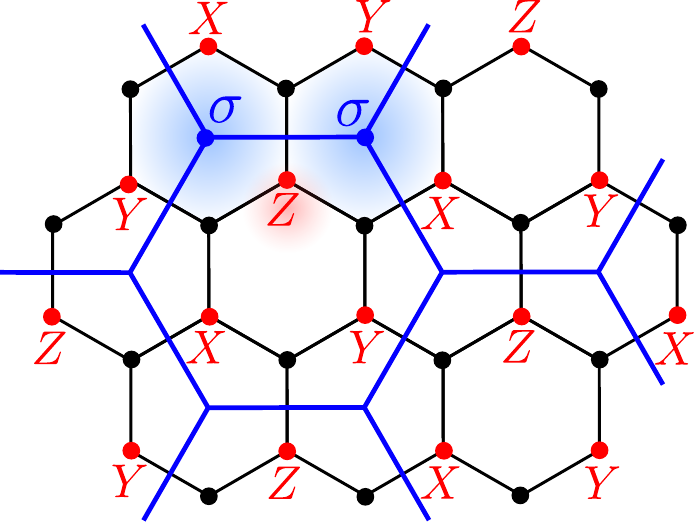}};
    \node at (-4.1,3.7) {(a)};
    \node at (0.2,3.7) {(b)};
    \node at (-4.1,0.1) {(c)};
    \node at (0.2,0.1) {(d)};
    \node[opacity=0.5] at (-3.75,3.35) {
        \begin{tikzpicture}
        \draw[-] (0,0) -- (0,0.3) node[right,xshift=-2]{\footnotesize $z$};
        \draw[-] (0,0) -- (0.866*0.3,-0.15) node[below,xshift=-1,yshift=1] {\footnotesize $y$};
        \draw[-] (0,0) -- (-0.866*0.3,-0.15) node[below,xshift=1,yshift=1] {\footnotesize $x$};
        \end{tikzpicture}
    };
    \end{tikzpicture}
    \caption{\textbf{Decohering non-Abelian gapped phase of Kitaev model.}  (a) Noise model (red) leading to super-honeycomb lattice (blue) on which non-Abelian visons hop. (b) Numerical validation of the scaling $|f(L)| \approx t_{\rm int}^{|L|} N^{C(L)}$ with $N=\sqrt{2}$ and $t_{\rm int}\approx 0.65$, \B{for $\kappa=0.2$}. Numerical results for the stat-mech model $\mathcal Z (t_{\rm ext}) = \sum_{L} t_{\rm ext}^{|L|} \; \left| f(L) \right|$ on a torus of size $\ell_x\ell_y$  are shown in panels (c) and (d) for the (normalized) variance of $|L|$ and the Binder cumulant $Q$, respectively.
    A similar analysis shows that the purity $\textrm{tr}(\rho^2)$ remains stable for all error rate $p$. 
    See additional details in Ref.~\cite{SM}.}
    \label{fig:main_Kitaev_mod}
\end{figure}
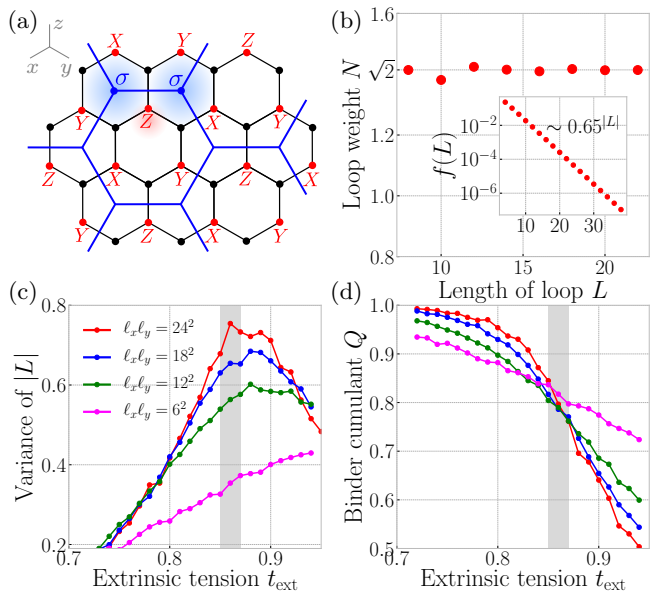

\textbf{Kitaev honeycomb model.} To test our framework for a non-bosonic anyon with non-integer quantum dimension and beyond fixed-point wavefunctions, we consider the chiral Ising anyon phase of the Kitaev honeycomb model \cite{Kitaev_2006}. We start with the free-fermion-solvable ground state of $H = \sum_{\alpha=x,y,z} \sum_{\langle i,j\rangle_{\alpha}} \sigma_i^\alpha \sigma_j^\alpha + \kappa\sum_{j,k,l} \sigma^x_j \sigma^y_k \sigma^z_l$ for $\kappa=0.2$. Here, a local Pauli, e.g., $\sigma^z$, creates two non-Abelian anyons (`visons') with $d=\sqrt{2}$ \cite{Kitaev_2006}. For this reason, we consider the pattern of Pauli decoherence shown in red in Fig.~\ref{fig:main_Kitaev_mod}a. This allows the vison to hop on a super-honeycomb lattice (shown in blue).

We numerically evaluate $f(L)$ \eqref{eq:fL} as a free-fermion determinant~\cite{SM}. In Fig.~\ref{fig:main_Kitaev_mod}b, we first validate its dependence on the topological weight $N^{C(L)}$ with $N=\sqrt{2}$. We extract this factor from loops $L$ (on the super-honeycomb lattice) of different lengths. The inset further shows that $f(L)$ decays exponentially with the length of the loop $L$: $|f(L)| \approx t_{\rm int}^{|L|} \sqrt{2}^{C(L)}$. Since $f(L)$ can be negative, \B{we first confirm in Figs.~\ref{fig:main_Kitaev_mod}(c,d) that $\mathcal Z (t_{\rm ext}) = \sum_{L} t_{\rm ext}^{|L|} \; \left| f(L) \right|$ is consistent with a O$(\sqrt{2})$ loop model} \B{where $t_{\rm ext} \in [0,1]$}. Using Monte Carlo simulations combined with Gaussian-states techniques, we numerically obtain the (normalized) variance $\textrm{Var}(|L|)$ of $|L|$ for various system sizes $\ell_x, \ell_y$; and the Binder cumulant $Q$ \B{reviewed in the supplemental material}~\cite{SM}. While the variance is not expected to diverge at this transition \cite{Liu_2011}, a non-analytic behavior is consistent with Fig.~\ref{fig:main_Kitaev_mod}c. 
More clearly, we find an approximate crossing in $Q$ for different system sizes for $t_{\textrm{ext}}\approx 0.86$ (Fig.~\ref{fig:main_Kitaev_mod}d), which signals a phase transition. \B{Consistently, the binder cumulant shows a transition at finite $t_{\textrm{ext}}$.} Similarly, the purity of the decohered state takes the approximate form $ \textrm{tr}(\rho^2) \approx  \sum_{L} t_{\rm eff}^{|L|} \sqrt{2}^{2C(L)}$, \B{$ t_{\rm eff}= t_{\rm int}  t_{\rm ext}$, and $ t_{\rm ext}=\frac{2p(1-p)}{p^2 + (1-p)^2}$}. Using the same method, we numerically find no phase transition for this O$(2)$ honeycomb loop model for any decoherence rate $p$~\cite{SM}, \B{as predicted assuming the expression for the critical tension as a function of the loop weight}.

\textbf{Conclusions and outlook.} We found that O$(N)$ loop models naturally describe the effect of decoherence on non-Abelian TO. Beyond exact results for Rényi-$n$ quantities, we found suggestive evidence of a remarkable stability of these non-Abelian quantum memories when anyons with large quantum dimension are proliferated (with the possibility of critical phases for smaller quantum dimension). This connects to random loop and spin models, providing a clear pathway for future work. \B{Moreover, this finding suggests that sometimes one only needs a smaller on-site Hilbert space to support the same TO.} We provided examples for quantum doubles and the Kitaev honeycomb model, with, e.g., the purity of the latter showing no transition, and quantum doubles such as for $G=S_3$ potentially having stable non-Abelian memory upon maximally decohering a flux of quantum dimension $3$---which we tied to the fate of a random-bond spin model via quantum fidelity measures. It would be exciting for future work to \B{generalize the quantum double construction to general lattices that allow for intersecting loops, and to} explore explicit error-correction protocols that can be informed by our stability results and stat-mech models. Moreover, in a companion work~\cite{long_paper} we apply this to $D_4$ TO which has been experimentally realized~\cite{iqbal2023creation}. It provides instances of net models consistent with fusion rules, as well as other loop models emerging when various types of anyons are simultaneously proliferated, a direction that is ripe for further exploration.

\textbf{Acknowledgments}
We are grateful to Jason Alicea for collaboration at the early stage of this project, and for his continued encouragement, as well as for collaboration on a closely related work~\cite{long_paper}. We also acknowledge helpful discussions and feedback from  Ehud Altman, Henrik Dreyer, Paul Fendley, Tarun Grover, Yujie Liu, Roger Mong, Lesik Motrunich, Benedikt Placke, John Preskill, Daniel Ranard, Ramanjit Sohal, Nathanan Tantivasadakarn and Sagar Vijay.
This work was partially conceived at the Aspen Center for Physics (P.S., R.V.), which is supported by National Science Foundation grant PHY-2210452 and Durand Fund. P.S. acknowledges support from the Caltech Institute for Quantum Information and Matter, an NSF Physics Frontiers Center (NSF Grant PHY-1733907), and the Walter Burke Institute for Theoretical Physics at Caltech.  
This research was supported in part by Perimeter Institute for Theoretical Physics. Research at Perimeter Institute is supported by the Government of Canada through the Department of Innovation, Science and Economic Development and by the Province of Ontario through the Ministry of Research, Innovation and Science.

\section*{End Matter}  \label{app:quantumdoubles}

\section{Decohering quantum doubles}
As stated in the main text we consider the ground state $\ket 0$ of the equal-weight superposition of all flux-free configurations, i.e., $B_p \equiv \prod_{n \in p} g_n^{s_n}=e$ for any plaquette $p$, where $s_n=\pm 1$ refers to bond orientations as in Fig.~\ref{fig:quantumdoubles}a
\begin{equation} \label{eq:gs_qd}
    \ket{0}\propto \sum_{\substack{\textrm{flux-free}\\ \{g_b\}}}\bigotimes_b\ket{g_b}.
\end{equation}
\B{ Here the proportionality factor is such that $\langle 0 | 0 \rangle = 1$.} 
A brief summary of the quantum double construction can be found in App.~\ref{app:Quant_doubles}.

Let us now subject this ground state to a decohering channel. In particular, we fix an order-$2$ non-trivial element of the group for which $g^2=e$, and on each bond $b$ of the system we apply 
\begin{equation}
    \mathcal{E}_b(\rho) = (1-p)\rho + p X_{g,b} \rho X_{g,b}.
\end{equation}
Here, we used the fact that when $g^2=e$, $X_g\equiv L^g_+$ becomes Hermitian ($X_g^\dagger=X_g$) and squares to the identity, and hence $X_g$ represents a ``Pauli'' error.  As in the main text, we denote $\ket{E}=\prod_{b\in E}X_{g,b}\ket{0}$ and $p_E=p^{|E|}(1-p)^{N-|E|}$ with $N$ the number of links on the triangular lattice. One finds that $f(E\oplus E') = \langle E|E'\rangle = \langle 0| \prod_{b\in E\oplus E'}X_{g,b}| 0\rangle$ is non-zero if and only if $L \equiv E\oplus E'$ is a closed loop. Consider a single triangle 
\begin{equation*}
\begin{tikzpicture}
     \draw[-] (-0.35,-0.3/1.46) -- (0.35,-0.3/1.46);
      \draw[->,line width=0] (-0.3,-0.3/1.46) -- (0.1,-0.3/1.46);
      \draw[-] (-0.35,-0.3/1.46) -- (0,0.41);
      \draw[->,line width=0] (-0.35,-0.3/1.46) -- (-.11,0.21);
      \draw[-](0.35,-0.3/1.46) -- (0,0.41);
      \draw[->,line width=0] (0.35,-0.3/1.46) -- (.11,0.21);
      \node at (0.,-.4) {$x$};
      \node at (0.34,0.26) {$y$};
      \node at (-0.34,0.26) {$z$};
\end{tikzpicture} 
\end{equation*}
for $x,y,z\in G$, which in the flux-free ground state satisfies $xy=z$. Suppose now that $X_g$ acts on a \emph{single} bond of this triangle (say, on the lower bond). Then for $\langle 0| \prod_{b\in E\oplus E'}X_{g,b}| 0\rangle$ to take a non-zero value, this configuration needs to satisfy the flux free condition given by $ (gx)y=z $, which together with the previous condition implies $g=e$. We thus see that $\langle E|E'\rangle$ can only get contributions from closed loops (where two bonds on a triangle are involved) or from branching (where all three bonds are involved).

We now show that the latter (i.e., branching) is not allowed when $g^2=e$. In the following we graphically represent the action of an error $E$ on a bond $b$ via a red line perpendicular to that bond. Hence branching on a triangle $p$ requires that three red lines meet at its center:
\begin{equation*}
\begin{tikzpicture}
     \draw[-] (-0.35,-0.3/1.46) -- (0.35,-0.3/1.46);
      \draw[->,line width=0] (-0.3,-0.3/1.46) -- (0.1,-0.3/1.46);
      \draw[-] (-0.35,-0.3/1.46) -- (0,0.41);
      \draw[->,line width=0] (-0.35,-0.3/1.46) -- (-.11,0.21);
      \draw[-](0.35,-0.3/1.46) -- (0,0.41);
      \draw[->,line width=0] (0.35,-0.3/1.46) -- (.11,0.21);
      \draw[-, red](0.,0.) -- (0.3,0.3);
      \draw[-, red](0.,0.) -- (-0.3,0.3);
      \draw[-, red](0.,0.) -- (0,-0.3);
      \node at (0.,-.5) {$\R{g}x$};
      \node at (0.47,0.41) {$\R{g}y$};
      \node at (-0.47,0.41) {$\R{g}z$};
\end{tikzpicture} 
\end{equation*}
Once again, for $\langle 0| \prod_{b\in L}X_{g,b}| 0\rangle$ to have a non-vanishing expectation value with respect to the ground state this configuration needs to satisfy the flux free condition (i.e., $B_p\prod_{b\in L}X_{g,b}| 0\rangle=\prod_{b\in L}X_{g,b}| 0\rangle$) given by $ (gx)(gy)=gz $, but together with $xy=z$ this leads to $gxgy = gxy$, i.e., $g=e$. Hence, we have derived that $\langle E|E'\rangle$ only takes values on closed loops on the honeycomb lattice (dual to the triangular lattice) without branching. 

\begin{figure}
    \centering
    \includegraphics[width=0.9\linewidth]{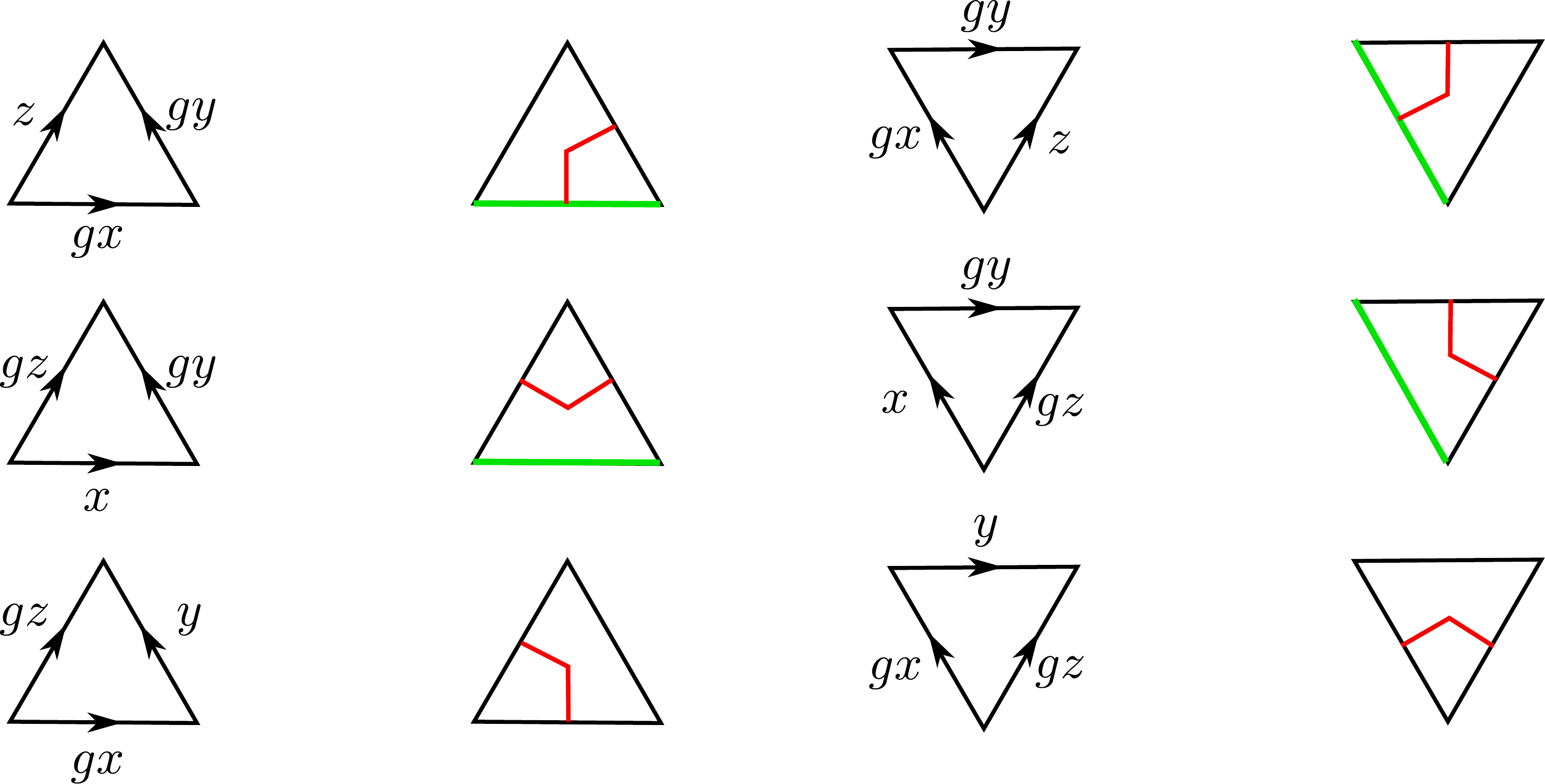}
    \caption{Triangle configurations leading to a non-zero overlap $\langle E|E'\rangle$.}
    \label{fig:conf_triang}
\end{figure}

Let us now derive when we can get nonzero contributions to $\langle E|E'\rangle$ from closed loop configurations, i.e., when two of the three bonds (of a triangle) are acted upon. There exist six possible such configurations (shown on the first and third columns of Fig.~\ref{fig:conf_triang}). 
For example, let us consider the case
\begin{equation*}
\begin{tikzpicture}
     \draw[-] (-0.35,-0.3/1.46) -- (0.35,-0.3/1.46);
      \draw[->,line width=0] (-0.3,-0.3/1.46) -- (0.1,-0.3/1.46);
      \draw[-] (-0.35,-0.3/1.46) -- (0,0.41);
      \draw[->,line width=0] (-0.35,-0.3/1.46) -- (-.11,0.21);
      \draw[-](0.35,-0.3/1.46) -- (0,0.41);
      \draw[->,line width=0] (0.35,-0.3/1.46) -- (.11,0.21);
      \draw[-, red](0.,0.) -- (0.3,0.3);
      
      \draw[-, red](0.,0.) -- (0,-0.3);
      \node at (0.,-.5) {$\R{g}x$};
      \node at (0.47,0.41) {$\R{g}y$};
      \node at (-0.38,0.34) {$z$};
\end{tikzpicture} 
\end{equation*}
The flux-free condition requires that $gxgy=z$ which together with $xy=z$ on the ground state, implies that $gxg=x$, namely that $x$ belongs to the normalizer of $g$ in $G$, i.e., $x\in N_G(g)=\{h\in G\,: \, hg=gh\}$. We highlight the corresponding bond $x$ satisfying the condition, namely
\begin{equation*}
\begin{tikzpicture}
     \draw[-] (-0.35,-0.3/1.46) -- (0.35,-0.3/1.46);
      \draw[->,line width=0] (-0.3,-0.3/1.46) -- (0.1,-0.3/1.46);
      \draw[-] (-0.35,-0.3/1.46) -- (0,0.41);
      \draw[->,line width=0] (-0.35,-0.3/1.46) -- (-.11,0.21);
      \draw[-](0.35,-0.3/1.46) -- (0,0.41);
      \draw[->,line width=0] (0.35,-0.3/1.46) -- (.11,0.21);
      \draw[-, red](0.,0.) -- (0.3,0.3);
      
      \draw[-, red](0.,0.) -- (0,-0.3);
      \node at (0.,-.5) {$\R{g}x$};
      \node at (0.47,0.41) {$\R{g}y$};
      \node at (-0.38,0.34) {$z$};
 
       \node at (2,0) {$\overset{x\in N_G(g)}{\longrightarrow}$};

     \draw[-, green] (-0.35+4,-0.3/1.46) -- (0.35+4,-0.3/1.46);
      \draw[-] (-0.35+4,-0.3/1.46) -- (0+4,0.41);
      \draw[-](0.35+4,-0.3/1.46) -- (0+4,0.41);
      \draw[-, red](0.+4,0.) -- (0.22/1.46+4,0.22/1.46);  
      \draw[-, red](0.+4,0.) -- (0+4,-0.3/1.46);

\end{tikzpicture} 
\end{equation*}
which will turn out to be a useful notation for keeping track of the stabilizer condition in future calculations. Fig.~\ref{fig:conf_triang} includes all possible configurations of a single triangle leading to a non-zero contribution in $\langle E|E'\rangle$.
 
Since only closed red loop configurations $L$ on the honeycomb lattice give a finite contribution in $\langle E|E'\rangle$, its green ``shadow'' (built up from those bonds $b$ where the corresponding group elements $g_b$ in any configuration of $\ket{0}$ need to lie within $N_g$) forms a closed loop on the underlying square lattice (with potential intersections). 
An example is shown in Fig.~\ref{fig:quantumdoubles}a.
%
Hence, we see that the weight of each configuration is given by the length of the red loop $(t^{|L|})$ and the constraints imposed on the green square lattice loop. We now turn to calculate $\langle E|E'\rangle$.

Let us consider a given configuration $L$, and refer as $\mathsf L$ to its shadow. For each bond $b$ of $\mathsf L$, we now define the projector onto the states which are in the normalizer of $g$ as
\begin{equation}
    P_{g,b}\equiv\sum_{h\in N_g}\ket{h}_b\bra{h}_b, \text{ and } P_{g,b}^\perp = \mathds{1}-P_{g,b},
\end{equation}
where $X_g$ and $P_g$ commute on every bond. Hence,
\begin{equation} \label{eq:overlap}
\begin{aligned}
    &\langle 0| \prod_{b\in L}X_{g,b}| 0\rangle 
    = \langle 0| \prod_{b\in L}X_{g,b} \prod_{b' \in \mathsf{L}} P_{g,b'}| 0\rangle.
\end{aligned}
\end{equation}
The next simplification we want to use is \color{black}
\begin{equation}
   \prod_{b\in L}X_{g,b} \prod_{b' \in \mathsf{L}} P_{g,b'}\ket{0}= \prod_{b' \in \mathsf{L}} P_{g,b'}\ket{0}. \label{eq:X=P}
\end{equation}
To see that this holds, first observe that the right-hand side of Eq.~\eqref{eq:X=P} is the equal-weight superposition of all flux-free states with the additional condition that all group elements appearing on the loop $\mathsf L$ must be in the centralizer of $g$; note that these properties uniquely identify a state. Since $X_{g,b}$ maps basis states to basis states, the left-hand side of Eq.~\eqref{eq:X=P} is \emph{also} an equal-weight superposition of basis states, with the same number of states as the right-hand side. Moreover, by virtue of our previous discussion, this state is also flux-free, and finally note that since $X_{g,b}$ and $P_{g,b'}$ commute, the left-hand side also satisfies the property that all group elements along the loop $\mathsf L$ are in the centralizer of $g$. Since these properties uniquely identify a state, the two states must be equal.
Combining Eq.~\eqref{eq:overlap} and Eq.~\eqref{eq:X=P}, we thus obtain

\begin{equation}
    \langle 0| \prod_{b\in L}X_{g,b}| 0\rangle  = \langle 0| \prod_{b \in \mathsf{L}} P_{g,b}| 0\rangle.
\end{equation}
\B{All that is left is computing the expectation value $\langle 0| \prod_{b \in \mathsf{L}} P_{g,b}| 0\rangle$. To this end, let us first consider a single bond $b$ and observe that we can write
\begin{equation}
    \ket{0}=\frac{1}{\sqrt{|G|}}\sum_{h\in G}\ket{h}_b\otimes \ket{\psi_h}. \label{eq:decomp0}
\end{equation}
Note that all $\langle \psi_h | \psi_{h'} \rangle = \delta_{h,h'}$: orthogonality follows from the flux-free condition of $\ket{0}$, and the fact that the norm is independent of $h$ follows from $\ket{h}_b\otimes \ket{\psi_h}$ and $\ket{h'}_b\otimes \ket{\psi_{h'}}$ being unitarily related via local `gauge' transformations.}  

{Acting then with $P_{g,b}$ on a bond, limits the sum in Eq.~\eqref{eq:decomp0} over group elements in the centralizer of $g$, i.e., $P_{g,b} \ket{0} = |G|^{-1/2} \sum_{h \in N_g} \ket{h}_b\otimes \ket{\psi_h}$. Hence, $\langle 0|P_{g,b}|0\rangle = \frac{|N_g|}{|G|} = \frac{1}{|[g]|}$. We obtain this factor for every bond $b$ along a closed loop \emph{except} the last bond. To see this,} consider a green loop $\mathsf{L}$ of length $|\mathsf L|$, and label the states of the bonds as $h_j$ with $j=1,\dots, |L|$. Apply now $P_{g}$ in all bonds except on bond $1$. Then, on any configuration $\ket{h_1,h_2,\dots,h_{|L|}}$ the free-flux condition implies that $h_1$ is given by the product of $h_j$'s (or their inverses). Hence, if $h_2,\dots, h_{|L|}\in N_g$, by construction $h_1$ also belongs to the centralizer, and hence the action of $P_{g,1}$ is trivial. Putting all together, we find that
\begin{equation} \label{eq:over_EE}
    f(L) =\frac{|[g]|^{C_{\mathsf{L}}}}{|[g]|^{|\mathsf{L}|}}
\end{equation}
is given by the topological factor on the underlying (green) square lattice rather than on the honeycomb lattice. Here, $C_{\mathsf{L}}$ is defined as the cyclomatic number (also known as cycle rank, or nullity) which corresponds to the minimum number of edges that must be removed from $\mathsf{L}$ lying on the square lattice to break all its cycles, making it into a tree. This agrees with the total number of faces in the planar graph (excluding the exterior face). As an example, the configuration in Fig.~\ref{fig:quantumdoubles}a has $C_{\mathsf{L}}=2$, since the green loop encloses two faces.  Notice that this loop model includes both contractible and non-contractible loops. 

\section{Quantum fidelity \label{app:fidelity}}

The quantum fidelity between two density matrices $\rho$ and $\sigma$ is defined as
\begin{equation}
    F(\rho, \sigma)\equiv \left(\textrm{tr}\sqrt{\sqrt{\rho}\sigma \sqrt{\rho}}\right)^2,
\end{equation}
and quantifies how distinguishable these density matrices are. Sometimes $F' \equiv \sqrt{F}$ is also used as a definition of quantum fidelity. Both definitions satisfy the so-called ``data processing inequality''~\cite{Alberti1983}, namely monotonicity with respect to any quantum channel $\mathcal{E}$. In particular, for quantum channels $\mathcal{E}_p$ resulting from the composition of local channels $\mathcal{E}^j_p(\rho)=(1-p)\rho + pO_j\rho O_j$ with $O_j^2=\mathds{1}$ and $p\in [0,1/2]$, this inequality implies that
\begin{equation}  \label{eq:DPI_Pauli}
 F(\mathcal{E}_p(\rho), \mathcal{E}_p(\sigma))\leq  F(\mathcal{E}_{p=\frac{1}{2}}(\rho), \mathcal{E}_{p=\frac{1}{2}}(\sigma)).
\end{equation}

When $p=\frac{1}{2}$, the local channel $\mathcal{E}^j_p$ can be rewritten as a random projector channel $\mathcal{E}^j_{p=\frac{1}{2}}(\rho) = P_{+,j}\rho P_{+,j} + P_{-,j} \rho P_{-,j}$,
with $P_{\pm,j}=(1\pm O_j)/2$. Let us consider two initial states $\ket{\psi}$ and $\ket{\varphi}$ and apply $\mathcal{E}_{p={\frac{1}{2}}}$ to them. Then, using $\eta_j=\pm 1$ to label this random sign we get the decohered states
\begin{align}
    &\rho_\psi=\sum_{\{\eta_j\}} \ket{\psi_{\bm{\eta}}}\bra{\psi_{\bm{\eta}}},\hspace{15pt} \rho_\varphi=\sum_{\{\eta_j\}} \ket{\varphi_{\bm{\eta}}}\bra{\varphi_{\bm{\eta}}}
\end{align}
where $\avg{\psi_{\bm{\eta}}|\psi_{\bm{\eta}'}}=\delta_{\bm{\eta},\bm{\eta}'}\avg{\psi_{\bm{\eta}}|\psi_{\bm{\eta}}}$. One then finds that the (square-root) fidelity equals the average overlap (see details in Ref.~\cite{SM})
\begin{equation} \label{eq:F_gen}
    F'(\rho_{\psi}, \rho_{\varphi}) =  \sum_{\{\eta_j\}}  p(\bm{\eta}) \frac{|\avg{\psi_{\bm{\eta}}|\varphi_{\bm{\eta}}}|}{\avg{\psi_{\bm{\eta}}|\psi_{\bm{\eta}}}}
\end{equation}
with $p(\bm{\eta})=\avg{\psi_{\bm{\eta}}|\psi_{\bm{\eta}}}$. 

When $\ket{\psi}=\ket{0}$ corresponds to the ground state in Eq.~\eqref{eq:gs_qd}, and $\ket{\varphi}=\ket{x,x'}$ includes two anyons at locations $x$ and $x'$ on top of this ground state, one finds that ${|\avg{\psi_{\bm{\eta}}|\varphi_{\bm{\eta}}}|}$ takes the form of a random O$(d)$ loop model as in Eq.~\eqref{eq:Zeta}, where loop configurations are restricted to those including an open string connecting sites $x$ and $x'$, with $d$ the quantum dimension of the proliferated anyon.

\color{black}


%

\appendix

\onecolumngrid
\pagebreak[4]

\leavevmode 
\section*{Supplementary material}

\section{From O$(N)$ loop models to Ising-like models} \label{app:from_ON_to_Ising}

\subsection{Equivalent representations of the Ising model} \label{app:Ising}
Let us consider a 2D Ising model with Ising variables $\sigma_i=\pm 1$ lying on the vertices of the honeycomb lattice 
\begin{equation} 
H_{\hexagon} = -\sum_{\langle i,j \rangle_{\hexagon} } \sigma_i \sigma_j.
\end{equation}
It turns out there are two alternative but equivalent representations of this model, which we briefly review here. The first one corresponds to an exact high-temperature expansion, namely, an expansion for $\beta \ll 1$. Using that $e^{\beta \sigma_i \sigma_j}= \cosh(\beta) + \sinh(\beta) \sigma_i \sigma_j$ for every bond $(i,j)$ of the honeycomb, we can rewrite its partition function as (up to an overall constant)
\begin{equation} \label{eq:2d_Ising_honey}
Z \propto \sum_{\{\sigma_j\}} \prod_{\langle i,j \rangle_{\hexagon} }\langle( 1 + \tanh(\beta)\sigma_i \sigma_j\rangle).
\end{equation}
To perform the sum over $\sigma_j$ we notice that $\sum_{\sigma_j=\pm 1} \sigma_j=0$, and then only configurations where two bonds overlap on every site lead to a finite contribution. Hence, non-vanishing contributions correspond to closed loop configurations on the honeycomb lattice $L$, with a probability given by $\tanh(\beta)^{|L|}$
\begin{equation} \label{eq:O1_loop}
Z  \propto  \sum_{L} \tanh(\beta)^{|L|}.
\end{equation}
Alternatively, these closed loop configurations can be understood as domain walls of a $2$D Ising model defined on the dual lattice, i.e., with Ising spins $\tau_i=\pm 1$ lying on the vertices of the triangular lattice. Specifically, a single bond belonging to $L$, corresponds to a domain wall between spins $\tau_i$ and $\tau_j$  with the quantity $1/2(1-\tau_i\tau_j)=1$, and vanishing otherwise. Hence, we can rewrite 
\begin{equation}
\sum_{L}\tanh(\beta)^{|L|} \to \sum_{\{\tau_i\}} \tanh(\beta)^{\sum_{\langle i,j \rangle_{\triangle} } \frac{1}{2}(1-\tau_i\tau_j) } =\sum_{\{\tau_i\}} e^{\tilde{\beta}\sum_{\langle i,j \rangle_{\triangle} } \tau_i\tau_j},
\end{equation}
with $e^{-2\tilde{\beta}}=\tanh(\beta)$. Overall, we find that Eq.~\eqref{eq:O1_loop} can be rewritten as the ferromagnetic Ising model on the triangular lattice
\begin{equation} \label{eq:2D_ising_dual}
Z  \propto  \sum_{\{\tau_i\}} e^{\tilde{\beta}\sum_{\langle i,j \rangle_{\triangle} } \tau_i\tau_j}.
\end{equation}

\subsection{From random loop model to random bond Ising model} \label{app:random_loop_to_RBIM}
In Eq.~\eqref{eq:Z_a} we found that the eigenvalues of the decohered density matrix when acting with a Pauli quantum channel on a toric code ground state is given by
\begin{equation}
     Z_a = \sum_{E: \partial E = a} p_E.
\end{equation}
Here we show that when $p_E=p^{|E|}(1-p)^{|\mathcal{N}|-|E|}$, $Z_a $ corresponds to the partition function of the random bond Ising model \cite{Dennis_2002}. First, recall the definition of the tension $t=p/(1-p)$. Then we can write
\begin{equation}
     Z_a = (1-p)^{\mathcal{N}}\sum_{E: \partial E = a} t^{|E|}.
\end{equation}
As a next step we note that any error chain $E$ with boundaries $\partial E = a$, can be expressed as the sum (mod $2$) of a reference error chain $E_{\textrm{ref}}$ with $\partial E_{\textrm{ref}}=a$ an a closed loop configuration~\footnote{Notice that while here we do not take care of the different homological classes of the closed loop configuration that appear for surfaces with finite genus, these can be taken care of}, namely $E=E_{\textrm{ref}}\oplus L$. Then the partition function becomes
\begin{equation} \label{eq:A_3}
\begin{aligned}
     Z_a &= (1-p)^{\mathcal{N}}\sum_{L: \partial E_{\textrm{ref}} = a} t^{|E_{\textrm{ref}}\oplus L|}=(1-p)^{\mathcal{N}}\sum_{L: \partial E_{\textrm{ref}} = a} \prod_{\substack{b\in L\\ b\notin E_{\textrm{ref}}}}t \prod_{\substack{b\in L\\ b\in E_{\textrm{ref}}}}1 \prod_{\substack{b\notin L\\ b\notin E_{\textrm{ref}}}}t \prod_{\substack{b\notin L\\ b\notin E_{\textrm{ref}}}}1 \\
     &= (1-p)^{\mathcal{N}}\sum_{L: \partial E_{\textrm{ref}} = a} \prod_{b\in L}\sqrt{t}^{1-\eta_b} \prod_{b\notin L}\sqrt{t}^{1+\eta_b} ,
\end{aligned}
\end{equation}
where $\eta_b=-1$ if the bond belongs to $E_{\textrm{ref}}$, and $\eta_b=+1$ if $b\notin E_{\textrm{ref}}$, i.e., we identify $E_{\textrm{ref}}=\{\eta_b\}$. Now we can introduce Ising variables $\sigma_i=\pm 1$ on the dual lattice, such that a bond $b$ occupied by the loop configuration $L$ and lying between $\sigma_i,\sigma_j$, corresponds to a domain wall, i.e., $\sigma_i\sigma_j=-1$. Then defining $t=e^{-2\beta}$, we find
\begin{equation}
    Z_a = (1-p)^{\mathcal{N}}\prod_b \sqrt{t}^{1-\eta_b}\sum_{\{\sigma_i\}} e^{-2\beta \sum_{\langle i,j \rangle} \eta_{i,j} \frac{1}{2}(1-\sigma_i \sigma_j) } = (1-p)^{\mathcal{N}}\sqrt{t}^{\mathcal{N}}\sum_{\{\sigma_i\}} e^{\beta \sum_{\langle i,j \rangle} \eta_{i,j} \sigma_i \sigma_j } 
\end{equation}
that we identify with the partition function of the random bond Ising model with disorder configuration $\{\eta_b\}$. Notice that the result is independent of the choice of reference $E_{\textrm{ref}}$ since two different configurations $\eta_b$ and $\eta_b'$ with the same flux configuration, i.e., $\prod_{b} \eta_b=\prod_b \eta_b'$ around every close loop, are related by a gauge transformation $\eta_{ij} \to  t_i \eta_{ij} t_j$, with $t_i=\pm 1$, which can be absorbed into a change of variables $\sigma_i \to \sigma_i t_i$. This is already clear from Eq.~\eqref{eq:A_3}, since changing $E_{\textrm{ref}}' = E_{\textrm{ref}} + L'$ can be absorbed into a change of variables of $L \to L + L'$.

\subsection{High-temperature expansion of O$(N)$ spin models.}

A  different class of models that can be thought as generalizations of the Ising model (e.g., on the honeycomb lattice), are the so-called O$(N)$ spin models. Here, the relevant variables are $N$-dimensional normalized vectors $S_i$ instead of the Ising variables, and the partition function is given by
\begin{equation}
Z = \int \prod_i d \bm{S}_i e^{\beta \sum_{\langle i,j \rangle_{\hexagon} } \bm{S}_i\cdot \bm{S}_j}.
\end{equation}
where $ d \bm{S}_i$ is the Haar measure over the $(N-1)$-dimensional sphere. The main difference with respect to the Ising model is that the high-temperature expansion is not exact anymore. Instead,
\begin{equation}
Z \approx \int \prod_i d \bm{S}_i \prod_{\langle i,j \rangle_{\hexagon}}\left(1 + \beta \sum_{\langle i,j \rangle_{\hexagon} } \bm{S}_i\cdot \bm{S}_j\right).
\end{equation}
only holds when $\beta \ll 1$. By expanding out this product and integrating over the spin variables (similar to Sec.~\ref{app:Ising}), one finds the O$(N)$ loop model on the same lattice; see e.g., Sec. 3.2 in Ref.~\cite{peled2019lectures}. As in the previous section, a similar relation holds even in the presence of bond disorder. Moreover, when $N$ is a positive integer, the loop O$(N)$ model admits a height
function representation~\cite{Nienhuis_81}, analogous to that of understanding loops as domain walls of Ising variables.

\subsection{Exact rewritings of O$(N)$ loop models via spin-$N$ local models} \label{app:ON_cubicmodels}

Here we provide two spin models (one on the square lattice, the other on the honeycomb) whose exact high-temperature expansions produce the O$(N)$ loop models.

\subsubsection{Square lattice}
Let us now start from an O$(N)$ loop model
\begin{equation}
    \mathcal{Z}(N)= \sum_{L} t^{|L|} N^{C(L)},
\end{equation}
on a square lattice for any value of the tension $t$, where $C(L)$ is the cyclomatic number of the graph $L$ (see main text). In the following we show that this can be equivalently expressed as
\begin{equation} \label{eq:Z_fc}
    \mathcal{Z}_{\textrm{FC}}(N)=\sum_{\{\bm{S}_j\}} \prod_{\langle i,j\rangle} \left( 1 + tN \bm{S}_i\cdot \bm{S}_j\right)
\end{equation}
which has manifestly non-negative weights for $t\leq 1/N$ \cite{BLOTE19841,Chayes_2000,Guo06,Deng_2007}. Here $\bm{S}_i$ is a $N$-dimensional vector where one and only one entry can take the value $+ 1$ or $-1$, and all the others vanish. Namely, $\bm{S}_i$ lies on the faces of an hypercube in $N$ dimensions. For example, for $N=2$, $\bm{S}_i$ lies on the sides of a square, while for $N=3$ $\bm{S}_i$ lies on the faces of a cube. The resulting stat-mech model is known as face-cubic (FC) model~\cite{Chayes_2000}. As a consequence of this definition one then finds that
\begin{equation}
      \langle S_i^\alpha\rangle_{t=0} \equiv \frac{ \sum_{\{\bm{S}_j\}} S_i^\alpha}{ \sum_{\{\bm{S}_j\}} 1}=  0,\hspace{20pt} \langle S_i^\alpha S_i^\beta\rangle_{t=0} =  \frac{\delta_{\alpha,\beta}}{N}, 
\end{equation}
and similarly for higher odd and even order correlations respectively. In particular, $ \langle (S_i^\alpha)^4\rangle_{t=0} =  \frac{1}{N}$. From here it follows that the only non-vanishing contributions to $ \mathcal{Z}_{\textrm{FC}}(N)$ come from multiplying the weights $tN S^\alpha_i\cdot S^\alpha_j$ around closed loops (with potential intersections) for each (fixed) component $\alpha=1,\dots,N$. The resulting Boltzmann weight for a single such closed loop configuration $L$ when summing over $\alpha$ is given by 
\begin{equation}
    \sum_{\alpha=1}^N (tN)^{|L|}\left(\frac{1}{N}\right)^{|L|-V(L)}=t^{|L|}N^{V(L)+1}
\end{equation}
where $V(L)$ is the number of four-valent vertices of the planar graph $L$. 
Now we can combine Euler's theorem for planar graphs together with the fact that the number of vertices is twice that of the number of edges to find that $F(L)=V(L)+2$. This allows us to write the previous weight as
\begin{equation}
    t^{|L|}N^{V(L)+1}=t^{|L|}N^{F(L)-1} =  t^{|L|}N^{C(L)}.
\end{equation}
Therefore, we find that $\mathcal{Z}_{\textrm{FC}}(N)$ in Eq.~\eqref{eq:Z_fc} equals the partition function of the O$(N)$ loop model on the square lattice (up to an overall factor). \\

\paragraph*{\textbf{Phase diagrams.}} For $N=2$, it is known that $\mathcal{Z}_{\textrm{FC}}(N)$ in Eq.~\eqref{eq:Z_fc} corresponds to the Ashkin-Teller (AT) model~\cite{AT_model}, with the $t=1$ lying on the self-dual line~\cite{selfdual_AT}. In fact, the $t=1$ point corresponds to the zero-temperature limit of the AT model along the self-dual line (with ferromagnetic two-spin interactions). It is known that this ferromagnetic region ($J > |U|$ where $U$ is the four-spin interaction) of the self-dual line is critical, by mapping it to the six-vertex model~\cite{six_vertex_model} as in Refs.~\cite{FAN1972136,Glazman_2023}. In fact this particular zero-temperature point is known to map to an integrable 19-vertex model and a dilute O$(2)$ Brauer loop model~\cite{Ikhlef_2012} (see also Ref.~\cite{Huang_2013}), and it gives rise to a compact boson CFT with Luttinger liquid parameter $K = 1/3$ or $K = 3/4$. It has been conjectured that the model is in the disordered ‘gapped’ phase for $t < 1$~\cite{Glazman_2023} and confirmed by numerical simulations.

For $N>2$, the face-cubic spin model is expected to still be in the disordered phase at $t=1$~\cite{BLOTE19841}, which was numerically confirmed for a variety of values in Ref.~\cite{Deng_2007}. The critical values of the tension $t$ for various loop weights $N$ are shown in Fig.~\ref{fig:loops}b.

\subsubsection{Honeycomb lattice}

Similarly to the previous section, one can equivalently rewrite O$(N)$ loop models on the honeycomb lattice via a local spin-$N$ model. Being the lattice bipartite, let us denote by $A$ and $B$ the two different sublattices and consider the \emph{mixed face-corner cubic} model
\begin{equation} \label{eq:Z_mix}
    Z_{\textrm{mix}}(N)=\sum_{\{\bm{S}_a\}, \{\bm{S}_b\}}\prod_{\langle a,b\rangle}(1+tN\bm{S}_a\cdot \bm{S}_b),
\end{equation}
with $a\in A$ and $b\in B$. Both $\bm{S}_a$ and $\bm{S}_b$ are $N$-dimensional binary normalized vectors with $\bm{S}_a$ as defined in the previous section, while 
\begin{equation}
    \bm{S}_b=\frac{1}{\sqrt{N}}(S_b^1,S^2_b,\dots, S^N_b)^T
\end{equation}
with $S^\alpha_b=\pm 1$. Namely, $ \bm{S}_b$ lies at the corners of an hypercube in $N$ dimensions, and hence the Boltzmann weights in Eq.~\eqref{eq:Z_mix} are non-negative if $t\leq 1/\sqrt{N}$. In this case $\langle S^\alpha_b \rangle_{t=0}=\langle  S^\alpha_b   S^\beta_b   S^\gamma_b  \rangle_{t=0} = 0$ while  $ \langle  S^\alpha_b S^\beta_b \rangle_{t=0}  =\delta_{\alpha,\beta} $ since $(S^\alpha_b)^{2k}=1$. It then follows that denoting by $L$ a closed loop configuration on the honeycomb lattice, the partition function becomes
\begin{equation} 
    Z_{\textrm{mix}}(N)=\sum_L \underbrace{(tN)^{|L|}\left(\frac{1}{\sqrt{N}}\right)^{|L|} \sum_{\{\bm{S}_a\}, \{\bm{S}_b\}} \prod_{\langle a,b\rangle  \in L }\left(\sum_{\alpha=1}^N S^\alpha_a S^\alpha_b\right)}_{=W_L}.
\end{equation}
The factor $(1/\sqrt{N})^{|L|}$ comes from the normalization of $\bm{S}_b$. Let us denote by $\ell$ the connected components of the loop configuration $L$. Then the Boltzmann weight for $L$ is given by (up to an overall factor)
\begin{equation}
    W_L = (tN)^{|L|} \left(\frac{1}{\sqrt{N}}\right)^{|L|} \prod_{\ell \in L}\left(\sum_{\alpha=1}^N \left(\frac{1}{N}\right)^{\frac{|\ell|}{2}}\right)=(tN)^{|L|} \left(\frac{1}{\sqrt{N}}\right)^{|L|} \left(\frac{1}{N}\right)^{\frac{|L|}{2}}N^{C(L)}=t^{|L|}N^{C(L)}, 
\end{equation}
where the factor  $(1/N)^{|\ell|/2}$ comes from the fact that $\langle (S_a^\alpha)^2 \rangle_{t=0} =  1/N$. All together, we find the partition function of the O$(N)$ loop model with tension $t$
\begin{equation} \label{eq:ON_honey}
    Z(N)=\sum_L t^{|L|}N^{C(L)}.
\end{equation}
The case $N=2$ is relevant to understand the decoherence transition for $D_4$ TO as discussed in the companion work~\cite{long_paper}.

\paragraph*{\textbf{Phase diagrams.}}  O$(N)$ loop models on the honeycomb lattice have been extensively studied in the literature (see review~\cite{peled2019lectures} ). In general, O$(N)$ loop models with \emph{loop weight} $N \in [-2,2]$ showcase two different phases separated by a critical point at the critical tension $t_c(N)=(2+\sqrt{2-N})^{-1/2}$: A dilute (or small loop) phase for $t<t_c(N)$, and a dense phase for $t>t_c(N)$~\cite{Nienhuis_82}. Both the critical and the latter phase are described by minimal unitary conformal field theories (CFT).  For example, the case $N=1$ corresponds to the 2D Ising model on the triangle lattice, where the critical point is described by an Ising CFT with central charge $c=1/2$.  For $N=2$ the critical point at $t_c(2)=1/\sqrt{2}$ is described by a Berezinskii-Kosterlitz-Thouless (BKT) transition which extends into an extended gapless phase described by a Luttinger liquid with central charge $c=1$.  In general, for O$(N)$ loop models with $N=2\cos(\pi/(k+2))$, the critical point is described by unitary CFT (a minimal unitary CFT with $p=k+3$ and $q=k+2$) and the dense phase by a minimal unitary CFT with $p=k+2$ and $q=k+1$  (see e.g., Appendix B in Ref.~\cite{nian2022nonunitary}). 

\section{Kitaev's model} \label{app:Kitaev_model}

\begin{figure}
    \centering
    \includegraphics[width=0.58\linewidth]{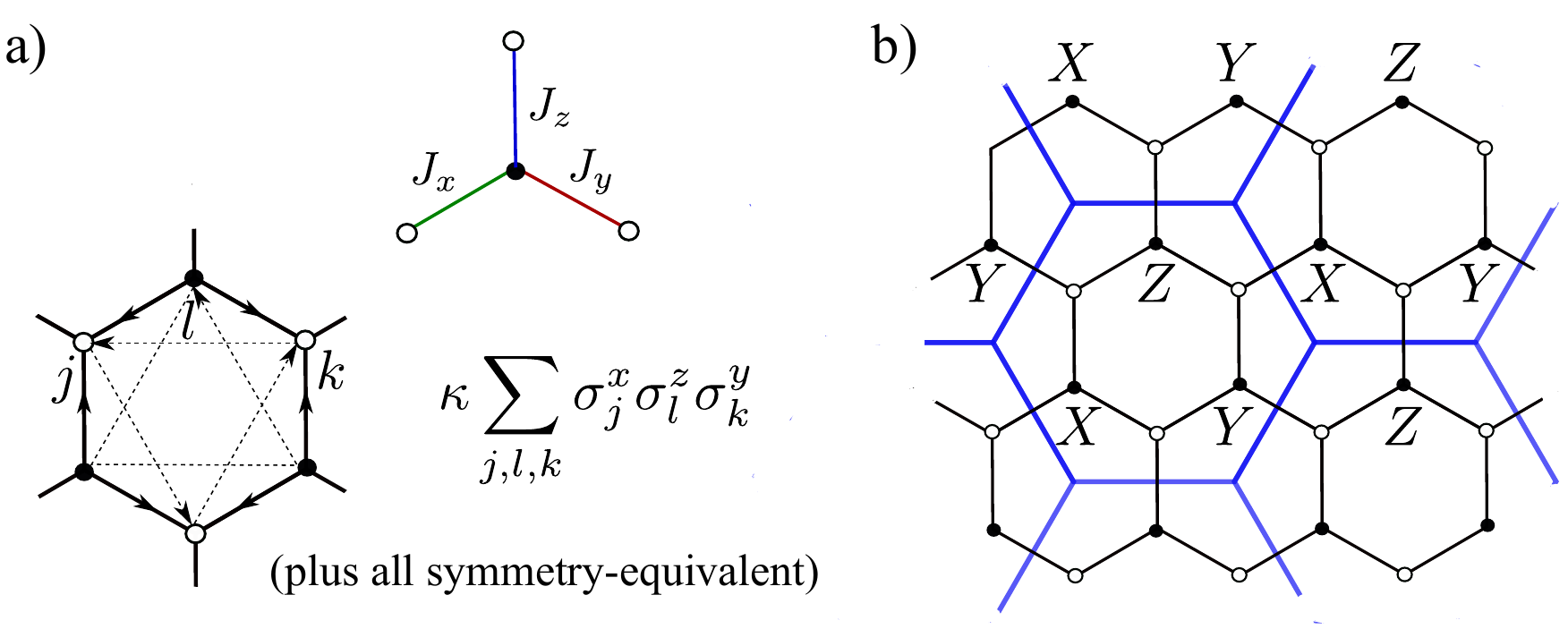}
    \caption{\textbf{Non-Abelian gapped phase of Kitaev model.} (panel (a) adapted from Ref.~\onlinecite{Kitaev_2006})}
    \label{fig:app_Kitaev_mod}
\end{figure}

We now consider the gapped non-Abelian topological ordered phase of the Kitaev model realized by the microscopic spin-$1/2$ Hamiltonian introduced in Ref.~\onlinecite{Kitaev_2006}. This appears in the presence of a weak external magnetic field which breaking time-reversal symmetry opens up a gap. The resulting Hamiltonian (defined on the honeycomb lattice ) combines Kitaev interactions with couplings $J_\alpha$ along the different directions $\alpha=x,y,z$ indicated in Fig.~\ref{fig:app_Kitaev_mod}a, plus an additional $3$-body term with tuning parameter $\kappa$
\begin{equation} \label{eq:H_spin}
    H = \sum_{\alpha=x,y,z}J_\alpha \sum_{\langle i,j\rangle_{\alpha}} \sigma_i^\alpha \sigma_j^\alpha + \kappa\sum_{j,k,l}\sigma^x_j \sigma^y_k \sigma_l^z,
\end{equation}
with $j,k,l$ corresponding to three nearest vertices lying on an hexagon, and for  $J_x\approx J_y \approx J_z$. In the following, we place the system on a torus with periodic boundary conditions (PBC). This Hamiltonian conserves the same plaquette operators
\begin{equation*}
\includegraphics[width=0.14\linewidth]{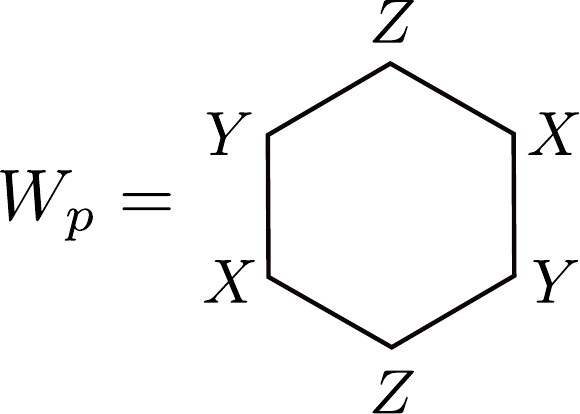}
\end{equation*}
as for $\kappa=0$ (namely in the absence of the $3$-body term), with the ground state lying within the sector $W_p=+1$ for all $p$. Excitations of such operators, i.e., plaquettes for which $W_p=-1$, correspond to ``vortices''. Each of this carries an unpaired Majorana mode and then correspond to non-Abelian $\sigma$ anyons with quantum dimension $d_\sigma=\sqrt{2}$. Notice that for PBC, the product of all plaqutte operators satisfy $\prod_p W_p=+1$, and whence Majorana anyons can only be created in pairs. The other two anyon sectors are the vacuum $1$, and an (Abelian) fermion $\varepsilon$. When two vortices fuse, they can either annihilate completely or leave a fermion behind: $\sigma \times \sigma=1+\varepsilon.$ The other non-trivial fusion rules are given by $\varepsilon\times \varepsilon = 1$, and $\varepsilon\times \sigma=\sigma$. This corresponds to the Ising topological order, and the remaining algebraic properties of its anyons can be found in Ref.~\onlinecite{Kitaev_2006}.

Following this reference we can rewrite Pauli matrices $\sigma_j^\alpha=ib^\alpha_j c_j$ in terms of two types of Majoranas with $\{c_i,c_j\}=2\delta_{ij}$. Fixing the gauge $u_{ij}=\avg{ib^\alpha_ib^{\alpha}_j}=+1$ along every link of the lattice, Hamiltonian \eqref{eq:H_spin} becomes quadratic in terms of Majoranas
\begin{equation} \label{eq:H_maj}
    H = -\frac{i}{2}\sum_{\alpha=x,y,z}\sum_{\avg{j,k}_\alpha} A^\alpha_{j,k} c_j c_k - \frac{i}{2}\sum_{\langle \langle j,k\rangle\rangle} B_{j,k} c_j c_k,
\end{equation}
with $A^\alpha_{j,k}=J_\alpha (\leftarrow)_{jk}$ and $B_{j,k}=\kappa(\dashleftarrow)_{jk}$. Here, the different types of arrows $\leftarrow$ and $\dashleftarrow$ indicate both the ordering of Majoranas in the product $c_jc_k$ as well as the involved sites as indicated in Fig.~\ref{fig:app_Kitaev_mod}a. In particular, $(\leftarrow)_{jk}$ is equal to $1$ if there is a solid arrow from $k$ to $j$, $-1$ if arrow goes in the opposite direction and $0$ otherwise. $\dashleftarrow$ is similarly defined. Without loss of generality in the following we consider $J_x=J_y=J_z\equiv J$ and denote by $ \ket{\kappa}$ the corresponding ground state. Being a quadratic Hamiltonian, ground state expectation values of a Majorana monomial $\langle i^{|\alpha|}c_1^{\alpha_1}c_2^{\alpha_2}\cdots c_{2\ell_x\ell_y}^{\alpha_{2\ell_x\ell_y}}\rangle$ for a system of size $\ell_x\ell_y$ with $\alpha_j=0,1$ and $|\alpha|=\sum_j \alpha_j$ (constrained to be even), can be directly computed~\cite{dias2023classical} from the covariance matrix
\begin{equation}
    G_{i,j}=\frac{i}{2}\langle \kappa | [c_i,c_j]|\kappa\rangle
\end{equation}
by restricting $G$ to a submatrix that includes all rows and columns within corresponding to non-vanishing $\alpha_j$'s, such that $\langle i^{|\alpha|}c_1^{\alpha_1}c_2^{\alpha_2}\cdots c_{2\ell_x\ell_y}^{\alpha_{2\ell_x\ell_y}}\rangle=\textrm{Pf}(\left.G\right|_{[\alpha]})$, where $\textrm{Pf}$ is the pfaffian of a matrix such that $\det(a)=\textrm{Pf}(A)^2$, such that $\textrm{Pf}(\left.G\right|_{[\alpha]})$ vanishes for odd-degree monomials. See additional details in Ref.~\cite{dias2023classical}.

In the following, we place Hamiltonian \eqref{eq:H_maj} on a torus of size $\ell_x \ell_y$ with both $\ell_x$ and $\ell_y$ even, and choose boundary conditions such that the ground state energy is minimal. Fixing $J=1$, we find that the ground state is attained when choosing anti-periodic boundary conditions around $x$ direction for $A_{j,k}^\alpha$ in Eq.~\eqref{eq:H_maj}, and periodic around the $y$ direction and also for $B_{j,k}$  around both the $x$ and $y$ directions. The sign of $\kappa$ does not enter in this minimization and in the following is taken to be positive $\kappa>0$. Notice that one can freely toggle among different combinations of the sings of $J$ and $\kappa$ by combining: (1) time reversal which maps $J\to J$ and $\kappa \to -\kappa$; and (2) the unitary transformation mapping $J_\alpha\to -J_\alpha$ and $\kappa \to \kappa$, by applying a $\pi$-rotation around an orthogonal axis to $\alpha$ on half of the sites. 

\subsection{Non-Abelian deformation }
\begin{figure}
    \centering
    \includegraphics[width=0.6\linewidth]{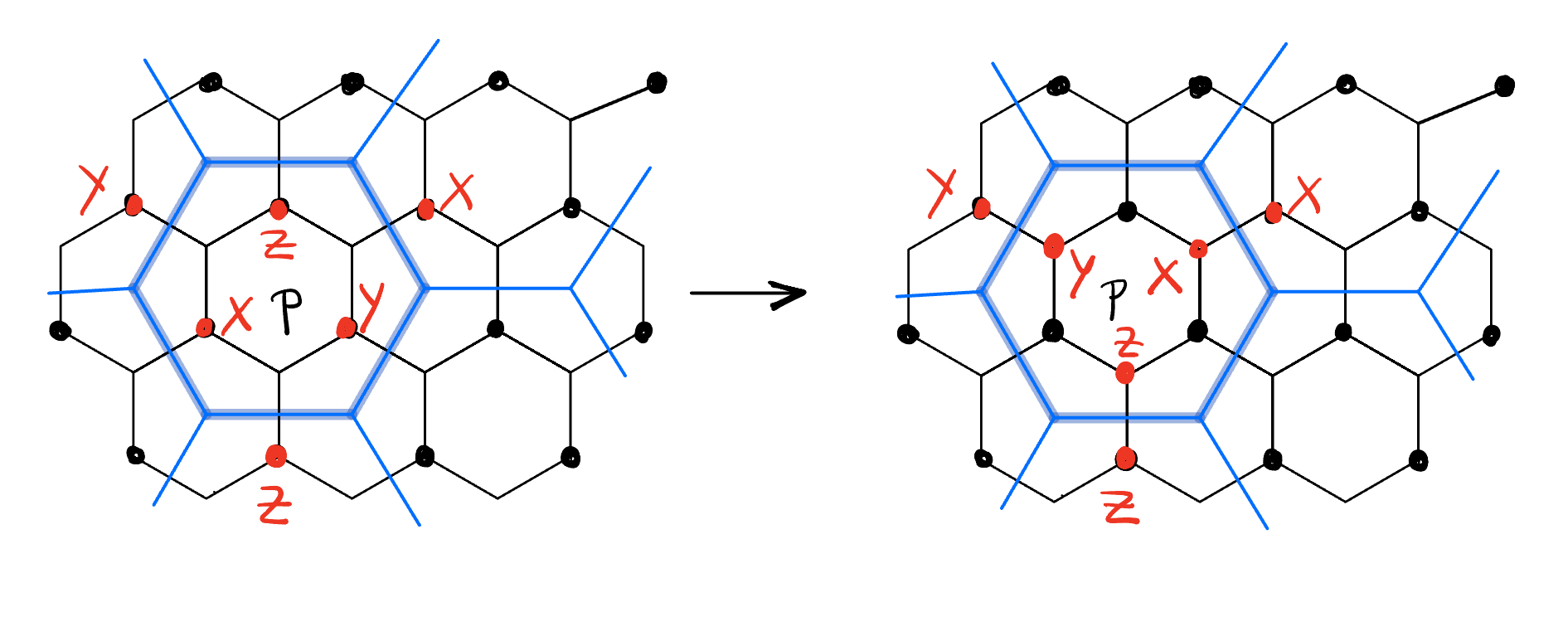}
    \caption{\textbf{Minimal loop configuration.}}
    \label{fig:loop_Kit}
\end{figure}

We first consider applying a deformation to the state $\ket{\kappa}$ in the form of an imaginary time evolution. 
In particular, for a given ground state $\ket{\kappa}$ of \eqref{eq:H_spin} with fixed $\kappa$ we consider the deformed pure wave function 
\begin{equation}  \label{eq:main_psi_beta_kappa}
|\tilde{\psi}(\beta,\kappa)\rangle=e^{\frac{\beta}{2}\sum_{a\in A}T_a}\ket{\kappa}
\end{equation}
where $T_a=\sigma^x,\sigma^y,\sigma^z$ are Pauli matrices only acting on one sublattice, denoted as $A$-sublattice in the following and corresponding to the full dotted vertices in Fig.~\ref{fig:main_Kitaev_mod}a. The functional form of $T_a$ is also specified in the figure. Its norm (up to an overall factor) reads
\begin{equation}
    \begin{aligned}
        \mathcal{Z}(\beta, \kappa)=\sum_{n_a=0}^{N_A}t_{\textrm{ext}}^{n_A}\sum_{a_1, \dots, a_{n_A}}\langle\kappa|\prod_{i=1}^{n_A}T_{a_i}|\kappa \rangle,
    \end{aligned}
\end{equation}
where we have defined $t_{\textrm{ext}}\equiv \tanh(\beta)$. In general, $\prod_{i=1}^{n_A}T_{a_i}$ does not commute with all plaquette operators $W_p$, then leading to a vanishing contribution. Indeed, it locally toggles the vortex charges from $W_p=+1\to -1$, and hence proliferate pairs of non-Abelian anyons $\sigma$. Therefore, to obtain a non-vanishing contribution these need to be paired up, leading to a non-vanishing contribution only when $\prod_{i=1}^{n_A} T_{a_i}$ forms a closed loop configuration $L$. This is defined on the dual honeycomb lattice of the triangular $A$-sublattice, namely the blue honeycomb lattice appearing in Fig.~\ref{fig:main_Kitaev_mod}a. An example of such non-vanishing contributions is shown in Fig.~\ref{fig:loop_Kit}. Notice that the expectation value $\langle\kappa|\prod_{a\in L}T_{a}|\kappa \rangle$ corresponds to the topological factor $f(L)$ defined in the main text, i.e.,
\begin{equation} \label{eq:fL_kitaev}
   f(L)=\langle\kappa|\prod_{a\in L}T_{a}|\kappa \rangle.
\end{equation}
Hence, the partition function takes the exact form
 \begin{equation} \label{eq:main_Zkit_psi}
    \begin{aligned}
        \mathcal{Z}_{\ket{\psi}}^{\textrm{Kit}}(\beta, \kappa)=\sum_{L}t_{\textrm{ext}}^{|L|}f(L).
    \end{aligned}
\end{equation}

Alternatively, we consider the composition of local quantum channels 
 \begin{equation}
     \mathcal{E}_a(\rho_0)=(1-p)\rho_0 + pT_a\rho_0T_a,
 \end{equation}
with $\rho_0=\ket{\kappa}\bra{\kappa}$ acting on the $A$-sublattice and subjected to the same pattern shown in Fig.~\ref{fig:main_Kitaev_mod}a. Once again, the expression of interest is the overlap $\langle E|E' \rangle$ between two error chains defined via $\ket{E}=\prod_{a\in E}T_a$. As in Eq.~\eqref{eq:fL_kitaev} this is given by
\begin{equation} \label{eq:EE_Kit}
    \langle E|E' \rangle = f(E\oplus E')=\langle\kappa|\prod_{a\in E\oplus E'}T_{a}|\kappa \rangle.
\end{equation}
For example, this allows us to compute the purity of the resulting decohered mixed state which (up to an overall factor) takes the form
\begin{equation} \label{eq:main_Zkit_rho}
    \begin{aligned}
        \mathcal{Z}_{\rho}^{\textrm{Kit}}(p, \kappa)=\sum_{L}t_{\textrm{ext}}^{|{L}|}f(L)^2,
    \end{aligned}
\end{equation}
where we have now defined $t_{\textrm{ext}}=\frac{2p(1-p)}{(1-p)^2 + p^2}$. 

While Eqs.~\eqref{eq:main_Zkit_psi}, \eqref{eq:EE_Kit} and \eqref{eq:main_Zkit_rho} are exact, we still need to evaluate $f(L)$. At this point, the fact that the ground state $\ket{\kappa}$ is not a fixed-point wave function like e.g., the Toric code or $D_4$ TO ground states, with zero correlation length, becomes important. While in this case no simple closed analytical expression can be derived, we can exploit that the Kitaev Hamiltonian  in Eq.~\eqref{eq:H_spin}, becomes quadratic in Majorana operators within the subspace $W_p=+1$~\cite{Kitaev_2006}, which is respected by the action of $\prod_{a\in L} T_a$ on the ground state.  
 Let us consider a loop configuration with a single connected component. When this is a contractible loop, we can write
\begin{equation} 
\begin{aligned}
f(L)=\langle\kappa|\prod_{a\in {L}}T_a|\kappa \rangle &= \langle\kappa|\prod_{\hexagon_p\in \textrm{int}({L})} \underbrace{\left(\prod_{a\in \hexagon_p}T_a\right)}_{=\mathcal{T}_p}|\kappa \rangle 
\end{aligned}
\end{equation}
i.e., namely as the product of plaquette $\mathcal{T}_p$ operators inside the loop, forming a membrane with its boundary given by $L$. Moreover, using that $W_p\ket{\kappa}=\ket{\kappa}$ one can write
\begin{equation}
\langle\kappa|\prod_{a\in {L}}T_a|\kappa \rangle = \langle\kappa|\prod_{\hexagon_p\in  \textrm{int}({L})} \mathcal{T}_pW_p|\kappa \rangle ,
\end{equation}
where for each plaquette the operator $\mathcal{T}_pW_p$ is a product of the three bilinears $S^\alpha S^\alpha$ on different links (see Fig.~\ref{fig:loop_Kit}b), and hence can be written completely in terms of $c$'s Majoranas after using the gauge $u_{jk}=+1$. This converts the boundary spin operator $\prod_{a\in {L}}T_a$ into a Majorana membrane operator lying inside $L$, leading to 
\begin{equation}
f(L) = \sigma(L) \times \textrm{Pf}\left(\left.G\right|_{{\textrm{int}(L)}}\right),
\end{equation}
where $\sigma(L)=\pm 1$ is a configuration-dependent overall sign that can be algorithmically obtained and is fixed by: (1) the number of elementary plaquette operators $\mathcal{T}_pW_p$, (2) a product of $u_{ij}$'s corresponding to different Majorana bilinears; and (3), an overall permutation to bring the product of $c$ Majoranas into the canonical ordering such that $\langle i^{|\alpha|}c_1^{\alpha_1}c_2^{\alpha_2}\cdots c_{2\ell_x\ell_y}^{\alpha_{2\ell_x\ell_y}}\rangle=\textrm{Pf}(\left.G\right|_{[\alpha]})$. 

Therefore, from here we obtain the partition functions
\begin{equation}
    \begin{aligned}
        \mathcal{Z}_{\ket{\psi}}^{\textrm{Kit}}(\beta, \kappa)=\sum_{L}t_{\textrm{ext}}^{|{L}|}\underbrace{\sigma(L) \textrm{Pf}\left(\left.G\right|_{{\textrm{int}(L)}}\right)}_{=f(L)},
    \end{aligned}
\end{equation}
for the pure wavefucntion deformation, and similarly for the purity 
\begin{equation}
    \begin{aligned}
        \mathcal{Z}_{\rho}^{\textrm{Kit}}(p, \kappa)=\sum_{{L}}t_{\textrm{ext}}^{|{L}|} \textrm{det}\left(\left.G\right|_{{\textrm{int}(L)}}\right),
    \end{aligned}
\end{equation}

At this point it is unclear whether the ``partition functions'' in Eqs.~\eqref{eq:main_Zkit_psi} and \eqref{eq:main_Zkit_rho} relate in any way to the physics of O$(N)$ loop models. In the following section, we will combine the universal properties of O$(N)$ loop models to show numerical evidence that indeed, the underlying physics is correctly captured by loop models with $N$ given by the quantum dimension (or its square for the purity calculation) of the $\sigma$ non-Abelian anyon.
First, notice that unlike for the case of quantum doubles and $D_4$ topological order on  the kagome lattice, the topological factor $f(L)$ has an $L$-dependent sign (although $\textrm{det}\left(\left.G\right|_{{\textrm{int}(L)}}\right)=\textrm{Pf}\left(\left.G\right|_{{\textrm{int}(L)}}\right)^2>0$). Unfortunately, while we can algorithmically compute this sign as well as $\sigma(L)$, we currently lack a simple rule for the overall sign of $f(L)$ to each loop configuration. Everything that remains is then evaluating the absolute value $|f(L)| $, and understand its dependence on the quantum dimension (in this case expected to be that of the $\sigma$ anyon with $d_\sigma=\sqrt{2}$), and on the length of the loop. However, unlike for the zero correlation-length ground state with $D_4$ topological order, this cannot be evaluated in closed-form, and we need to resort to its numerical evaluation that we accomplish using Gaussian-state techniques. We expect that for single components of $L$ (assume $L=\cup_j \gamma_j$) whose length is much larger than the correlation length $\xi$, this contribution factors out as $\textrm{Pf}\left(\left.G\right|_{{\textrm{int}(L)}}\right)\approx \prod_j \textrm{Pf}\left(\left.G\right|_{{\textrm{int}(\gamma_j)}}\right)$.

\subsection{Numerical evaluation of weights $\textrm{Pf}\left(\left.G\right|_{{\textrm{int}(L)}}\right)$ using Gaussian states} \label{app:Num_eval}

\begin{figure}
    \centering
    \includegraphics[width=0.65\linewidth]{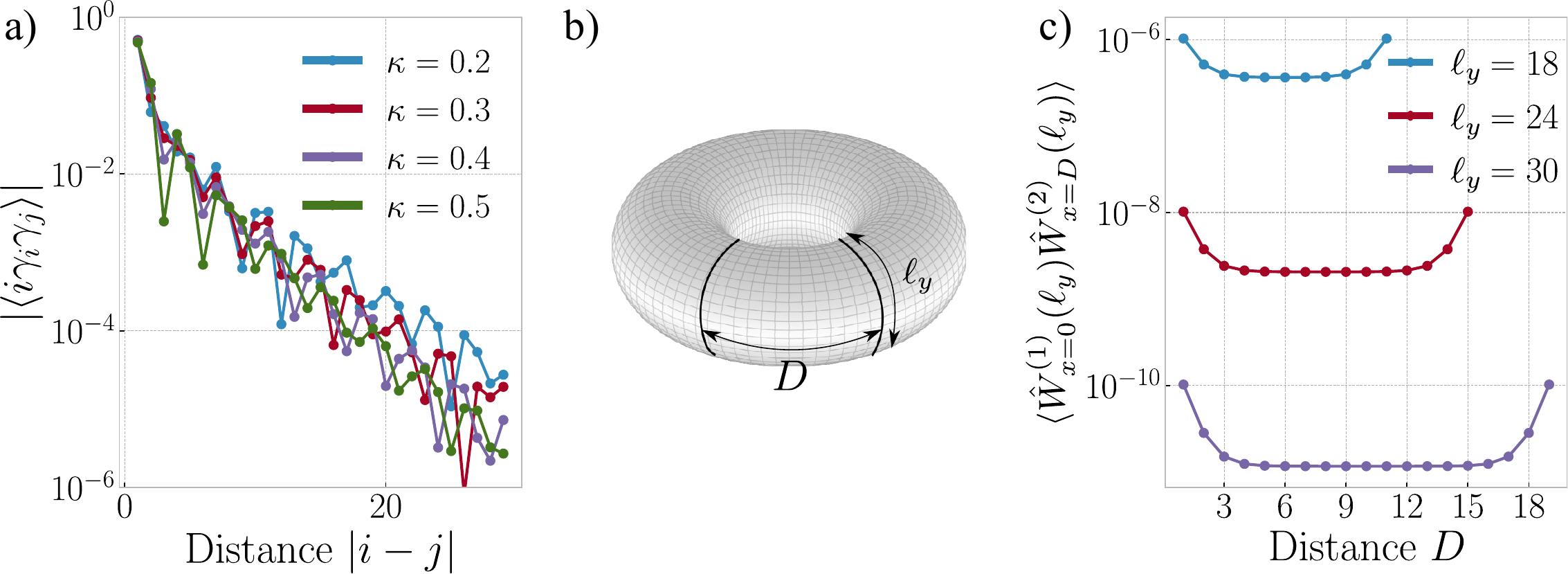}
    \caption{\textbf{Dependence of the weight $\langle \hat{W}_{x=0}^{(1)}(\ell_y) \hat{W}_{x=D}^{(2)}(\ell_y) \rangle$ with the distance $D$.} }
    \label{fig:torus}
\end{figure}

In this section we will numerically show that 
\begin{equation} \label{eq:num_check}
    |\textrm{Pf}\left(\left.G\right|_{{\textrm{int}(L)}}\right)|\approx t_{\textrm{int}}^{|L|} \sqrt{2}^{C(L)}
\end{equation}
for sufficiently long loops, where the value of the intrinsic tension $t_{\textrm{int}}$ depends on $\kappa$. First, we use the two-point Majorana correlation functions $\langle i\gamma_i \gamma_j\rangle $ to extract the correlation length $\xi$. These simply correspond to matrix elements of $G$. We calculate the covariance matrix $G$ for the ground state of the quadratic Hamiltonian in Eq.~\eqref{eq:H_maj} on a torus of size $\ell_x\times \ell_y$ assuming anti-periodic boundary conditions (as for this we find the lowest possible ground state energy). We then obtain $G$ by diagonalizing $H=U\textrm{diag}(E_i)U^\dagger$ such that the covariance matrix is given by $G=iU\textrm{sign}(\textrm{diag}(E_i))U^\dagger$ (see e.g., Ref.~\cite{Kitaev_2006}). Fig.~\ref{fig:torus}a shows the fast exponential decay of these correlations for different values of $\kappa$ with $\xi\approx 9$ lattice sites. In the following, unless otherwise stated, we assume $\kappa=0.2$.

To extract the tension $t$ and validate the behavior in Eq.~\eqref{eq:num_check} we proceed as follows. First, we place two non-contractible loops of length $t_{\textrm{int}}$ and at distance $D$ from each other wrapping around the handle of the torus along the $\hat{y}$ direction as shown in Fig.~\ref{fig:torus}b. We take $\ell_x=2\ell_y$ with $\ell_y$ even. Call $\hat{W}_{x=0}^{(1)}(\ell_y) \hat{W}_{x=D}^{(2)}((\ell_y)$ the spin operator that creates such configuration. Then following the same steps as in the previous section, $\hat{W}_{\ell_y}^{(1)} \hat{W}_{\ell_y}^{(2)}$ can be rewritten as a membrane Majorana operator inside the region delimited by the two non-contractible loops at $x$ coordinates $x=0,D$. Then the expectation value $\langle\hat{W}_{x=0}^{(1)}(\ell_y) \hat{W}_{x=D}^{(2)}(\ell_y) \rangle $ can be efficiently computed. Since $\langle i\gamma_i \gamma_j\rangle $ decay exponentially fast with the distance, we expect that for sufficiently large distance $D$, correlations among sufficiently far Majoranas are negligeable and hence $\langle\hat{W}_{x=0}^{(1)}(\ell_y) \hat{W}_{x=D}^{(2)}(\ell_y)\rangle $ approximately factorizes as the product of the weights  $f(L)$ of each non-contractible loop (one at $x=0$ and the other at $x=D$), i.e., $\langle \hat{W}_{x=0}^{(1)}(\ell_y) \hat{W}_{x=D}^{(2)}(\ell_y) \rangle \approx (f(L))^2$. Fig.~\ref{fig:torus}c, shows that for various values of $\ell_y$, this expectation value becomes indeed independent of $D$ and positive, hence consistent with this expectation. In the following, for a given system size $\ell_x,\ell_y$, we fix $D=\ell_x/2$.

\begin{figure}
    \centering
    \includegraphics[width=0.7\linewidth]{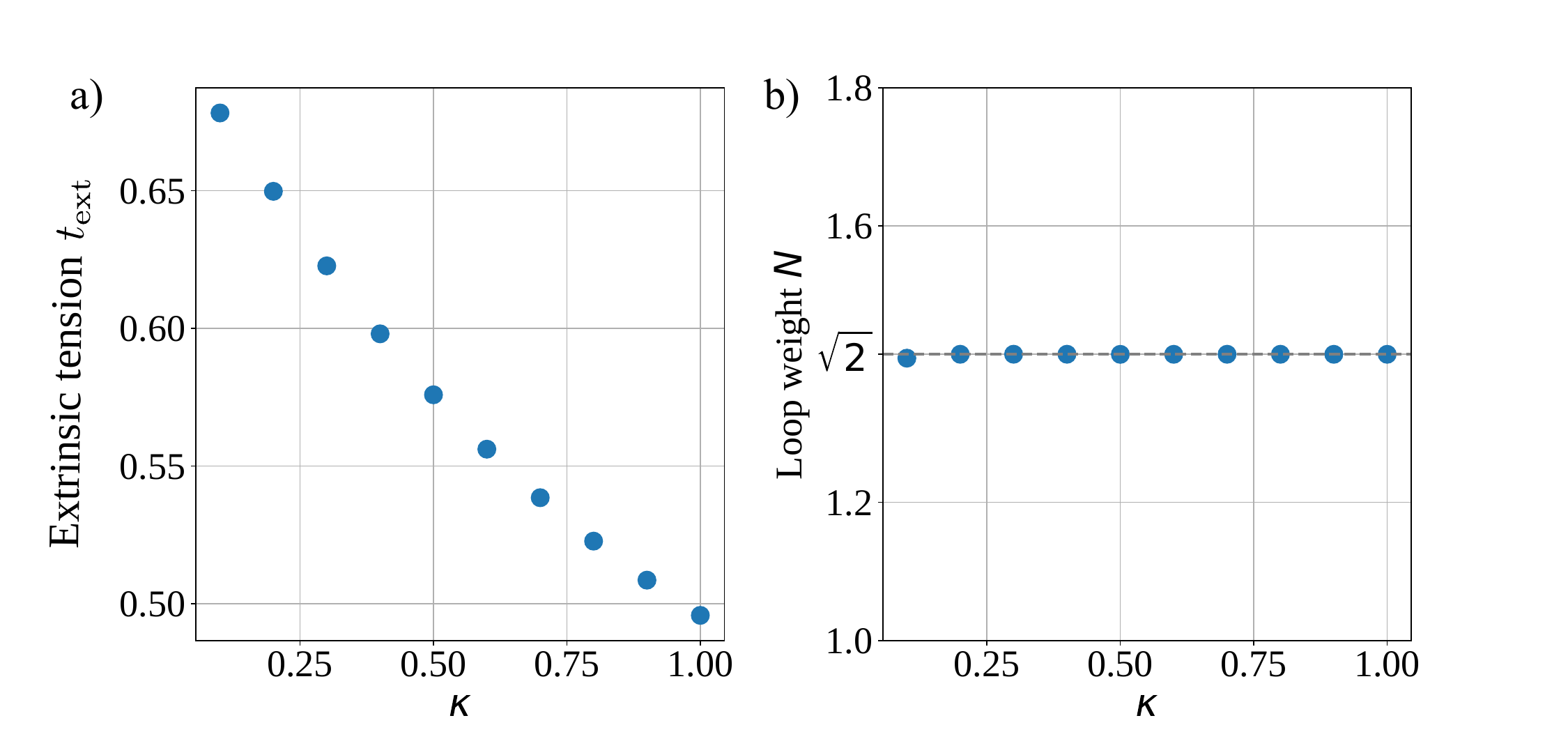}
    \caption{\textbf{Dependence of the extrinsic tension $t_{\textrm{ext}}$ and loop weight $N$ on $\kappa$.} }
    \label{fig:QD_dep_kappa}
\end{figure}

Second, in the inset of Fig.~\ref{fig:main_Kitaev_mod}b we verify that $f(L)$ decays exponentially with the length of the loop $L$, from where we can directly extract the tension $t_{\textrm{int}}\approx 0.65$ at $\kappa=0.2$. Finally, we verify the dependence on the topological factor $N^{C(L)}$ with $N=\sqrt{2}$. The configuration shown in Fig.~\ref{fig:torus}a has two non-contractable loops around the torus, and then the number of components $C(L)$ of the loop configuration is $2$. Therefore, if $f_{\ell_x,\ell_y}(L) \approx t_{\textrm{int}}^{|L|} N^2$ for a system size $\ell_x\ell_y$, we can obtain the loop weight $N$ via the ratio
\begin{equation} \label{eq:W_ratio}
    \frac{(f_{\ell_x,\ell_y}(L))^2}{f_{\ell_x,2\ell_y}(L)} \approx \frac{(t_{\textrm{int}}^{\ell_y} N)^2}{t_{\textrm{int}}^{2\ell_y}N} = N.
\end{equation}
In Fig.~\ref{fig:main_Kitaev_mod}b (main panel) we show that this ratio is rather constant as a function of the loop length and equals to $N=\sqrt{2}$, agreeing with the quantum dimension of the corresponding non-Abelian anyon. Finally, Fig.~\ref{fig:QD_dep_kappa} shows the dependence of the intrinsic tension $t_{\textrm{int}}$ (panel a) and the loop weight $N$ (panel b) on $\kappa$. Notice that the previous analysis has been performed for $\kappa=0.2$, and that for $\kappa> 0.2$ the tension decreases. On the other hand, panel b shows the dependence of $N$, as obtained from the ratio in Eq.~\eqref{eq:W_ratio}, on $\kappa$. We see that the maximum tension is attained for small values of $\kappa$. Overall the conclusion of this section is that the topological factor is approximately given by
\begin{equation}
    f(L)\approx t_{\textrm{int}}^{|L|}\sqrt{2}^{C(L)}
\end{equation}
where $t_{\textrm{int}}$ depends on the value of $\kappa$. We have also computed this dependence for different loop configurations (not necessarily wrapping around the torus) and finding consistent results.

\subsection{Monte-Carlo simulations} \label{app:MC}

\begin{figure}
    \centering
    \includegraphics[width=0.65\linewidth]{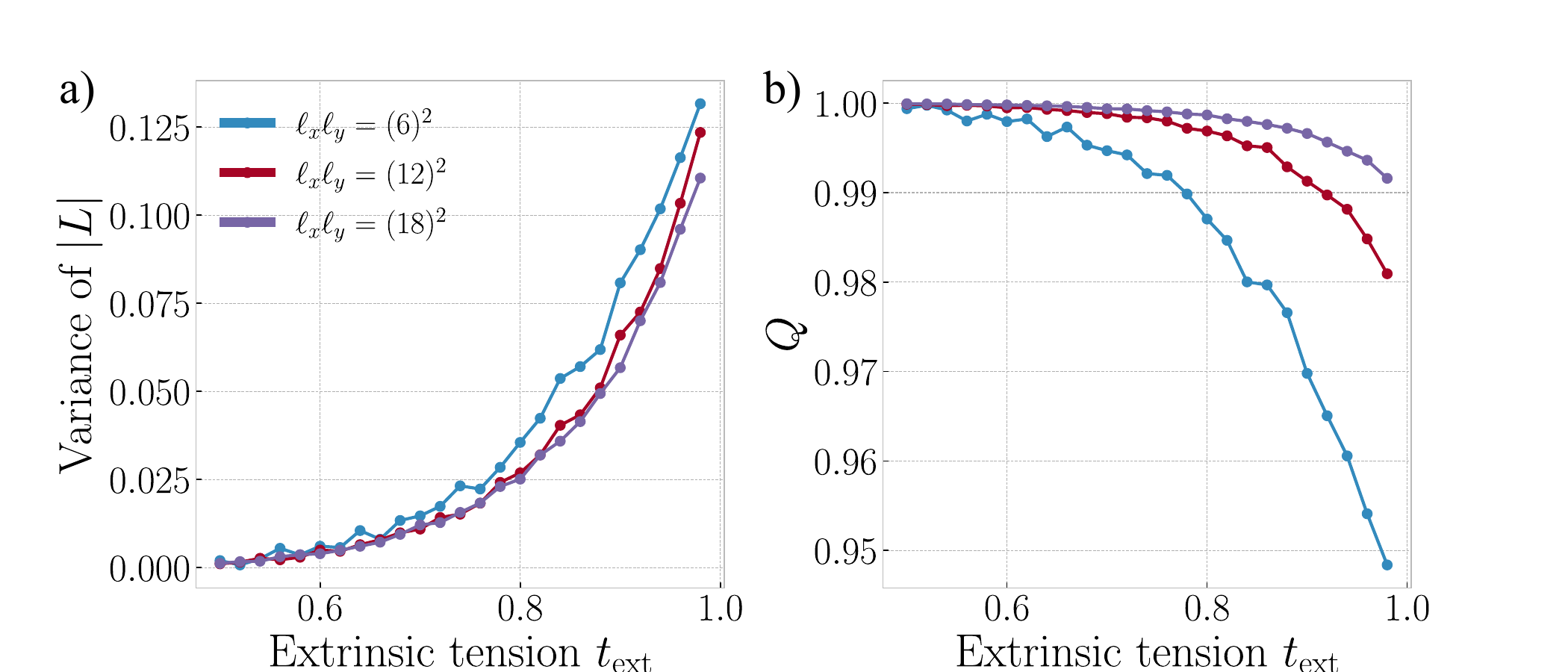}
    \caption{\textbf{Monte-Carlo simulations for decohered Kitaev model via the purity $\textrm{tr}(\rho^2)$.}}
    \label{fig:O2_kitaev}
\end{figure}

In this appendix we provide details about the Monte-Carlo simulations that we use to obtain the numerical results shown in Figs.~\ref{fig:main_Kitaev_mod}(c,d) in the main text as well as Fig.~\ref{fig:O2_kitaev}. To simplify the numerical implementation, we coordinate each honeycomb layer with a brick wall structure shown below

\begin{center}
    \includegraphics[width=0.3\linewidth]{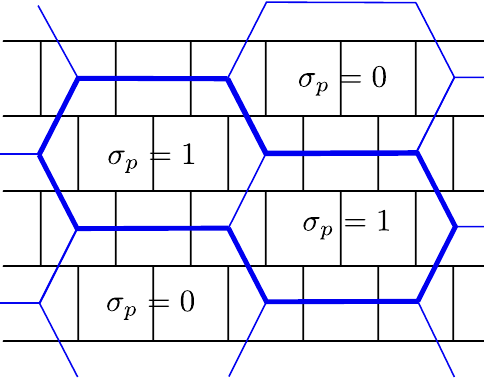}
\end{center}

Our numerical implementation of the Metropolis algorithm runs as follows:
\begin{enumerate}
    \item Fix parameters $\kappa$, system sizes $\ell_x=6n$ and even $\ell_y$, and find the covariance matrix $G$ on the ground state.
    \item Fix error strength $\beta$ (for pure wavefunction deformations) or $p$ (for decohered density matrix) which fixes the extrinsic tension $t_{\textrm{ext}}$.
    \item Initialize a random configuration of $\{\sigma_p=0,1\}$ per plaquette in the blue superlattice. The corresponding loop configuration $L(\{\sigma_p\})$ is given by the non-overlaping boundaries of plaquettes for which $\sigma_p=1$.
    \item Compute weight $W(L)=t_{\textrm{ext}}^{|L|}|f(L)|$ (or $W(L)=t_{\textrm{ext}}^{|L|}f(L)^2$ when characterizing the purity for the decohered mixed state). This involves the computation of a pfaffian (or a determinant). 
    \item Perform $\textrm{eqSteps}$ number of Metropolis steps --each of them involving $ \ell_x\times \ell_y$ single plaquette updates $\sigma_p \to 1-\sigma_p$. We choose $\textrm{eqSteps}=5000 (3000)$ for the analysis of the deformed wavefunction (purity). The acceptance ratio is given by the logarithm of the ratio $W(L|\sigma_p\to 1-\sigma_p)/W(L) $. Notice that we assume that single-site updates lead to non-reducible dynamics.
    \item Compute average quantities by performing additional $\textrm{mcSteps}=3000$ Metropolis-steps.
\end{enumerate}

In particular, we numerically calculate the Binder cumulant $Q$ defined via
\begin{equation}
    Q\equiv \frac{\langle (M)^2\rangle^2}{\langle (M)^4\rangle}
\end{equation}
where $M= \sum_{p} (2\sigma_p -1)$. Namely, $M$ is the magnetization of the Ising variables $s_p\equiv 2\sigma_p-1$ lying on the triangular lattice described by the center of the plaquettes. Deep in the small loop phase (i.e., for $t=t_{\textrm{ext}}t_{\textrm{int}}\ll t_c$) all spins $s_p$ are aligned and hence $Q\to 1 $ in the thermodynamic limit. On the other hand, deep in the dense loop phase, $Q$ is controlled by the fluctuations of the magnetization. For example, for the $2$D Ising model $Q\to 1/3$ due to the Gaussian form of the distribution of the magnetization~\cite{Selke_2007}. We also compute the (normalized) variance of the loop length $|L|$
\begin{equation}
    \textrm{Var}(|L|)\equiv \frac{\langle |L|^2\rangle - \langle |L|\rangle^2 }{\ell_x\ell_y}.
\end{equation}
This one is predicted to scale with system size $\ell=\sqrt{\ell_x \ell_y}$ as $\textrm{Var}(|L|)\sim a + b\ell^{2(3-16/g)}$ with $a,b$ constants and the Coulomb parameter $g$ fixed by the equation $N=-2\cos(\pi g/4)$ with $g\in [4,6]$~\cite{Liu_2011}. In particular, one finds that
\begin{equation}
    \textrm{Var}(|L|)\sim \left\{ \begin{array}{ll} a + b\ln(\ell) &  \text{if } N=1, \\  
    a + b\ell^{-2/5} & \text{if } N=\sqrt{2}, \\ a + b\ell^{-2} &\text{if } N=2.
    \end{array}  \right. 
\end{equation}
Hence, while $\textrm{Var}(|L|)$ diverges with system size for $N=1$, it  approaches a constant value in the thermodynamic limit for $N>1$. 
While we find that our data is consistent with this scaling, we are limited by the number of data points and the large fluctuations of the numerical data. In Fig.~\ref{fig:O2_kitaev} we show the dependence of these two quantities for the decohered density matrix as characterized by the purity. Based on these results, we conclude that no transition occurs for the purity.

\section{Review of quantum double construction} \label{app:Quant_doubles}

The quantum double construction of topological order was introduced by Kitaev in Ref.~\onlinecite{Kitaev_2003}. Here, we follow this reference and we review the necessary structure that will be used when considering the decohered density matrix. Consider a finite group $G$ with identity element denoted by $1$. The local Hilbert space on an edge for an arbitrary lattice corresponds to the group algebra $\mathbb{C}[G]$ given by the space of linear combinations of group elements with complex coefficients. Hence, the state of a qudit is spanned by the $|G|$-dimensional orthonormal basis $\{\ket{g} : g\in G\}$. In the following, we denote by $\mathcal{H}$ the global Hilbert space. We here consider a triangular lattice (such that maximally three fluxons can \emph{ a priori} fuse on a single plaquette) \\
\begin{center}
    \includegraphics[width=0.3\linewidth]{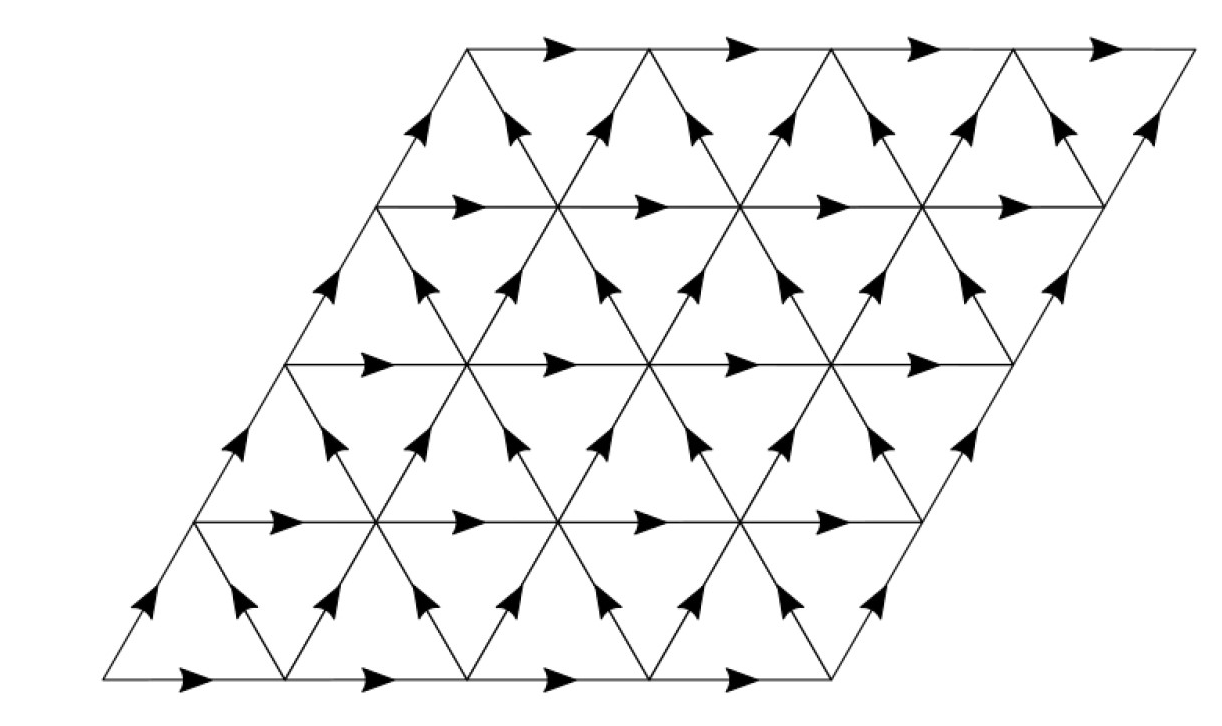}
 \end{center}
where a particular orientation along the edges is fixed. Changing the direction of an arrow represents the basis transformation $\ket{g}\to \ket{g^{-1}}$ for the corresponding qudit.

To describe the model, we need to introduce $4$ types of linear operators $L_+^g, L_-^g, T_+^h, T_-^h$ acting on $\mathcal{H}$. They act as follows
\begin{equation}
    \begin{aligned}
    L^g_+\ket{z}&=\ket{gz}, \hspace{35pt} T_+^h\ket{z}=\delta_{h,z}\ket{z}\\
    L^g_-\ket{z}&=\ket{zg^{-1}}, \hspace{35pt} T_+^h\ket{z}=\delta_{h^{-1},z}\ket{z}.
    \end{aligned}
\end{equation}
Let us now consider an oriented edge of the triangular lattice $b$ with $s$ one of its boundaries. Then we define the operator $L^q(b,s)=L_{-}^q(b)$ if $s$ is the origin of the arrow, and  $L^q(b,j)=L_{+}^q(b)$ otherwise. Similarly, we define $T^h(b,p)$ as $T_-^h$ ($T_+^h$) acting on bond $b$, if $p$ is the right (left) face adjacent to bond $b$. From here, we can then define local gauge transformations $A_g(j,p) $ and magnetic charge operators $B_h(s,p)$ for a vertex $s$ and an adjacent triangular plaquette $p$ as
\begin{equation}
    \begin{aligned}
    &A_g(s,p)=\prod_{b\in \textrm{star}(j)}L^g(b,s) \\
   & B_h(s,p)=\sum_{\substack{h_1, h_2,h_3 \\ \textrm{st } h_1 h_2 h_3=h}} T^{h_1}(b_1, p)  T^{h_2}(b_2, p) T^{h_3}(b_3, p)
    \end{aligned}
\end{equation}
where $b_1, b_2,b_3$ are the boundary bonds of $p$ listed in the counterclockwise order, starting from and ending at vertex $s$. Notice that the order of the multiplications in $h_1 h_2 h_3=h$ is important. These operators generate an algebra $\bm{D}(G)$ known as Drinfield's quantum double of the group algebra $\mathbb{C}[G]$. Finally, the Hamiltonian of the system will be expressed in terms of the symmetric combinations
\begin{equation}
    A_s=\frac{1}{|G|}\sum_{g\in G}A_g(s,p), \hspace{20pt} B_p=B_1(s,p),
\end{equation}
which are projection operators that commute with each other, and on a basis state $\ket{g_1,g_2,g_3}$ on a triangular act as
\usetikzlibrary {shapes.geometric}
\tikzset{>=latex}

\begin{equation}
\begin{tikzpicture}
    \node at (-1.,0.) {$B_p|$};

      \draw[-] (-0.35,-0.3/1.46) -- (0.35,-0.3/1.46);
      \draw[->,line width=0] (-0.3,-0.3/1.46) -- (0.1,-0.3/1.46);
      \draw[-] (-0.35,-0.3/1.46) -- (0,0.41);
      \draw[->,line width=0] (-0.35,-0.3/1.46) -- (-.11,0.21);
      \draw[-](0.35,-0.3/1.46) -- (0,0.41);
      \draw[->,line width=0] (0.35,-0.3/1.46) -- (.11,0.21);
      
      \node at (0.,0.) {$p$};
      \node at (0.,-.4) {$g_1$};
      \node at (0.34,0.26) {$g_2$};
      \node at (-0.34,0.26) {$g_3$};
      \node at (0.6,0.) {$\rangle $};
      \node at (1.8,0.) {$=\delta_{g_1g_2g_3^{-1},e} $};
      \node at (2.8,0.) {$|$};
      
      \draw[-] (3.4-0.35,-0.3/1.46) -- (3.4+0.35,-0.3/1.46);
      \draw[->,line width=0] (3.4-0.3,-0.3/1.46) -- (3.4+0.1,-0.3/1.46);
      \draw[-] (3.4-0.35,-0.3/1.46) -- (3.4+0,0.41);
      \draw[->,line width=0] (3.4-0.35,-0.3/1.46) -- (3.4-.11,0.21);
      \draw[-](3.4+0.35,-0.3/1.46) -- (3.4,0.41);
      \draw[->,line width=0] (3.4+0.35,-0.3/1.46) -- (3.4+.11,0.21);
      
      \node at (3.4,0.) {$p$};
      \node at (3.4,-.4) {$g_1$};
      \node at (3.74,0.26) {$g_2$};
      \node at (3.06,0.26) {$g_3$};
      \node at (4.,0.) {$\rangle, $};
      
\end{tikzpicture} 
\end{equation}
\begin{equation*}
\begin{tikzpicture}
    \node at (-1.3,0.) {$A_g(s,p)|$};
     \draw[-] (-0.35,-0.3/1.46) -- (0.35,-0.3/1.46);
      \draw[->,line width=0] (-0.3,-0.3/1.46) -- (0.1,-0.3/1.46);
      \draw[-] (-0.35,-0.3/1.46) -- (0,0.41);
      \draw[->,line width=0] (-0.35,-0.3/1.46) -- (-.11,0.21);
      \draw[-](0.35,-0.3/1.46) -- (0,0.41);
      \draw[->,line width=0] (0.35,-0.3/1.46) -- (.11,0.21);
      \node at (-.5,-.3) {$s$};
      \node at (0.,0.) {$p$};
      \node at (0.,-.4) {$g_1$};
      \node at (0.34,0.26) {$g_2$};
      \node at (-0.34,0.26) {$g_3$};
      \node at (0.6,0.) {$\rangle $};
      \node at (1.2,0.) {$=$};
      \node at (1.9,0.) {$|$};

      \draw[-] (3.1-0.35,-0.3/1.46) -- (3.1+0.35,-0.3/1.46);
      \draw[->,line width=0] (3.1-0.3,-0.3/1.46) -- (3.1+0.1,-0.3/1.46);
      \draw[-] (3.1-0.35,-0.3/1.46) -- (3.1+0,0.41);
      \draw[->,line width=0] (3.1-0.35,-0.3/1.46) -- (3.1-.11,0.21);
      \draw[-](3.1+0.35,-0.3/1.46) -- (3.1,0.41);
      \draw[->,line width=0] (3.1+0.35,-0.3/1.46) -- (3.1+.11,0.21);
      
      \node at (3.1,0.) {$p$};
      \node at (3.2,-.45) {$g_1g^{-1}$};
      \node at (3.54,0.26) {$g_2$};
      \node at (2.6,0.26) {$g_3 g^{-1}$};
      \node at (3.8,0.) {$\rangle $};
      
\end{tikzpicture} 
\end{equation*}
From here one can define the Hamiltonian
\begin{equation} \label{eq:H_QD}
    H=\sum_s (1-A_s) + \sum_p (1-B_p)
\end{equation}
resembling that of the Toric code Hamiltonian (in fact, this corresponds to the group $G=\mathbb{Z}_2$). The space of (topological) ground states is then given by gauge invariant states with zero magnetic flux on every plaquette, namely
\begin{equation}
    \mathcal{L}=\{ \ket{\psi} \in \mathcal{H}: \,\, A_s\ket{\psi}=\ket{\psi}, \,\, B_p\ket{\psi}=\ket{\psi} \text{ for all } s,p  \}.
\end{equation}
While the ground state is unique on the sphere, it is degenerate on a finite genus surface.

\subsection{Anyon content}
Let $g\in G$ be an element of the group, $[g]=\{hgh^{-1}|h\in G\}$ its conjugacy class and $N_g=\{h\in G: hg=gh\}$ its centralizer. The latter has a group structure and notice that if $a,b\in [g]$ then their centralizers are isomorphic. Hence, it does not matter which element of $[g]$ is used to define $N_g$. Then the anyons of the theory are labelled by the pair $([g],\chi) $ where $\chi$ is an irreducible representation of $N_g$. The conjugacy class $[g]$ can be interpreted as a mangetic flux, while $\chi$ corresponds to the electric charge (see also Ref.~\onlinecite{Preskill_LN}). If the flux is trivial, i.e., the conjugacy class of the identity (i.e., $[e]=\{e\}$) such that $N_e=G$, then the charge can be any of the irreducible representations of $G$. These have quantum dimension $d=\textrm{dim}(\chi)$, namely the dimension of the irreducible representation. On the other hand if the charge is trivial, i.e., $\chi$ is the identity representation, then the corresponding anyon only carries mangetic flux but no electric charge and it has quantum dimension $d=|[g]|$. In general, the quantum dimension of an anyon $([g],\chi)$ is given by $d_{([g],\chi)}=|[g]|\times \textrm{dim}(\chi)$.


\section{Quantum fidelity \label{app:fidelity}}

The  (Uhlmann-Josza) quantum fidelity between two density matrices $\rho$ and $\sigma$ is defined as
\begin{equation}
    F(\rho, \sigma)\equiv \left(\textrm{tr}\sqrt{\sqrt{\rho}\sigma \sqrt{\rho}}\right)^2,
\end{equation}
and quantifies how distinguishable these density matrices are. Sometimes $F' \equiv \sqrt{F}$ is also used as a definition of quantum fidelity. For example, when $\rho=\ket{\psi_\rho}\bra{\psi_\rho}$ and $\sigma = \ket{\psi_\sigma}\bra{\psi_\sigma}$ are projectors on normalized pure states, then the quantum fidelity agrees with the overlap
\begin{equation}
    F(\rho, \sigma) = |\langle  \psi_\rho |\psi_\sigma \rangle |^2.
\end{equation}
This vanishes when $\ket{\psi_\rho}$ is orthogonal to $\ket{\psi_\sigma}$. The quantum fidelity $F$ satisfies the so-called ``data processing inequality''~\cite{Alberti1983}
\begin{equation} \label{eq:DPI}
    F(\rho,\sigma)\leq F(\mathcal{E}(\rho), \mathcal{E}(\sigma)),
\end{equation}
namely monotonicity with respect to any quantum channel $\mathcal{E}$. In particular, we are here interested in quantum channels $\mathcal{E}_p=\mathcal{E}^1_p\circ \mathcal{E}^2_p \circ \cdots \circ \mathcal{E}^L_p$ resulting from the composition of local channels  of the form $\mathcal{E}^j_p(\rho)=(1-p)\rho + pO_j\rho O_j$ with $O_j^2=\mathds{1}$ and $p\in [0,1/2]$. For these, Eq.~\eqref{eq:DPI} implies that
\begin{equation}  \label{eq:DPI_Pauli}
 F(\mathcal{E}_p(\rho), \mathcal{E}_p(\sigma))\leq  F(\mathcal{E}_{p=\frac{1}{2}}(\rho), \mathcal{E}_{p=\frac{1}{2}}(\sigma)).
\end{equation}
In the following we make use of this property to draw conclusions from the behavior of the decohered density matrix at maximum error rate $p=\frac{1}{2}$. 

Let us set $p=\frac{1}{2}$. Then, every local channel $\mathcal{E}^j_p$ can be rewritten as a random projector channel
\begin{equation}
    \mathcal{E}^j_{p=\frac{1}{2}}(\rho)=\frac{1}{2}\left(\rho + O_j \rho O_j\right) = P_{+,j}\rho P_{+,j} + P_{-,j} \rho P_{-,j},
\end{equation}
with $P_{\pm,j}=(1\pm O_j)/2$. Let us consider two initial states $\ket{\psi}$ and $\ket{\varphi}$ and apply $\mathcal{E}_{p={\frac{1}{2}}}$ to them. Then, using $s_j$ to label this random sign we get that
\begin{equation}
    \rho_\psi=\mathcal{E}_{p=\frac{1}{2}} (\ket{\psi}\bra{\psi})=\sum_{\{s_j\}} \ket{\psi_{\bm{s}}}\bra{\psi_{\bm{s}}},\hspace{15pt} \rho_\varphi=\mathcal{E}_{p=\frac{1}{2}} (\ket{\varphi}\bra{\varphi})=\sum_{\{s_j\}} \ket{\varphi_{\bm{s}}}\bra{\varphi_{\bm{s}}}
\end{equation}
where $\avg{\psi_{\bm{s}}|\psi_{\bm{s}'}}=\delta_{\bm{s},\bm{s}'}\avg{\psi_{\bm{s}}|\psi_{\bm{s}}}$ due to the orthogonality of the projectors $P_{\pm, j}$. Hence,
\begin{equation}
    \sqrt{\rho_{\psi}} = \sum_{\{s_j\}} \frac{\ket{\psi_{\bm{s}}}\bra{\psi_{\bm{s}}}}{\sqrt{\avg{\psi_{\bm{s}}|\psi_{\bm{s}}}}}
\end{equation}
and we thus have
\begin{equation}
    \sqrt{\rho_{\psi}} \rho_{\varphi} \sqrt{\rho_{\psi}}= \sum_{\{s_j\}} \frac{|\avg{\psi_{\bm{s}}|\varphi_{\bm{s}}}|^2}{{\avg{\psi_{\bm{s}}|\psi_{\bm{s}}}}}\ket{\psi_{\bm{s}}}\bra{\psi_{\bm{s}}},
\end{equation}
which implies 
\begin{equation}
    \sqrt{\sqrt{\rho_{\psi}} \rho_{\varphi} \sqrt{\rho_{\psi}}}= \sum_{\{s_j\}} \frac{|\avg{\psi_{\bm{s}}|\varphi_{\bm{s}}}|}{{\avg{\psi_{\bm{s}}|\psi_{\bm{s}}}}}\ket{\psi_{\bm{s}}}\bra{\psi_{\bm{s}}}.
\end{equation}
One then finds that the (square-root) fidelity equals the average overlap
\begin{equation} \label{eq:F_gen}
    F'(\rho_{\psi}, \rho_{\varphi}) = \sum_{\{s_j\}}  |\avg{\psi_{\bm{s}}|\varphi_{\bm{s}}}| =  \sum_{\{s_j\}}  p(\bm{s}) \frac{|\avg{\psi_{\bm{s}}|\varphi_{\bm{s}}}|}{\avg{\psi_{\bm{s}}|\psi_{\bm{s}}}}
\end{equation}
with $p(\bm{s})=\avg{\psi_{\bm{s}}|\psi_{\bm{s}}}$.


\begin{thebibliography}{85}%
\makeatletter
\providecommand \@ifxundefined [1]{%
 \@ifx{#1\undefined}
}%
\providecommand \@ifnum [1]{%
 \ifnum #1\expandafter \@firstoftwo
 \else \expandafter \@secondoftwo
 \fi
}%
\providecommand \@ifx [1]{%
 \ifx #1\expandafter \@firstoftwo
 \else \expandafter \@secondoftwo
 \fi
}%
\providecommand \natexlab [1]{#1}%
\providecommand \enquote  [1]{``#1''}%
\providecommand \bibnamefont  [1]{#1}%
\providecommand \bibfnamefont [1]{#1}%
\providecommand \citenamefont [1]{#1}%
\providecommand \href@noop [0]{\@secondoftwo}%
\providecommand \href [0]{\begingroup \@sanitize@url \@href}%
\providecommand \@href[1]{\@@startlink{#1}\@@href}%
\providecommand \@@href[1]{\endgroup#1\@@endlink}%
\providecommand \@sanitize@url [0]{\catcode `\\12\catcode `\$12\catcode `\&12\catcode `\#12\catcode `\^12\catcode `\_12\catcode `\%12\relax}%
\providecommand \@@startlink[1]{}%
\providecommand \@@endlink[0]{}%
\providecommand \url  [0]{\begingroup\@sanitize@url \@url }%
\providecommand \@url [1]{\endgroup\@href {#1}{\urlprefix }}%
\providecommand \urlprefix  [0]{URL }%
\providecommand \Eprint [0]{\href }%
\providecommand \doibase [0]{https://doi.org/}%
\providecommand \selectlanguage [0]{\@gobble}%
\providecommand \bibinfo  [0]{\@secondoftwo}%
\providecommand \bibfield  [0]{\@secondoftwo}%
\providecommand \translation [1]{[#1]}%
\providecommand \BibitemOpen [0]{}%
\providecommand \bibitemStop [0]{}%
\providecommand \bibitemNoStop [0]{.\EOS\space}%
\providecommand \EOS [0]{\spacefactor3000\relax}%
\providecommand \BibitemShut  [1]{\csname bibitem#1\endcsname}%
\let\auto@bib@innerbib\@empty
\bibitem [{\citenamefont {Wen}(2007)}]{Wen_book}%
  \BibitemOpen
  \bibfield  {author} {\bibinfo {author} {\bibfnamefont {X.-G.}\ \bibnamefont {Wen}},\ }\href {https://doi.org/10.1093/acprof:oso/9780199227259.001.0001} {\emph {\bibinfo {title} {{Quantum Field Theory of Many-Body Systems: From the Origin of Sound to an Origin of Light and Electrons}}}}\ (\bibinfo  {publisher} {Oxford University Press},\ \bibinfo {year} {2007})\BibitemShut {NoStop}%
\bibitem [{\citenamefont {Sachdev}(2023)}]{Sachdev_2023}%
  \BibitemOpen
  \bibfield  {author} {\bibinfo {author} {\bibfnamefont {S.}~\bibnamefont {Sachdev}},\ }\href@noop {} {\emph {\bibinfo {title} {Quantum Phases of Matter}}}\ (\bibinfo  {publisher} {Cambridge University Press},\ \bibinfo {year} {2023})\BibitemShut {NoStop}%
\bibitem [{\citenamefont {Leinaas}\ and\ \citenamefont {Myrheim}(1977)}]{Leinaas_77}%
  \BibitemOpen
  \bibfield  {author} {\bibinfo {author} {\bibfnamefont {J.~M.}\ \bibnamefont {Leinaas}}\ and\ \bibinfo {author} {\bibfnamefont {J.}~\bibnamefont {Myrheim}},\ }\bibfield  {title} {\bibinfo {title} {On the theory of identical particles},\ }\href {https://doi.org/10.1007/BF02727953} {\bibfield  {journal} {\bibinfo  {journal} {Il Nuovo Cimento B (1971-1996)}\ }\textbf {\bibinfo {volume} {37}},\ \bibinfo {pages} {1} (\bibinfo {year} {1977})}\BibitemShut {NoStop}%
\bibitem [{\citenamefont {Goldin}\ \emph {et~al.}(1981)\citenamefont {Goldin}, \citenamefont {Menikoff},\ and\ \citenamefont {Sharp}}]{Goldin_81}%
  \BibitemOpen
  \bibfield  {author} {\bibinfo {author} {\bibfnamefont {G.~A.}\ \bibnamefont {Goldin}}, \bibinfo {author} {\bibfnamefont {R.}~\bibnamefont {Menikoff}},\ and\ \bibinfo {author} {\bibfnamefont {D.~H.}\ \bibnamefont {Sharp}},\ }\bibfield  {title} {\bibinfo {title} {{Representations of a local current algebra in nonsimply connected space and the Aharonov–Bohm effect}},\ }\href {https://doi.org/10.1063/1.525110} {\bibfield  {journal} {\bibinfo  {journal} {Journal of Mathematical Physics}\ }\textbf {\bibinfo {volume} {22}},\ \bibinfo {pages} {1664} (\bibinfo {year} {1981})},\ \Eprint {https://arxiv.org/abs/https://pubs.aip.org/aip/jmp/article-pdf/22/8/1664/19078813/1664\_1\_online.pdf} {https://pubs.aip.org/aip/jmp/article-pdf/22/8/1664/19078813/1664\_1\_online.pdf} \BibitemShut {NoStop}%
\bibitem [{\citenamefont {Wilczek}(1982)}]{Wilczek_82}%
  \BibitemOpen
  \bibfield  {author} {\bibinfo {author} {\bibfnamefont {F.}~\bibnamefont {Wilczek}},\ }\bibfield  {title} {\bibinfo {title} {Quantum mechanics of fractional-spin particles},\ }\href {https://doi.org/10.1103/PhysRevLett.49.957} {\bibfield  {journal} {\bibinfo  {journal} {Phys. Rev. Lett.}\ }\textbf {\bibinfo {volume} {49}},\ \bibinfo {pages} {957} (\bibinfo {year} {1982})}\BibitemShut {NoStop}%
\bibitem [{\citenamefont {Einarsson}(1990)}]{Einarsson90}%
  \BibitemOpen
  \bibfield  {author} {\bibinfo {author} {\bibfnamefont {T.}~\bibnamefont {Einarsson}},\ }\bibfield  {title} {\bibinfo {title} {Fractional statistics on a torus},\ }\href {https://doi.org/10.1103/PhysRevLett.64.1995} {\bibfield  {journal} {\bibinfo  {journal} {Phys. Rev. Lett.}\ }\textbf {\bibinfo {volume} {64}},\ \bibinfo {pages} {1995} (\bibinfo {year} {1990})}\BibitemShut {NoStop}%
\bibitem [{\citenamefont {Kitaev}(2003)}]{Kitaev_2003}%
  \BibitemOpen
  \bibfield  {author} {\bibinfo {author} {\bibfnamefont {A.}~\bibnamefont {Kitaev}},\ }\bibfield  {title} {\bibinfo {title} {Fault-tolerant quantum computation by anyons},\ }\bibfield  {journal} {\bibinfo  {journal} {Annals of Physics}\ }\textbf {\bibinfo {volume} {303}},\ \href {https://doi.org/10.1016/s0003-4916(02)00018-0} {10.1016/s0003-4916(02)00018-0} (\bibinfo {year} {2003})\BibitemShut {NoStop}%
\bibitem [{\citenamefont {Freedman}\ \emph {et~al.}(2002)\citenamefont {Freedman}, \citenamefont {Larsen},\ and\ \citenamefont {Wang}}]{Freedman_2000gwh}%
  \BibitemOpen
  \bibfield  {author} {\bibinfo {author} {\bibfnamefont {M.~H.}\ \bibnamefont {Freedman}}, \bibinfo {author} {\bibfnamefont {M.}~\bibnamefont {Larsen}},\ and\ \bibinfo {author} {\bibfnamefont {Z.}~\bibnamefont {Wang}},\ }\bibfield  {title} {\bibinfo {title} {{A Modular Functor Which is Universal for Quantum Computation}},\ }\href {https://doi.org/10.1007/s002200200645} {\bibfield  {journal} {\bibinfo  {journal} {Commun. Math. Phys.}\ }\textbf {\bibinfo {volume} {227}},\ \bibinfo {pages} {605} (\bibinfo {year} {2002})},\ \Eprint {https://arxiv.org/abs/quant-ph/0001108} {arXiv:quant-ph/0001108} \BibitemShut {NoStop}%
\bibitem [{\citenamefont {Freedman}\ \emph {et~al.}(2006)\citenamefont {Freedman}, \citenamefont {Nayak},\ and\ \citenamefont {Walker}}]{Freedman_2006}%
  \BibitemOpen
  \bibfield  {author} {\bibinfo {author} {\bibfnamefont {M.}~\bibnamefont {Freedman}}, \bibinfo {author} {\bibfnamefont {C.}~\bibnamefont {Nayak}},\ and\ \bibinfo {author} {\bibfnamefont {K.}~\bibnamefont {Walker}},\ }\bibfield  {title} {\bibinfo {title} {Towards universal topological quantum computation in the $\nu=5/2$ fractional quantum hall state},\ }\bibfield  {journal} {\bibinfo  {journal} {Physical Review B}\ }\textbf {\bibinfo {volume} {73}},\ \href {https://doi.org/10.1103/physrevb.73.245307} {10.1103/physrevb.73.245307} (\bibinfo {year} {2006})\BibitemShut {NoStop}%
\bibitem [{\citenamefont {Nayak}\ \emph {et~al.}(2008)\citenamefont {Nayak}, \citenamefont {Simon}, \citenamefont {Stern}, \citenamefont {Freedman},\ and\ \citenamefont {Das~Sarma}}]{Nayak_08}%
  \BibitemOpen
  \bibfield  {author} {\bibinfo {author} {\bibfnamefont {C.}~\bibnamefont {Nayak}}, \bibinfo {author} {\bibfnamefont {S.~H.}\ \bibnamefont {Simon}}, \bibinfo {author} {\bibfnamefont {A.}~\bibnamefont {Stern}}, \bibinfo {author} {\bibfnamefont {M.}~\bibnamefont {Freedman}},\ and\ \bibinfo {author} {\bibfnamefont {S.}~\bibnamefont {Das~Sarma}},\ }\bibfield  {title} {\bibinfo {title} {Non-abelian anyons and topological quantum computation},\ }\href {https://doi.org/10.1103/RevModPhys.80.1083} {\bibfield  {journal} {\bibinfo  {journal} {Rev. Mod. Phys.}\ }\textbf {\bibinfo {volume} {80}},\ \bibinfo {pages} {1083} (\bibinfo {year} {2008})}\BibitemShut {NoStop}%
\bibitem [{\citenamefont {Terhal}(2015)}]{Terhal_15}%
  \BibitemOpen
  \bibfield  {author} {\bibinfo {author} {\bibfnamefont {B.~M.}\ \bibnamefont {Terhal}},\ }\bibfield  {title} {\bibinfo {title} {Quantum error correction for quantum memories},\ }\href {https://doi.org/10.1103/RevModPhys.87.307} {\bibfield  {journal} {\bibinfo  {journal} {Rev. Mod. Phys.}\ }\textbf {\bibinfo {volume} {87}},\ \bibinfo {pages} {307} (\bibinfo {year} {2015})}\BibitemShut {NoStop}%
\bibitem [{\citenamefont {Dennis}\ \emph {et~al.}(2002)\citenamefont {Dennis}, \citenamefont {Kitaev}, \citenamefont {Landahl},\ and\ \citenamefont {Preskill}}]{Dennis_2002}%
  \BibitemOpen
  \bibfield  {author} {\bibinfo {author} {\bibfnamefont {E.}~\bibnamefont {Dennis}}, \bibinfo {author} {\bibfnamefont {A.}~\bibnamefont {Kitaev}}, \bibinfo {author} {\bibfnamefont {A.}~\bibnamefont {Landahl}},\ and\ \bibinfo {author} {\bibfnamefont {J.}~\bibnamefont {Preskill}},\ }\bibfield  {title} {\bibinfo {title} {Topological quantum memory},\ }\href {https://doi.org/10.1063/1.1499754} {\bibfield  {journal} {\bibinfo  {journal} {Journal of Mathematical Physics}\ }\textbf {\bibinfo {volume} {43}},\ \bibinfo {pages} {4452–4505} (\bibinfo {year} {2002})}\BibitemShut {NoStop}%
\bibitem [{\citenamefont {Fan}\ \emph {et~al.}(2024)\citenamefont {Fan}, \citenamefont {Bao}, \citenamefont {Altman},\ and\ \citenamefont {Vishwanath}}]{fan2023diagnostics}%
  \BibitemOpen
  \bibfield  {author} {\bibinfo {author} {\bibfnamefont {R.}~\bibnamefont {Fan}}, \bibinfo {author} {\bibfnamefont {Y.}~\bibnamefont {Bao}}, \bibinfo {author} {\bibfnamefont {E.}~\bibnamefont {Altman}},\ and\ \bibinfo {author} {\bibfnamefont {A.}~\bibnamefont {Vishwanath}},\ }\bibfield  {title} {\bibinfo {title} {Diagnostics of mixed-state topological order and breakdown of quantum memory},\ }\href {https://doi.org/10.1103/PRXQuantum.5.020343} {\bibfield  {journal} {\bibinfo  {journal} {PRX Quantum}\ }\textbf {\bibinfo {volume} {5}},\ \bibinfo {pages} {020343} (\bibinfo {year} {2024})}\BibitemShut {NoStop}%
\bibitem [{\citenamefont {Bao}\ \emph {et~al.}(2023)\citenamefont {Bao}, \citenamefont {Fan}, \citenamefont {Vishwanath},\ and\ \citenamefont {Altman}}]{bao2023mixedstate}%
  \BibitemOpen
  \bibfield  {author} {\bibinfo {author} {\bibfnamefont {Y.}~\bibnamefont {Bao}}, \bibinfo {author} {\bibfnamefont {R.}~\bibnamefont {Fan}}, \bibinfo {author} {\bibfnamefont {A.}~\bibnamefont {Vishwanath}},\ and\ \bibinfo {author} {\bibfnamefont {E.}~\bibnamefont {Altman}},\ }\href@noop {} {\bibinfo {title} {Mixed-state topological order and the errorfield double formulation of decoherence-induced transitions}} (\bibinfo {year} {2023}),\ \Eprint {https://arxiv.org/abs/2301.05687} {arXiv:2301.05687 [quant-ph]} \BibitemShut {NoStop}%
\bibitem [{\citenamefont {Lee}\ \emph {et~al.}(2023)\citenamefont {Lee}, \citenamefont {Jian},\ and\ \citenamefont {Xu}}]{LeeYouXu2022}%
  \BibitemOpen
  \bibfield  {author} {\bibinfo {author} {\bibfnamefont {J.~Y.}\ \bibnamefont {Lee}}, \bibinfo {author} {\bibfnamefont {C.-M.}\ \bibnamefont {Jian}},\ and\ \bibinfo {author} {\bibfnamefont {C.}~\bibnamefont {Xu}},\ }\bibfield  {title} {\bibinfo {title} {Quantum criticality under decoherence or weak measurement},\ }\href {https://doi.org/10.1103/PRXQuantum.4.030317} {\bibfield  {journal} {\bibinfo  {journal} {PRX Quantum}\ }\textbf {\bibinfo {volume} {4}},\ \bibinfo {pages} {030317} (\bibinfo {year} {2023})}\BibitemShut {NoStop}%
\bibitem [{\citenamefont {Chen}\ and\ \citenamefont {Grover}(2024{\natexlab{a}})}]{chen2023separability}%
  \BibitemOpen
  \bibfield  {author} {\bibinfo {author} {\bibfnamefont {Y.-H.}\ \bibnamefont {Chen}}\ and\ \bibinfo {author} {\bibfnamefont {T.}~\bibnamefont {Grover}},\ }\bibfield  {title} {\bibinfo {title} {Separability transitions in topological states induced by local decoherence},\ }\href {https://doi.org/10.1103/PhysRevLett.132.170602} {\bibfield  {journal} {\bibinfo  {journal} {Phys. Rev. Lett.}\ }\textbf {\bibinfo {volume} {132}},\ \bibinfo {pages} {170602} (\bibinfo {year} {2024}{\natexlab{a}})}\BibitemShut {NoStop}%
\bibitem [{\citenamefont {{Sang}}\ \emph {et~al.}(2023)\citenamefont {{Sang}}, \citenamefont {{Zou}},\ and\ \citenamefont {{Hsieh}}}]{Renorm_QECC_23}%
  \BibitemOpen
  \bibfield  {author} {\bibinfo {author} {\bibfnamefont {S.}~\bibnamefont {{Sang}}}, \bibinfo {author} {\bibfnamefont {Y.}~\bibnamefont {{Zou}}},\ and\ \bibinfo {author} {\bibfnamefont {T.~H.}\ \bibnamefont {{Hsieh}}},\ }\bibfield  {title} {\bibinfo {title} {{Mixed-state Quantum Phases: Renormalization and Quantum Error Correction}},\ }\href {https://doi.org/10.48550/arXiv.2310.08639} {\bibfield  {journal} {\bibinfo  {journal} {arXiv e-prints}\ ,\ \bibinfo {eid} {arXiv:2310.08639}} (\bibinfo {year} {2023})},\ \Eprint {https://arxiv.org/abs/2310.08639} {arXiv:2310.08639 [quant-ph]} \BibitemShut {NoStop}%
\bibitem [{\citenamefont {Wang}\ \emph {et~al.}(2023)\citenamefont {Wang}, \citenamefont {Wu},\ and\ \citenamefont {Wang}}]{wang2023intrinsic}%
  \BibitemOpen
  \bibfield  {author} {\bibinfo {author} {\bibfnamefont {Z.}~\bibnamefont {Wang}}, \bibinfo {author} {\bibfnamefont {Z.}~\bibnamefont {Wu}},\ and\ \bibinfo {author} {\bibfnamefont {Z.}~\bibnamefont {Wang}},\ }\href@noop {} {\bibinfo {title} {Intrinsic mixed-state topological order without quantum memory}} (\bibinfo {year} {2023}),\ \Eprint {https://arxiv.org/abs/2307.13758} {arXiv:2307.13758 [quant-ph]} \BibitemShut {NoStop}%
\bibitem [{\citenamefont {{Li}}\ and\ \citenamefont {{Mong}}(2024)}]{Mong_24}%
  \BibitemOpen
  \bibfield  {author} {\bibinfo {author} {\bibfnamefont {Z.}~\bibnamefont {{Li}}}\ and\ \bibinfo {author} {\bibfnamefont {R.~S.~K.}\ \bibnamefont {{Mong}}},\ }\bibfield  {title} {\bibinfo {title} {{Replica topological order in quantum mixed states and quantum error correction}},\ }\href {https://doi.org/10.48550/arXiv.2402.09516} {\bibfield  {journal} {\bibinfo  {journal} {arXiv e-prints}\ ,\ \bibinfo {eid} {arXiv:2402.09516}} (\bibinfo {year} {2024})},\ \Eprint {https://arxiv.org/abs/2402.09516} {arXiv:2402.09516 [quant-ph]} \BibitemShut {NoStop}%
\bibitem [{\citenamefont {Ellison}\ and\ \citenamefont {Cheng}(2024)}]{ellison2024classificationmixedstatetopologicalorders}%
  \BibitemOpen
  \bibfield  {author} {\bibinfo {author} {\bibfnamefont {T.}~\bibnamefont {Ellison}}\ and\ \bibinfo {author} {\bibfnamefont {M.}~\bibnamefont {Cheng}},\ }\href {https://arxiv.org/abs/2405.02390} {\bibinfo {title} {Towards a classification of mixed-state topological orders in two dimensions}} (\bibinfo {year} {2024}),\ \Eprint {https://arxiv.org/abs/2405.02390} {arXiv:2405.02390 [cond-mat.str-el]} \BibitemShut {NoStop}%
\bibitem [{\citenamefont {{Sohal}}\ and\ \citenamefont {{Prem}}(2024)}]{sohal_24}%
  \BibitemOpen
  \bibfield  {author} {\bibinfo {author} {\bibfnamefont {R.}~\bibnamefont {{Sohal}}}\ and\ \bibinfo {author} {\bibfnamefont {A.}~\bibnamefont {{Prem}}},\ }\bibfield  {title} {\bibinfo {title} {{A Noisy Approach to Intrinsically Mixed-State Topological Order}},\ }\href {https://doi.org/10.48550/arXiv.2403.13879} {\bibfield  {journal} {\bibinfo  {journal} {arXiv e-prints}\ ,\ \bibinfo {eid} {arXiv:2403.13879}} (\bibinfo {year} {2024})},\ \Eprint {https://arxiv.org/abs/2403.13879} {arXiv:2403.13879 [cond-mat.str-el]} \BibitemShut {NoStop}%
\bibitem [{\citenamefont {Su}\ \emph {et~al.}(2024)\citenamefont {Su}, \citenamefont {Yang},\ and\ \citenamefont {Jian}}]{tapestry_24}%
  \BibitemOpen
  \bibfield  {author} {\bibinfo {author} {\bibfnamefont {K.}~\bibnamefont {Su}}, \bibinfo {author} {\bibfnamefont {Z.}~\bibnamefont {Yang}},\ and\ \bibinfo {author} {\bibfnamefont {C.-M.}\ \bibnamefont {Jian}},\ }\bibfield  {title} {\bibinfo {title} {Tapestry of dualities in decohered quantum error correction codes},\ }\href {https://doi.org/10.1103/PhysRevB.110.085158} {\bibfield  {journal} {\bibinfo  {journal} {Phys. Rev. B}\ }\textbf {\bibinfo {volume} {110}},\ \bibinfo {pages} {085158} (\bibinfo {year} {2024})}\BibitemShut {NoStop}%
\bibitem [{\citenamefont {Chen}\ and\ \citenamefont {Grover}(2024{\natexlab{b}})}]{chen2024unconventional}%
  \BibitemOpen
  \bibfield  {author} {\bibinfo {author} {\bibfnamefont {Y.-H.}\ \bibnamefont {Chen}}\ and\ \bibinfo {author} {\bibfnamefont {T.}~\bibnamefont {Grover}},\ }\href@noop {} {\bibinfo {title} {Unconventional topological mixed-state transition and critical phase induced by self-dual coherent errors}} (\bibinfo {year} {2024}{\natexlab{b}}),\ \Eprint {https://arxiv.org/abs/2403.06553} {arXiv:2403.06553 [quant-ph]} \BibitemShut {NoStop}%
\bibitem [{\citenamefont {{Hauser}}\ \emph {et~al.}(2024)\citenamefont {{Hauser}}, \citenamefont {{Bao}}, \citenamefont {{Sang}}, \citenamefont {{Lavasani}}, \citenamefont {{Agrawal}},\ and\ \citenamefont {{Fisher}}}]{Hauser_24}%
  \BibitemOpen
  \bibfield  {author} {\bibinfo {author} {\bibfnamefont {J.}~\bibnamefont {{Hauser}}}, \bibinfo {author} {\bibfnamefont {Y.}~\bibnamefont {{Bao}}}, \bibinfo {author} {\bibfnamefont {S.}~\bibnamefont {{Sang}}}, \bibinfo {author} {\bibfnamefont {A.}~\bibnamefont {{Lavasani}}}, \bibinfo {author} {\bibfnamefont {U.}~\bibnamefont {{Agrawal}}},\ and\ \bibinfo {author} {\bibfnamefont {M.~P.~A.}\ \bibnamefont {{Fisher}}},\ }\bibfield  {title} {\bibinfo {title} {{Information dynamics in decohered quantum memory with repeated syndrome measurements: a dual approach}},\ }\href {https://doi.org/10.48550/arXiv.2407.07882} {\bibfield  {journal} {\bibinfo  {journal} {arXiv e-prints}\ ,\ \bibinfo {eid} {arXiv:2407.07882}} (\bibinfo {year} {2024})},\ \Eprint {https://arxiv.org/abs/2407.07882} {arXiv:2407.07882 [quant-ph]} \BibitemShut {NoStop}%
\bibitem [{\citenamefont {Lyons}(2024)}]{lyons24}%
  \BibitemOpen
  \bibfield  {author} {\bibinfo {author} {\bibfnamefont {A.}~\bibnamefont {Lyons}},\ }\href {https://arxiv.org/abs/2403.03955} {\bibinfo {title} {Understanding stabilizer codes under local decoherence through a general statistical mechanics mapping}} (\bibinfo {year} {2024}),\ \Eprint {https://arxiv.org/abs/2403.03955} {arXiv:2403.03955 [quant-ph]} \BibitemShut {NoStop}%
\bibitem [{\citenamefont {Sala}\ \emph {et~al.}(2024{\natexlab{a}})\citenamefont {Sala}, \citenamefont {Gopalakrishnan}, \citenamefont {Oshikawa},\ and\ \citenamefont {You}}]{2024_sala_SSSB}%
  \BibitemOpen
  \bibfield  {author} {\bibinfo {author} {\bibfnamefont {P.}~\bibnamefont {Sala}}, \bibinfo {author} {\bibfnamefont {S.}~\bibnamefont {Gopalakrishnan}}, \bibinfo {author} {\bibfnamefont {M.}~\bibnamefont {Oshikawa}},\ and\ \bibinfo {author} {\bibfnamefont {Y.}~\bibnamefont {You}},\ }\href@noop {} {\bibinfo {title} {Spontaneous strong symmetry breaking in open systems: Purification perspective}} (\bibinfo {year} {2024}{\natexlab{a}}),\ \Eprint {https://arxiv.org/abs/2405.02402} {arXiv:2405.02402 [quant-ph]} \BibitemShut {NoStop}%
\bibitem [{\citenamefont {{Sang}}\ and\ \citenamefont {{Hsieh}}(2024)}]{Markov_length_24}%
  \BibitemOpen
  \bibfield  {author} {\bibinfo {author} {\bibfnamefont {S.}~\bibnamefont {{Sang}}}\ and\ \bibinfo {author} {\bibfnamefont {T.~H.}\ \bibnamefont {{Hsieh}}},\ }\bibfield  {title} {\bibinfo {title} {{Stability of mixed-state quantum phases via finite Markov length}},\ }\href {https://doi.org/10.48550/arXiv.2404.07251} {\bibfield  {journal} {\bibinfo  {journal} {arXiv e-prints}\ ,\ \bibinfo {eid} {arXiv:2404.07251}} (\bibinfo {year} {2024})},\ \Eprint {https://arxiv.org/abs/2404.07251} {arXiv:2404.07251 [quant-ph]} \BibitemShut {NoStop}%
\bibitem [{\citenamefont {{Lu}}(2024)}]{TshungCheng_24}%
  \BibitemOpen
  \bibfield  {author} {\bibinfo {author} {\bibfnamefont {T.-C.}\ \bibnamefont {{Lu}}},\ }\bibfield  {title} {\bibinfo {title} {{Disentangling transitions in topological order induced by boundary decoherence}},\ }\href {https://doi.org/10.48550/arXiv.2404.06514} {\bibfield  {journal} {\bibinfo  {journal} {arXiv e-prints}\ ,\ \bibinfo {eid} {arXiv:2404.06514}} (\bibinfo {year} {2024})},\ \Eprint {https://arxiv.org/abs/2404.06514} {arXiv:2404.06514 [quant-ph]} \BibitemShut {NoStop}%
\bibitem [{\citenamefont {Lee}(2024)}]{lee2024exactcalculationscoherentinformation}%
  \BibitemOpen
  \bibfield  {author} {\bibinfo {author} {\bibfnamefont {J.~Y.}\ \bibnamefont {Lee}},\ }\href {https://arxiv.org/abs/2402.16937} {\bibinfo {title} {Exact calculations of coherent information for toric codes under decoherence: Identifying the fundamental error threshold}} (\bibinfo {year} {2024}),\ \Eprint {https://arxiv.org/abs/2402.16937} {arXiv:2402.16937 [cond-mat.stat-mech]} \BibitemShut {NoStop}%
\bibitem [{\citenamefont {Goldin}\ \emph {et~al.}(1985)\citenamefont {Goldin}, \citenamefont {Menikoff},\ and\ \citenamefont {Sharp}}]{Goldin_85}%
  \BibitemOpen
  \bibfield  {author} {\bibinfo {author} {\bibfnamefont {G.~A.}\ \bibnamefont {Goldin}}, \bibinfo {author} {\bibfnamefont {R.}~\bibnamefont {Menikoff}},\ and\ \bibinfo {author} {\bibfnamefont {D.~H.}\ \bibnamefont {Sharp}},\ }\bibfield  {title} {\bibinfo {title} {Comments on "general theory for quantum statistics in two dimensions"},\ }\href {https://doi.org/10.1103/PhysRevLett.54.603} {\bibfield  {journal} {\bibinfo  {journal} {Phys. Rev. Lett.}\ }\textbf {\bibinfo {volume} {54}},\ \bibinfo {pages} {603} (\bibinfo {year} {1985})}\BibitemShut {NoStop}%
\bibitem [{\citenamefont {Wen}(1991)}]{Wen_91_FQH}%
  \BibitemOpen
  \bibfield  {author} {\bibinfo {author} {\bibfnamefont {X.~G.}\ \bibnamefont {Wen}},\ }\bibfield  {title} {\bibinfo {title} {Non-abelian statistics in the fractional quantum hall states},\ }\href {https://doi.org/10.1103/PhysRevLett.66.802} {\bibfield  {journal} {\bibinfo  {journal} {Phys. Rev. Lett.}\ }\textbf {\bibinfo {volume} {66}},\ \bibinfo {pages} {802} (\bibinfo {year} {1991})}\BibitemShut {NoStop}%
\bibitem [{\citenamefont {Moore}\ and\ \citenamefont {Read}(1991)}]{MOORE1991362}%
  \BibitemOpen
  \bibfield  {author} {\bibinfo {author} {\bibfnamefont {G.}~\bibnamefont {Moore}}\ and\ \bibinfo {author} {\bibfnamefont {N.}~\bibnamefont {Read}},\ }\bibfield  {title} {\bibinfo {title} {Nonabelions in the fractional quantum hall effect},\ }\href {https://doi.org/https://doi.org/10.1016/0550-3213(91)90407-O} {\bibfield  {journal} {\bibinfo  {journal} {Nuclear Physics B}\ }\textbf {\bibinfo {volume} {360}},\ \bibinfo {pages} {362} (\bibinfo {year} {1991})}\BibitemShut {NoStop}%
\bibitem [{\citenamefont {Moore}\ and\ \citenamefont {Seiberg}(1989)}]{Moore1989}%
  \BibitemOpen
  \bibfield  {author} {\bibinfo {author} {\bibfnamefont {G.}~\bibnamefont {Moore}}\ and\ \bibinfo {author} {\bibfnamefont {N.}~\bibnamefont {Seiberg}},\ }\bibfield  {title} {\bibinfo {title} {Classical and quantum conformal field theory},\ }\href {https://doi.org/10.1007/BF01238857} {\bibfield  {journal} {\bibinfo  {journal} {Communications in Mathematical Physics}\ }\textbf {\bibinfo {volume} {123}},\ \bibinfo {pages} {177} (\bibinfo {year} {1989})}\BibitemShut {NoStop}%
\bibitem [{\citenamefont {Mochon}(2003)}]{Mochon03}%
  \BibitemOpen
  \bibfield  {author} {\bibinfo {author} {\bibfnamefont {C.}~\bibnamefont {Mochon}},\ }\bibfield  {title} {\bibinfo {title} {Anyons from nonsolvable finite groups are sufficient for universal quantum computation},\ }\href {https://doi.org/10.1103/PhysRevA.67.022315} {\bibfield  {journal} {\bibinfo  {journal} {Phys. Rev. A}\ }\textbf {\bibinfo {volume} {67}},\ \bibinfo {pages} {022315} (\bibinfo {year} {2003})}\BibitemShut {NoStop}%
\bibitem [{\citenamefont {Mochon}(2004)}]{Mochon04}%
  \BibitemOpen
  \bibfield  {author} {\bibinfo {author} {\bibfnamefont {C.}~\bibnamefont {Mochon}},\ }\bibfield  {title} {\bibinfo {title} {Anyon computers with smaller groups},\ }\href {https://doi.org/10.1103/PhysRevA.69.032306} {\bibfield  {journal} {\bibinfo  {journal} {Phys. Rev. A}\ }\textbf {\bibinfo {volume} {69}},\ \bibinfo {pages} {032306} (\bibinfo {year} {2004})}\BibitemShut {NoStop}%
\bibitem [{\citenamefont {Wootton}\ \emph {et~al.}(2014)\citenamefont {Wootton}, \citenamefont {Burri}, \citenamefont {Iblisdir},\ and\ \citenamefont {Loss}}]{Wootton_14}%
  \BibitemOpen
  \bibfield  {author} {\bibinfo {author} {\bibfnamefont {J.~R.}\ \bibnamefont {Wootton}}, \bibinfo {author} {\bibfnamefont {J.}~\bibnamefont {Burri}}, \bibinfo {author} {\bibfnamefont {S.}~\bibnamefont {Iblisdir}},\ and\ \bibinfo {author} {\bibfnamefont {D.}~\bibnamefont {Loss}},\ }\bibfield  {title} {\bibinfo {title} {Error correction for non-abelian topological quantum computation},\ }\href {https://doi.org/10.1103/PhysRevX.4.011051} {\bibfield  {journal} {\bibinfo  {journal} {Phys. Rev. X}\ }\textbf {\bibinfo {volume} {4}},\ \bibinfo {pages} {011051} (\bibinfo {year} {2014})}\BibitemShut {NoStop}%
\bibitem [{\citenamefont {Brell}\ \emph {et~al.}(2014)\citenamefont {Brell}, \citenamefont {Burton}, \citenamefont {Dauphinais}, \citenamefont {Flammia},\ and\ \citenamefont {Poulin}}]{Brell_14}%
  \BibitemOpen
  \bibfield  {author} {\bibinfo {author} {\bibfnamefont {C.~G.}\ \bibnamefont {Brell}}, \bibinfo {author} {\bibfnamefont {S.}~\bibnamefont {Burton}}, \bibinfo {author} {\bibfnamefont {G.}~\bibnamefont {Dauphinais}}, \bibinfo {author} {\bibfnamefont {S.~T.}\ \bibnamefont {Flammia}},\ and\ \bibinfo {author} {\bibfnamefont {D.}~\bibnamefont {Poulin}},\ }\bibfield  {title} {\bibinfo {title} {Thermalization, error correction, and memory lifetime for ising anyon systems},\ }\href {https://doi.org/10.1103/PhysRevX.4.031058} {\bibfield  {journal} {\bibinfo  {journal} {Phys. Rev. X}\ }\textbf {\bibinfo {volume} {4}},\ \bibinfo {pages} {031058} (\bibinfo {year} {2014})}\BibitemShut {NoStop}%
\bibitem [{\citenamefont {Wootton}\ and\ \citenamefont {Hutter}(2016)}]{Wootton_16}%
  \BibitemOpen
  \bibfield  {author} {\bibinfo {author} {\bibfnamefont {J.~R.}\ \bibnamefont {Wootton}}\ and\ \bibinfo {author} {\bibfnamefont {A.}~\bibnamefont {Hutter}},\ }\bibfield  {title} {\bibinfo {title} {Active error correction for abelian and non-abelian anyons},\ }\href {https://doi.org/10.1103/PhysRevA.93.022318} {\bibfield  {journal} {\bibinfo  {journal} {Phys. Rev. A}\ }\textbf {\bibinfo {volume} {93}},\ \bibinfo {pages} {022318} (\bibinfo {year} {2016})}\BibitemShut {NoStop}%
\bibitem [{\citenamefont {Burton}\ \emph {et~al.}(2017)\citenamefont {Burton}, \citenamefont {Brell},\ and\ \citenamefont {Flammia}}]{Burton_2017}%
  \BibitemOpen
  \bibfield  {author} {\bibinfo {author} {\bibfnamefont {S.}~\bibnamefont {Burton}}, \bibinfo {author} {\bibfnamefont {C.~G.}\ \bibnamefont {Brell}},\ and\ \bibinfo {author} {\bibfnamefont {S.~T.}\ \bibnamefont {Flammia}},\ }\bibfield  {title} {\bibinfo {title} {Classical simulation of quantum error correction in a fibonacci anyon code},\ }\bibfield  {journal} {\bibinfo  {journal} {Physical Review A}\ }\textbf {\bibinfo {volume} {95}},\ \href {https://doi.org/10.1103/physreva.95.022309} {10.1103/physreva.95.022309} (\bibinfo {year} {2017})\BibitemShut {NoStop}%
\bibitem [{\citenamefont {Dauphinais}\ and\ \citenamefont {Poulin}(2017)}]{Dauphinais_2017}%
  \BibitemOpen
  \bibfield  {author} {\bibinfo {author} {\bibfnamefont {G.}~\bibnamefont {Dauphinais}}\ and\ \bibinfo {author} {\bibfnamefont {D.}~\bibnamefont {Poulin}},\ }\bibfield  {title} {\bibinfo {title} {Fault-tolerant quantum error correction for non-abelian anyons},\ }\href {https://doi.org/10.1007/s00220-017-2923-9} {\bibfield  {journal} {\bibinfo  {journal} {Communications in Mathematical Physics}\ }\textbf {\bibinfo {volume} {355}},\ \bibinfo {pages} {519–560} (\bibinfo {year} {2017})}\BibitemShut {NoStop}%
\bibitem [{\citenamefont {{Schotte}}\ \emph {et~al.}(2022)\citenamefont {{Schotte}}, \citenamefont {{Burgelman}},\ and\ \citenamefont {{Zhu}}}]{Schotte_22a}%
  \BibitemOpen
  \bibfield  {author} {\bibinfo {author} {\bibfnamefont {A.}~\bibnamefont {{Schotte}}}, \bibinfo {author} {\bibfnamefont {L.}~\bibnamefont {{Burgelman}}},\ and\ \bibinfo {author} {\bibfnamefont {G.}~\bibnamefont {{Zhu}}},\ }\bibfield  {title} {\bibinfo {title} {{Fault-tolerant error correction for a universal non-Abelian topological quantum computer at finite temperature}},\ }\href {https://doi.org/10.48550/arXiv.2301.00054} {\bibfield  {journal} {\bibinfo  {journal} {arXiv e-prints}\ ,\ \bibinfo {eid} {arXiv:2301.00054}} (\bibinfo {year} {2022})},\ \Eprint {https://arxiv.org/abs/2301.00054} {arXiv:2301.00054 [quant-ph]} \BibitemShut {NoStop}%
\bibitem [{\citenamefont {Schotte}\ \emph {et~al.}(2022)\citenamefont {Schotte}, \citenamefont {Zhu}, \citenamefont {Burgelman},\ and\ \citenamefont {Verstraete}}]{Schotte_22b}%
  \BibitemOpen
  \bibfield  {author} {\bibinfo {author} {\bibfnamefont {A.}~\bibnamefont {Schotte}}, \bibinfo {author} {\bibfnamefont {G.}~\bibnamefont {Zhu}}, \bibinfo {author} {\bibfnamefont {L.}~\bibnamefont {Burgelman}},\ and\ \bibinfo {author} {\bibfnamefont {F.}~\bibnamefont {Verstraete}},\ }\bibfield  {title} {\bibinfo {title} {Quantum error correction thresholds for the universal fibonacci turaev-viro code},\ }\href {https://doi.org/10.1103/PhysRevX.12.021012} {\bibfield  {journal} {\bibinfo  {journal} {Phys. Rev. X}\ }\textbf {\bibinfo {volume} {12}},\ \bibinfo {pages} {021012} (\bibinfo {year} {2022})}\BibitemShut {NoStop}%
\bibitem [{\citenamefont {Iqbal}\ \emph {et~al.}(2024)\citenamefont {Iqbal}, \citenamefont {Tantivasadakarn}, \citenamefont {Verresen}, \citenamefont {Campbell}, \citenamefont {Dreiling}, \citenamefont {Figgatt}, \citenamefont {Gaebler}, \citenamefont {Johansen}, \citenamefont {Mills}, \citenamefont {Moses}, \citenamefont {Pino}, \citenamefont {Ransford}, \citenamefont {Rowe}, \citenamefont {Siegfried}, \citenamefont {Stutz}, \citenamefont {Foss-Feig}, \citenamefont {Vishwanath},\ and\ \citenamefont {Dreyer}}]{iqbal2023creation}%
  \BibitemOpen
  \bibfield  {author} {\bibinfo {author} {\bibfnamefont {M.}~\bibnamefont {Iqbal}}, \bibinfo {author} {\bibfnamefont {N.}~\bibnamefont {Tantivasadakarn}}, \bibinfo {author} {\bibfnamefont {R.}~\bibnamefont {Verresen}}, \bibinfo {author} {\bibfnamefont {S.~L.}\ \bibnamefont {Campbell}}, \bibinfo {author} {\bibfnamefont {J.~M.}\ \bibnamefont {Dreiling}}, \bibinfo {author} {\bibfnamefont {C.}~\bibnamefont {Figgatt}}, \bibinfo {author} {\bibfnamefont {J.~P.}\ \bibnamefont {Gaebler}}, \bibinfo {author} {\bibfnamefont {J.}~\bibnamefont {Johansen}}, \bibinfo {author} {\bibfnamefont {M.}~\bibnamefont {Mills}}, \bibinfo {author} {\bibfnamefont {S.~A.}\ \bibnamefont {Moses}}, \bibinfo {author} {\bibfnamefont {J.~M.}\ \bibnamefont {Pino}}, \bibinfo {author} {\bibfnamefont {A.}~\bibnamefont {Ransford}}, \bibinfo {author} {\bibfnamefont {M.}~\bibnamefont {Rowe}}, \bibinfo {author} {\bibfnamefont {P.}~\bibnamefont {Siegfried}}, \bibinfo {author} {\bibfnamefont {R.~P.}\ \bibnamefont {Stutz}}, \bibinfo {author}
  {\bibfnamefont {M.}~\bibnamefont {Foss-Feig}}, \bibinfo {author} {\bibfnamefont {A.}~\bibnamefont {Vishwanath}},\ and\ \bibinfo {author} {\bibfnamefont {H.}~\bibnamefont {Dreyer}},\ }\bibfield  {title} {\bibinfo {title} {Non-abelian topological order and anyons on a trapped-ion processor},\ }\href {https://doi.org/10.1038/s41586-023-06934-4} {\bibfield  {journal} {\bibinfo  {journal} {Nature}\ }\textbf {\bibinfo {volume} {626}},\ \bibinfo {pages} {505} (\bibinfo {year} {2024})}\BibitemShut {NoStop}%
\bibitem [{\citenamefont {Minev}\ \emph {et~al.}(2024)\citenamefont {Minev}, \citenamefont {Najafi}, \citenamefont {Majumder}, \citenamefont {Wang}, \citenamefont {Stern}, \citenamefont {Kim}, \citenamefont {Jian},\ and\ \citenamefont {Zhu}}]{minev24}%
  \BibitemOpen
  \bibfield  {author} {\bibinfo {author} {\bibfnamefont {Z.~K.}\ \bibnamefont {Minev}}, \bibinfo {author} {\bibfnamefont {K.}~\bibnamefont {Najafi}}, \bibinfo {author} {\bibfnamefont {S.}~\bibnamefont {Majumder}}, \bibinfo {author} {\bibfnamefont {J.}~\bibnamefont {Wang}}, \bibinfo {author} {\bibfnamefont {A.}~\bibnamefont {Stern}}, \bibinfo {author} {\bibfnamefont {E.-A.}\ \bibnamefont {Kim}}, \bibinfo {author} {\bibfnamefont {C.-M.}\ \bibnamefont {Jian}},\ and\ \bibinfo {author} {\bibfnamefont {G.}~\bibnamefont {Zhu}},\ }\href {https://arxiv.org/abs/2406.12820} {\bibinfo {title} {Realizing string-net condensation: Fibonacci anyon braiding for universal gates and sampling chromatic polynomials}} (\bibinfo {year} {2024}),\ \Eprint {https://arxiv.org/abs/2406.12820} {arXiv:2406.12820 [quant-ph]} \BibitemShut {NoStop}%
\bibitem [{\citenamefont {Domany}\ \emph {et~al.}(1981)\citenamefont {Domany}, \citenamefont {Mukamel}, \citenamefont {Nienhuis},\ and\ \citenamefont {Schwimmer}}]{Nienhuis_81}%
  \BibitemOpen
  \bibfield  {author} {\bibinfo {author} {\bibfnamefont {E.}~\bibnamefont {Domany}}, \bibinfo {author} {\bibfnamefont {D.}~\bibnamefont {Mukamel}}, \bibinfo {author} {\bibfnamefont {B.}~\bibnamefont {Nienhuis}},\ and\ \bibinfo {author} {\bibfnamefont {A.}~\bibnamefont {Schwimmer}},\ }\bibfield  {title} {\bibinfo {title} {Duality relations and equivalences for models with o(n) and cubic symmetry},\ }\href {https://doi.org/https://doi.org/10.1016/0550-3213(81)90559-9} {\bibfield  {journal} {\bibinfo  {journal} {Nuclear Physics B}\ }\textbf {\bibinfo {volume} {190}},\ \bibinfo {pages} {279} (\bibinfo {year} {1981})}\BibitemShut {NoStop}%
\bibitem [{\citenamefont {Nienhuis}(1982)}]{Nienhuis_82}%
  \BibitemOpen
  \bibfield  {author} {\bibinfo {author} {\bibfnamefont {B.}~\bibnamefont {Nienhuis}},\ }\bibfield  {title} {\bibinfo {title} {Exact critical point and critical exponents of $\mathrm{O}(n)$ models in two dimensions},\ }\href {https://doi.org/10.1103/PhysRevLett.49.1062} {\bibfield  {journal} {\bibinfo  {journal} {Phys. Rev. Lett.}\ }\textbf {\bibinfo {volume} {49}},\ \bibinfo {pages} {1062} (\bibinfo {year} {1982})}\BibitemShut {NoStop}%
\bibitem [{\citenamefont {Peled}\ and\ \citenamefont {Spinka}(2019)}]{peled2019lectures}%
  \BibitemOpen
  \bibfield  {author} {\bibinfo {author} {\bibfnamefont {R.}~\bibnamefont {Peled}}\ and\ \bibinfo {author} {\bibfnamefont {Y.}~\bibnamefont {Spinka}},\ }\href@noop {} {\bibinfo {title} {Lectures on the spin and loop $o(n)$ models}} (\bibinfo {year} {2019}),\ \Eprint {https://arxiv.org/abs/1708.00058} {arXiv:1708.00058 [math-ph]} \BibitemShut {NoStop}%
\bibitem [{\citenamefont {Bais}\ and\ \citenamefont {Slingerland}(2009)}]{Bais09}%
  \BibitemOpen
  \bibfield  {author} {\bibinfo {author} {\bibfnamefont {F.~A.}\ \bibnamefont {Bais}}\ and\ \bibinfo {author} {\bibfnamefont {J.~K.}\ \bibnamefont {Slingerland}},\ }\bibfield  {title} {\bibinfo {title} {Condensate-induced transitions between topologically ordered phases},\ }\href {https://doi.org/10.1103/PhysRevB.79.045316} {\bibfield  {journal} {\bibinfo  {journal} {Phys. Rev. B}\ }\textbf {\bibinfo {volume} {79}},\ \bibinfo {pages} {045316} (\bibinfo {year} {2009})}\BibitemShut {NoStop}%
\bibitem [{\citenamefont {Burnell}(2018)}]{Burnell_2018}%
  \BibitemOpen
  \bibfield  {author} {\bibinfo {author} {\bibfnamefont {F.}~\bibnamefont {Burnell}},\ }\bibfield  {title} {\bibinfo {title} {Anyon condensation and its applications},\ }\href {https://doi.org/10.1146/annurev-conmatphys-033117-054154} {\bibfield  {journal} {\bibinfo  {journal} {Annual Review of Condensed Matter Physics}\ }\textbf {\bibinfo {volume} {9}},\ \bibinfo {pages} {307–327} (\bibinfo {year} {2018})}\BibitemShut {NoStop}%
\bibitem [{\citenamefont {Sala}\ \emph {et~al.}(2024{\natexlab{b}})\citenamefont {Sala}, \citenamefont {Alicea},\ and\ \citenamefont {Verresen}}]{long_paper}%
  \BibitemOpen
  \bibfield  {author} {\bibinfo {author} {\bibfnamefont {P.}~\bibnamefont {Sala}}, \bibinfo {author} {\bibfnamefont {J.}~\bibnamefont {Alicea}},\ and\ \bibinfo {author} {\bibfnamefont {R.}~\bibnamefont {Verresen}},\ }\href {https://arxiv.org/abs/2409.12948} {\bibinfo {title} {{Decoherence and wavefunction deformation of $D_4$ non-Abelian topological order [to appear in the same arXiv posting]}}} (\bibinfo {year} {2024}{\natexlab{b}}),\ \Eprint {https://arxiv.org/abs/2409.12948} {arXiv:2409.12948 [cond-mat.str-el]} \BibitemShut {NoStop}%
\bibitem [{\citenamefont {Deng}\ \emph {et~al.}(2007)\citenamefont {Deng}, \citenamefont {Garoni}, \citenamefont {Guo}, \citenamefont {Blöte},\ and\ \citenamefont {Sokal}}]{Deng_2007}%
  \BibitemOpen
  \bibfield  {author} {\bibinfo {author} {\bibfnamefont {Y.}~\bibnamefont {Deng}}, \bibinfo {author} {\bibfnamefont {T.~M.}\ \bibnamefont {Garoni}}, \bibinfo {author} {\bibfnamefont {W.}~\bibnamefont {Guo}}, \bibinfo {author} {\bibfnamefont {H.~W.~J.}\ \bibnamefont {Blöte}},\ and\ \bibinfo {author} {\bibfnamefont {A.~D.}\ \bibnamefont {Sokal}},\ }\bibfield  {title} {\bibinfo {title} {Cluster simulations of loop models on two-dimensional lattices},\ }\bibfield  {journal} {\bibinfo  {journal} {Physical Review Letters}\ }\textbf {\bibinfo {volume} {98}},\ \href {https://doi.org/10.1103/physrevlett.98.120601} {10.1103/physrevlett.98.120601} (\bibinfo {year} {2007})\BibitemShut {NoStop}%
\bibitem [{\citenamefont {ESSAM}\ and\ \citenamefont {FISHER}(1970)}]{Essam70}%
  \BibitemOpen
  \bibfield  {author} {\bibinfo {author} {\bibfnamefont {J.~W.}\ \bibnamefont {ESSAM}}\ and\ \bibinfo {author} {\bibfnamefont {M.~E.}\ \bibnamefont {FISHER}},\ }\bibfield  {title} {\bibinfo {title} {Some basic definitions in graph theory},\ }\href {https://doi.org/10.1103/RevModPhys.42.271} {\bibfield  {journal} {\bibinfo  {journal} {Rev. Mod. Phys.}\ }\textbf {\bibinfo {volume} {42}},\ \bibinfo {pages} {271} (\bibinfo {year} {1970})}\BibitemShut {NoStop}%
\bibitem [{\citenamefont {Chayes}\ \emph {et~al.}(2000)\citenamefont {Chayes}, \citenamefont {Pryadko},\ and\ \citenamefont {Shtengel}}]{Chayes_2000}%
  \BibitemOpen
  \bibfield  {author} {\bibinfo {author} {\bibfnamefont {L.}~\bibnamefont {Chayes}}, \bibinfo {author} {\bibfnamefont {L.~P.}\ \bibnamefont {Pryadko}},\ and\ \bibinfo {author} {\bibfnamefont {K.}~\bibnamefont {Shtengel}},\ }\bibfield  {title} {\bibinfo {title} {Intersecting loop models on $z^d$: rigorous results},\ }\href {https://doi.org/10.1016/s0550-3213(99)00780-4} {\bibfield  {journal} {\bibinfo  {journal} {Nuclear Physics B}\ }\textbf {\bibinfo {volume} {570}},\ \bibinfo {pages} {590–614} (\bibinfo {year} {2000})}\BibitemShut {NoStop}%
\bibitem [{SM()}]{SM}%
  \BibitemOpen
  \href@noop {} {}\bibinfo {note} {Supplemental Material containing: (1) A reformulation of O$(N)$ loop models as Ising-like models and various equivalent rewritings in App.A; (2) additional details of our analysis of the Kitaev honeycomb model under decoherence (App. B); (3) a brief review of the quantum double construction in App. C; and finally, App.D provides additional details of the calculation of the quantum fidelity between two mixed states at maximum error rate.}\BibitemShut {Stop}%
\bibitem [{\citenamefont {Nienhuis}(1984)}]{Nienhuis_84}%
  \BibitemOpen
  \bibfield  {author} {\bibinfo {author} {\bibfnamefont {B.}~\bibnamefont {Nienhuis}},\ }\bibfield  {title} {\bibinfo {title} {Critical behavior of two-dimensional spin models and charge asymmetry in the coulomb gas},\ }\href {https://doi.org/10.1007/BF01009437} {\bibfield  {journal} {\bibinfo  {journal} {Journal of Statistical Physics}\ }\textbf {\bibinfo {volume} {34}},\ \bibinfo {pages} {731 } (\bibinfo {year} {1984})}\BibitemShut {NoStop}%
\bibitem [{\citenamefont {Blöte}\ and\ \citenamefont {Nightingale}(1984)}]{BLOTE19841}%
  \BibitemOpen
  \bibfield  {author} {\bibinfo {author} {\bibfnamefont {H.}~\bibnamefont {Blöte}}\ and\ \bibinfo {author} {\bibfnamefont {M.}~\bibnamefont {Nightingale}},\ }\bibfield  {title} {\bibinfo {title} {The temperature exponent of the n-component cubic model},\ }\href {https://doi.org/https://doi.org/10.1016/0378-4371(84)90018-9} {\bibfield  {journal} {\bibinfo  {journal} {Physica A: Statistical Mechanics and its Applications}\ }\textbf {\bibinfo {volume} {129}},\ \bibinfo {pages} {1} (\bibinfo {year} {1984})}\BibitemShut {NoStop}%
\bibitem [{\citenamefont {Guo}\ \emph {et~al.}(2006)\citenamefont {Guo}, \citenamefont {Qian}, \citenamefont {Bl\"ote},\ and\ \citenamefont {Wu}}]{Guo06}%
  \BibitemOpen
  \bibfield  {author} {\bibinfo {author} {\bibfnamefont {W.}~\bibnamefont {Guo}}, \bibinfo {author} {\bibfnamefont {X.}~\bibnamefont {Qian}}, \bibinfo {author} {\bibfnamefont {H.~W.~J.}\ \bibnamefont {Bl\"ote}},\ and\ \bibinfo {author} {\bibfnamefont {F.~Y.}\ \bibnamefont {Wu}},\ }\bibfield  {title} {\bibinfo {title} {Critical line of an $n$-component cubic model},\ }\href {https://doi.org/10.1103/PhysRevE.73.026104} {\bibfield  {journal} {\bibinfo  {journal} {Phys. Rev. E}\ }\textbf {\bibinfo {volume} {73}},\ \bibinfo {pages} {026104} (\bibinfo {year} {2006})}\BibitemShut {NoStop}%
\bibitem [{\citenamefont {Freedman}(2003)}]{Freedman2003}%
  \BibitemOpen
  \bibfield  {author} {\bibinfo {author} {\bibfnamefont {M.~H.}\ \bibnamefont {Freedman}},\ }\bibfield  {title} {\bibinfo {title} {A magnetic model with a possible chern-simons phase},\ }\href {https://doi.org/10.1007/s00220-002-0785-1} {\bibfield  {journal} {\bibinfo  {journal} {Communications in Mathematical Physics}\ }\textbf {\bibinfo {volume} {234}},\ \bibinfo {pages} {129} (\bibinfo {year} {2003})}\BibitemShut {NoStop}%
\bibitem [{\citenamefont {Freedman}\ \emph {et~al.}(2004)\citenamefont {Freedman}, \citenamefont {Nayak}, \citenamefont {Shtengel}, \citenamefont {Walker},\ and\ \citenamefont {Wang}}]{Freedman04}%
  \BibitemOpen
  \bibfield  {author} {\bibinfo {author} {\bibfnamefont {M.}~\bibnamefont {Freedman}}, \bibinfo {author} {\bibfnamefont {C.}~\bibnamefont {Nayak}}, \bibinfo {author} {\bibfnamefont {K.}~\bibnamefont {Shtengel}}, \bibinfo {author} {\bibfnamefont {K.}~\bibnamefont {Walker}},\ and\ \bibinfo {author} {\bibfnamefont {Z.}~\bibnamefont {Wang}},\ }\bibfield  {title} {\bibinfo {title} {A class of p,t-invariant topological phases of interacting electrons},\ }\href {https://doi.org/https://doi.org/10.1016/j.aop.2004.01.006} {\bibfield  {journal} {\bibinfo  {journal} {Annals of Physics}\ }\textbf {\bibinfo {volume} {310}},\ \bibinfo {pages} {428} (\bibinfo {year} {2004})}\BibitemShut {NoStop}%
\bibitem [{\citenamefont {Freedman}\ \emph {et~al.}(2005{\natexlab{a}})\citenamefont {Freedman}, \citenamefont {Nayak},\ and\ \citenamefont {Shtengel}}]{Freedman05a}%
  \BibitemOpen
  \bibfield  {author} {\bibinfo {author} {\bibfnamefont {M.}~\bibnamefont {Freedman}}, \bibinfo {author} {\bibfnamefont {C.}~\bibnamefont {Nayak}},\ and\ \bibinfo {author} {\bibfnamefont {K.}~\bibnamefont {Shtengel}},\ }\bibfield  {title} {\bibinfo {title} {Line of critical points in $2+1$ dimensions: Quantum critical loop gases and non-abelian gauge theory},\ }\href {https://doi.org/10.1103/PhysRevLett.94.147205} {\bibfield  {journal} {\bibinfo  {journal} {Phys. Rev. Lett.}\ }\textbf {\bibinfo {volume} {94}},\ \bibinfo {pages} {147205} (\bibinfo {year} {2005}{\natexlab{a}})}\BibitemShut {NoStop}%
\bibitem [{\citenamefont {Freedman}\ \emph {et~al.}(2005{\natexlab{b}})\citenamefont {Freedman}, \citenamefont {Nayak},\ and\ \citenamefont {Shtengel}}]{Freedman05b}%
  \BibitemOpen
  \bibfield  {author} {\bibinfo {author} {\bibfnamefont {M.}~\bibnamefont {Freedman}}, \bibinfo {author} {\bibfnamefont {C.}~\bibnamefont {Nayak}},\ and\ \bibinfo {author} {\bibfnamefont {K.}~\bibnamefont {Shtengel}},\ }\bibfield  {title} {\bibinfo {title} {Extended hubbard model with ring exchange: A route to a non-abelian topological phase},\ }\href {https://doi.org/10.1103/PhysRevLett.94.066401} {\bibfield  {journal} {\bibinfo  {journal} {Phys. Rev. Lett.}\ }\textbf {\bibinfo {volume} {94}},\ \bibinfo {pages} {066401} (\bibinfo {year} {2005}{\natexlab{b}})}\BibitemShut {NoStop}%
\bibitem [{\citenamefont {Fidkowski}\ \emph {et~al.}(2006)\citenamefont {Fidkowski}, \citenamefont {Freedman}, \citenamefont {Nayak}, \citenamefont {Walker},\ and\ \citenamefont {Wang}}]{Fidkowski06}%
  \BibitemOpen
  \bibfield  {author} {\bibinfo {author} {\bibfnamefont {L.}~\bibnamefont {Fidkowski}}, \bibinfo {author} {\bibfnamefont {M.}~\bibnamefont {Freedman}}, \bibinfo {author} {\bibfnamefont {C.}~\bibnamefont {Nayak}}, \bibinfo {author} {\bibfnamefont {K.}~\bibnamefont {Walker}},\ and\ \bibinfo {author} {\bibfnamefont {Z.}~\bibnamefont {Wang}},\ }\href {https://arxiv.org/abs/cond-mat/0610583} {\bibinfo {title} {From string nets to nonabelions}} (\bibinfo {year} {2006}),\ \Eprint {https://arxiv.org/abs/cond-mat/0610583} {arXiv:cond-mat/0610583 [cond-mat.str-el]} \BibitemShut {NoStop}%
\bibitem [{\citenamefont {Fendley}(2008)}]{Fendley_2008}%
  \BibitemOpen
  \bibfield  {author} {\bibinfo {author} {\bibfnamefont {P.}~\bibnamefont {Fendley}},\ }\bibfield  {title} {\bibinfo {title} {Topological order from quantum loops and nets},\ }\href {https://doi.org/10.1016/j.aop.2008.04.011} {\bibfield  {journal} {\bibinfo  {journal} {Annals of Physics}\ }\textbf {\bibinfo {volume} {323}},\ \bibinfo {pages} {3113–3136} (\bibinfo {year} {2008})}\BibitemShut {NoStop}%
\bibitem [{\citenamefont {Troyer}\ \emph {et~al.}(2008)\citenamefont {Troyer}, \citenamefont {Trebst}, \citenamefont {Shtengel},\ and\ \citenamefont {Nayak}}]{Troyer08}%
  \BibitemOpen
  \bibfield  {author} {\bibinfo {author} {\bibfnamefont {M.}~\bibnamefont {Troyer}}, \bibinfo {author} {\bibfnamefont {S.}~\bibnamefont {Trebst}}, \bibinfo {author} {\bibfnamefont {K.}~\bibnamefont {Shtengel}},\ and\ \bibinfo {author} {\bibfnamefont {C.}~\bibnamefont {Nayak}},\ }\bibfield  {title} {\bibinfo {title} {Local interactions and non-abelian quantum loop gases},\ }\href {https://doi.org/10.1103/PhysRevLett.101.230401} {\bibfield  {journal} {\bibinfo  {journal} {Phys. Rev. Lett.}\ }\textbf {\bibinfo {volume} {101}},\ \bibinfo {pages} {230401} (\bibinfo {year} {2008})}\BibitemShut {NoStop}%
\bibitem [{\citenamefont {Fendley}\ \emph {et~al.}(2013)\citenamefont {Fendley}, \citenamefont {Isakov},\ and\ \citenamefont {Troyer}}]{Fendley13}%
  \BibitemOpen
  \bibfield  {author} {\bibinfo {author} {\bibfnamefont {P.}~\bibnamefont {Fendley}}, \bibinfo {author} {\bibfnamefont {S.~V.}\ \bibnamefont {Isakov}},\ and\ \bibinfo {author} {\bibfnamefont {M.}~\bibnamefont {Troyer}},\ }\bibfield  {title} {\bibinfo {title} {Fibonacci topological order from quantum nets},\ }\href {https://doi.org/10.1103/PhysRevLett.110.260408} {\bibfield  {journal} {\bibinfo  {journal} {Phys. Rev. Lett.}\ }\textbf {\bibinfo {volume} {110}},\ \bibinfo {pages} {260408} (\bibinfo {year} {2013})}\BibitemShut {NoStop}%
\bibitem [{\citenamefont {Calderbank}\ and\ \citenamefont {Shor}(1996)}]{CS_96}%
  \BibitemOpen
  \bibfield  {author} {\bibinfo {author} {\bibfnamefont {A.~R.}\ \bibnamefont {Calderbank}}\ and\ \bibinfo {author} {\bibfnamefont {P.~W.}\ \bibnamefont {Shor}},\ }\bibfield  {title} {\bibinfo {title} {Good quantum error-correcting codes exist},\ }\href {https://doi.org/10.1103/PhysRevA.54.1098} {\bibfield  {journal} {\bibinfo  {journal} {Phys. Rev. A}\ }\textbf {\bibinfo {volume} {54}},\ \bibinfo {pages} {1098} (\bibinfo {year} {1996})}\BibitemShut {NoStop}%
\bibitem [{S_9(1996)}]{S_96}%
  \BibitemOpen
  \href {https://doi.org/10.1098/rspa.1996.0136} {\bibfield  {journal} {\bibinfo  {journal} {Proceedings of the Royal Society of London. Series A: Mathematical, Physical and Engineering Sciences}\ }\textbf {\bibinfo {volume} {452}},\ \bibinfo {pages} {2551–2577} (\bibinfo {year} {1996})}\BibitemShut {NoStop}%
\bibitem [{\citenamefont {Nielsen}\ and\ \citenamefont {Chuang}(2010)}]{nielsen2010quantum}%
  \BibitemOpen
  \bibfield  {author} {\bibinfo {author} {\bibfnamefont {M.}~\bibnamefont {Nielsen}}\ and\ \bibinfo {author} {\bibfnamefont {I.}~\bibnamefont {Chuang}},\ }\href {https://books.google.com/books?id=-s4DEy7o-a0C} {\emph {\bibinfo {title} {Quantum Computation and Quantum Information: 10th Anniversary Edition}}}\ (\bibinfo  {publisher} {Cambridge University Press},\ \bibinfo {year} {2010})\BibitemShut {NoStop}%
\bibitem [{\citenamefont {Kitaev}(2006)}]{Kitaev_2006}%
  \BibitemOpen
  \bibfield  {author} {\bibinfo {author} {\bibfnamefont {A.}~\bibnamefont {Kitaev}},\ }\bibfield  {title} {\bibinfo {title} {Anyons in an exactly solved model and beyond},\ }\href {https://doi.org/10.1016/j.aop.2005.10.005} {\bibfield  {journal} {\bibinfo  {journal} {Annals of Physics}\ }\textbf {\bibinfo {volume} {321}},\ \bibinfo {pages} {2–111} (\bibinfo {year} {2006})}\BibitemShut {NoStop}%
\bibitem [{\citenamefont {Shi}\ \emph {et~al.}(2020)\citenamefont {Shi}, \citenamefont {Kato},\ and\ \citenamefont {Kim}}]{Shi_2020}%
  \BibitemOpen
  \bibfield  {author} {\bibinfo {author} {\bibfnamefont {B.}~\bibnamefont {Shi}}, \bibinfo {author} {\bibfnamefont {K.}~\bibnamefont {Kato}},\ and\ \bibinfo {author} {\bibfnamefont {I.~H.}\ \bibnamefont {Kim}},\ }\bibfield  {title} {\bibinfo {title} {Fusion rules from entanglement},\ }\href {https://doi.org/10.1016/j.aop.2020.168164} {\bibfield  {journal} {\bibinfo  {journal} {Annals of Physics}\ }\textbf {\bibinfo {volume} {418}},\ \bibinfo {pages} {168164} (\bibinfo {year} {2020})}\BibitemShut {NoStop}%
\bibitem [{\citenamefont {Preskill}(2004)}]{Preskill_LN}%
  \BibitemOpen
  \bibfield  {author} {\bibinfo {author} {\bibfnamefont {J.}~\bibnamefont {Preskill}},\ }\href {http://theory.caltech.edu/~preskill/ph219/topological.pdf} {\bibinfo {title} {Chapter 9. topological quantum computation}} (\bibinfo {year} {2004})\BibitemShut {NoStop}%
\bibitem [{\citenamefont {Shi}(2019)}]{Shi_2019}%
  \BibitemOpen
  \bibfield  {author} {\bibinfo {author} {\bibfnamefont {B.}~\bibnamefont {Shi}},\ }\bibfield  {title} {\bibinfo {title} {Seeing topological entanglement through the information convex},\ }\bibfield  {journal} {\bibinfo  {journal} {Physical Review Research}\ }\textbf {\bibinfo {volume} {1}},\ \href {https://doi.org/10.1103/physrevresearch.1.033048} {10.1103/physrevresearch.1.033048} (\bibinfo {year} {2019})\BibitemShut {NoStop}%
\bibitem [{\citenamefont {Aharonov}\ and\ \citenamefont {Bohm}(1959)}]{Aharonov59}%
  \BibitemOpen
  \bibfield  {author} {\bibinfo {author} {\bibfnamefont {Y.}~\bibnamefont {Aharonov}}\ and\ \bibinfo {author} {\bibfnamefont {D.}~\bibnamefont {Bohm}},\ }\bibfield  {title} {\bibinfo {title} {Significance of electromagnetic potentials in the quantum theory},\ }\href {https://doi.org/10.1103/PhysRev.115.485} {\bibfield  {journal} {\bibinfo  {journal} {Phys. Rev.}\ }\textbf {\bibinfo {volume} {115}},\ \bibinfo {pages} {485} (\bibinfo {year} {1959})}\BibitemShut {NoStop}%
\bibitem [{\citenamefont {Liu}\ \emph {et~al.}(2011)\citenamefont {Liu}, \citenamefont {Deng},\ and\ \citenamefont {Garoni}}]{Liu_2011}%
  \BibitemOpen
  \bibfield  {author} {\bibinfo {author} {\bibfnamefont {Q.}~\bibnamefont {Liu}}, \bibinfo {author} {\bibfnamefont {Y.}~\bibnamefont {Deng}},\ and\ \bibinfo {author} {\bibfnamefont {T.~M.}\ \bibnamefont {Garoni}},\ }\bibfield  {title} {\bibinfo {title} {Worm monte carlo study of the honeycomb-lattice loop model},\ }\href {https://doi.org/10.1016/j.nuclphysb.2011.01.003} {\bibfield  {journal} {\bibinfo  {journal} {Nuclear Physics B}\ }\textbf {\bibinfo {volume} {846}},\ \bibinfo {pages} {283–315} (\bibinfo {year} {2011})}\BibitemShut {NoStop}%
\bibitem [{\citenamefont {Alberti}\ and\ \citenamefont {Uhlmann}(1983)}]{Alberti1983}%
  \BibitemOpen
  \bibfield  {author} {\bibinfo {author} {\bibfnamefont {P.~M.}\ \bibnamefont {Alberti}}\ and\ \bibinfo {author} {\bibfnamefont {A.}~\bibnamefont {Uhlmann}},\ }\bibfield  {title} {\bibinfo {title} {Stochastic linear maps and transition probability},\ }\href {https://doi.org/10.1007/BF00419927} {\bibfield  {journal} {\bibinfo  {journal} {Letters in Mathematical Physics}\ }\textbf {\bibinfo {volume} {7}},\ \bibinfo {pages} {107 } (\bibinfo {year} {1983})}\BibitemShut {NoStop}%
\bibitem [{\citenamefont {Ashkin}\ and\ \citenamefont {Teller}(1943)}]{AT_model}%
  \BibitemOpen
  \bibfield  {author} {\bibinfo {author} {\bibfnamefont {J.}~\bibnamefont {Ashkin}}\ and\ \bibinfo {author} {\bibfnamefont {E.}~\bibnamefont {Teller}},\ }\bibfield  {title} {\bibinfo {title} {Statistics of two-dimensional lattices with four components},\ }\href {https://doi.org/10.1103/PhysRev.64.178} {\bibfield  {journal} {\bibinfo  {journal} {Phys. Rev.}\ }\textbf {\bibinfo {volume} {64}},\ \bibinfo {pages} {178} (\bibinfo {year} {1943})}\BibitemShut {NoStop}%
\bibitem [{\citenamefont {Mittag}\ and\ \citenamefont {Stephen}(1971)}]{selfdual_AT}%
  \BibitemOpen
  \bibfield  {author} {\bibinfo {author} {\bibfnamefont {L.}~\bibnamefont {Mittag}}\ and\ \bibinfo {author} {\bibfnamefont {M.~J.}\ \bibnamefont {Stephen}},\ }\bibfield  {title} {\bibinfo {title} {{Dual Transformations in Many‐Component Ising Models}},\ }\href {https://doi.org/10.1063/1.1665606} {\bibfield  {journal} {\bibinfo  {journal} {Journal of Mathematical Physics}\ }\textbf {\bibinfo {volume} {12}},\ \bibinfo {pages} {441} (\bibinfo {year} {1971})},\ \Eprint {https://arxiv.org/abs/https://pubs.aip.org/aip/jmp/article-pdf/12/3/441/19155845/441\_1\_online.pdf} {https://pubs.aip.org/aip/jmp/article-pdf/12/3/441/19155845/441\_1\_online.pdf} \BibitemShut {NoStop}%
\bibitem [{\citenamefont {Baxter}(1971)}]{six_vertex_model}%
  \BibitemOpen
  \bibfield  {author} {\bibinfo {author} {\bibfnamefont {R.~J.}\ \bibnamefont {Baxter}},\ }\bibfield  {title} {\bibinfo {title} {Generalized ferroelectric model on a square lattice},\ }\href {https://doi.org/https://doi.org/10.1002/sapm197150151} {\bibfield  {journal} {\bibinfo  {journal} {Studies in Applied Mathematics}\ }\textbf {\bibinfo {volume} {50}},\ \bibinfo {pages} {51} (\bibinfo {year} {1971})},\ \Eprint {https://arxiv.org/abs/https://onlinelibrary.wiley.com/doi/pdf/10.1002/sapm197150151} {https://onlinelibrary.wiley.com/doi/pdf/10.1002/sapm197150151} \BibitemShut {NoStop}%
\bibitem [{\citenamefont {Fan}(1972)}]{FAN1972136}%
  \BibitemOpen
  \bibfield  {author} {\bibinfo {author} {\bibfnamefont {C.}~\bibnamefont {Fan}},\ }\bibfield  {title} {\bibinfo {title} {On critical properties of the ashkin-teller model},\ }\href {https://doi.org/https://doi.org/10.1016/0375-9601(72)91051-1} {\bibfield  {journal} {\bibinfo  {journal} {Physics Letters A}\ }\textbf {\bibinfo {volume} {39}},\ \bibinfo {pages} {136} (\bibinfo {year} {1972})}\BibitemShut {NoStop}%
\bibitem [{\citenamefont {Glazman}\ and\ \citenamefont {Peled}(2023)}]{Glazman_2023}%
  \BibitemOpen
  \bibfield  {author} {\bibinfo {author} {\bibfnamefont {A.}~\bibnamefont {Glazman}}\ and\ \bibinfo {author} {\bibfnamefont {R.}~\bibnamefont {Peled}},\ }\bibfield  {title} {\bibinfo {title} {On the transition between the disordered and antiferroelectric phases of the 6-vertex model},\ }\bibfield  {journal} {\bibinfo  {journal} {Electronic Journal of Probability}\ }\textbf {\bibinfo {volume} {28}},\ \href {https://doi.org/10.1214/23-ejp980} {10.1214/23-ejp980} (\bibinfo {year} {2023})\BibitemShut {NoStop}%
\bibitem [{\citenamefont {Ikhlef}\ and\ \citenamefont {Rajabpour}(2012)}]{Ikhlef_2012}%
  \BibitemOpen
  \bibfield  {author} {\bibinfo {author} {\bibfnamefont {Y.}~\bibnamefont {Ikhlef}}\ and\ \bibinfo {author} {\bibfnamefont {M.~A.}\ \bibnamefont {Rajabpour}},\ }\bibfield  {title} {\bibinfo {title} {Spin interfaces in the ashkin–teller model and sle},\ }\href {https://doi.org/10.1088/1742-5468/2012/01/p01012} {\bibfield  {journal} {\bibinfo  {journal} {Journal of Statistical Mechanics: Theory and Experiment}\ }\textbf {\bibinfo {volume} {2012}},\ \bibinfo {pages} {P01012} (\bibinfo {year} {2012})}\BibitemShut {NoStop}%
\bibitem [{\citenamefont {Huang}\ \emph {et~al.}(2013)\citenamefont {Huang}, \citenamefont {Deng}, \citenamefont {Jacobsen},\ and\ \citenamefont {Salas}}]{Huang_2013}%
  \BibitemOpen
  \bibfield  {author} {\bibinfo {author} {\bibfnamefont {Y.}~\bibnamefont {Huang}}, \bibinfo {author} {\bibfnamefont {Y.}~\bibnamefont {Deng}}, \bibinfo {author} {\bibfnamefont {J.~L.}\ \bibnamefont {Jacobsen}},\ and\ \bibinfo {author} {\bibfnamefont {J.}~\bibnamefont {Salas}},\ }\bibfield  {title} {\bibinfo {title} {The hintermann–merlini–baxter–wu and the infinite-coupling-limit ashkin–teller models},\ }\href {https://doi.org/10.1016/j.nuclphysb.2012.11.015} {\bibfield  {journal} {\bibinfo  {journal} {Nuclear Physics B}\ }\textbf {\bibinfo {volume} {868}},\ \bibinfo {pages} {492–538} (\bibinfo {year} {2013})}\BibitemShut {NoStop}%
\bibitem [{\citenamefont {Nian}\ \emph {et~al.}(2022)\citenamefont {Nian}, \citenamefont {Yu},\ and\ \citenamefont {Ye}}]{nian2022nonunitary}%
  \BibitemOpen
  \bibfield  {author} {\bibinfo {author} {\bibfnamefont {J.}~\bibnamefont {Nian}}, \bibinfo {author} {\bibfnamefont {X.}~\bibnamefont {Yu}},\ and\ \bibinfo {author} {\bibfnamefont {J.}~\bibnamefont {Ye}},\ }\href@noop {} {\bibinfo {title} {A non-unitary conformal field theory approach to two-dimensional turbulence}} (\bibinfo {year} {2022}),\ \Eprint {https://arxiv.org/abs/2210.06762} {arXiv:2210.06762 [hep-th]} \BibitemShut {NoStop}%
\bibitem [{\citenamefont {Dias}\ and\ \citenamefont {Koenig}(2023)}]{dias2023classical}%
  \BibitemOpen
  \bibfield  {author} {\bibinfo {author} {\bibfnamefont {B.}~\bibnamefont {Dias}}\ and\ \bibinfo {author} {\bibfnamefont {R.}~\bibnamefont {Koenig}},\ }\href@noop {} {\bibinfo {title} {Classical simulation of non-gaussian fermionic circuits}} (\bibinfo {year} {2023}),\ \Eprint {https://arxiv.org/abs/2307.12912} {arXiv:2307.12912 [quant-ph]} \BibitemShut {NoStop}%
\bibitem [{\citenamefont {Selke}(2007)}]{Selke_2007}%
  \BibitemOpen
  \bibfield  {author} {\bibinfo {author} {\bibfnamefont {W.}~\bibnamefont {Selke}},\ }\bibfield  {title} {\bibinfo {title} {The critical binder cumulant for isotropic ising models on square and triangular lattices},\ }\href {https://doi.org/10.1088/1742-5468/2007/04/p04008} {\bibfield  {journal} {\bibinfo  {journal} {Journal of Statistical Mechanics: Theory and Experiment}\ }\textbf {\bibinfo {volume} {2007}},\ \bibinfo {pages} {P04008–P04008} (\bibinfo {year} {2007})}\BibitemShut {NoStop}%
\end{thebibliography}
\end{document}